\documentclass[10pt,journal,cspaper,compsoc]{IEEEtran}
\usepackage{color}

\ifCLASSOPTIONcompsoc
  \usepackage[nocompress]{cite}
\else
  \usepackage{cite}
\fi

\usepackage{graphicx}
\graphicspath{{Figs/},{Figs/reconstruction/}}

\usepackage[cmex10]{amsmath}

\usepackage{array}

\usepackage{mdwmath}
\usepackage{mdwtab}
\usepackage{subfig}
\usepackage{url}

\usepackage{multirow}
\usepackage{booktabs}
\usepackage{amssymb}
\usepackage{lscape}

\newcommand{\ra}[1]{\renewcommand{\arraystretch}{#1}}

\newtheorem{myalgo}{Algorithm}

\newcommand{\real}{\ensuremath{\mathbb{R}}}

\newcommand{\stwo}{\ensuremath{\mathbf{S}^2}}
\newcommand{\s}{\ensuremath{\mathbf{S}}}
\newcommand{\ltwo}{\ensuremath{\mathbb{L}^2}}

\newcommand{\Rn}[1]{\real^{#1}}
\newcommand{\rthree}{\Rn{3}}
\newcommand{\inner}[2]{\langle #1, #2 \rangle }

\newcommand{\norm}[1]{\left\lVert #1 \right\rVert}
\newcommand{\normtwo}[1]{\left\lVert #1 \right\rVert_{2}^2}
\newcommand{\Deps}{\frac{d}{d\epsilon}|_{\epsilon=0}}

\newcommand{\Abs}[1]{\left\vert #1 \right\vert}
\newcommand{\Space}[1]{\ensuremath{\mathcal{#1}}}

\newcommand{\noi}{\noindent}
\newcommand{\non}{\nonumber}

\newcommand{\laspace}{\;}

\newcommand{\eg}{\emph{e.g.\ }}
\newcommand{\ie}{\emph{i.e.\ }}

\newcommand{\etalnospace}{\emph{et al.}}

\newcommand{\eqcomma}{\laspace,}
\newcommand{\eqstop}{\laspace.}

\newcommand{\ntilde}{\tilde{n}}
\newcommand{\area}{\omega}

\usepackage{todonotes}

\DeclareGraphicsExtensions{.pdf}

\begin{document}
\bstctlcite{IEEEexample:BSTcontrol}

%


\title{Numerical Inversion of SRNF Maps for Elastic Shape Analysis of
Genus-Zero Surfaces}

\author{Hamid Laga, Qian Xie, Ian H. Jermyn, and Anuj Srivastava}

\markboth{Numerical Inversion of SRNF Maps}%
{Laga \MakeLowercase{\etalnospace}: SRNF Inversion}
%

\IEEEcompsoctitleabstractindextext{
\begin{abstract}
Recent developments in elastic shape analysis (ESA) are motivated by the fact that it provides comprehensive frameworks for simultaneous registration, deformation, and comparison of shapes. These methods achieve computational efficiency using certain square-root representations that transform invariant elastic metrics into Euclidean metrics, allowing for applications of standard algorithms and statistical tools. For analyzing shapes of embeddings of $\s^2$ in $\real^3$, Jermyn et al.~\cite{jermyn:2012} introduced square-root normal fields (SRNFs) that transformed an elastic metric, with desirable invariant properties, into the $\ltwo$ metric. These SRNFs are essentially surface normals scaled by square-roots of infinitesimal area elements. A critical need in shape analysis is to invert solutions (deformations, averages, modes of variations, etc) computed in the SRNF space, back to the original surface space for visualizations and inferences.
Due to the lack of theory for understanding SRNFs maps and their inverses, we take a numerical approach and derive an efficient multiresolution algorithm, based on solving an optimization problem in the surface space, that estimates surfaces corresponding to given SRNFs. This solution is found effective, even for complex shapes, e.g.  human bodies and animals, that undergo significant deformations including bending and stretching. Specifically, we use this inversion for computing elastic shape deformations, transferring  deformations, summarizing shapes, and for finding modes of variability in a given collection, while simultaneously registering the surfaces. We demonstrate  the proposed algorithms using a statistical analysis of human body shapes, classification of generic surfaces and analysis of brain structures.
\end{abstract}

}

\maketitle

\IEEEdisplaynotcompsoctitleabstractindextext

%
\IEEEpeerreviewmaketitle


\section{Introduction}
\label{sec:introduction} 

Shape analysis is an important area of research with a wide variety of applications. The need for shape analysis arises in many branches of science, including anatomy, bioinformatics, computer graphics, and 3D printing and prototyping. A comprehensive method for shape analysis provides: (i) a {\bf shape metric}, for pairwise quantification of shape differences between objects; (ii) a {\bf shape summary}, a compact representation of a class of objects in terms of the  center (mean or median) and covariance of their shapes; and (iii) {\bf modes of variability}, the dominant modes of variation around the center, analogous to PCA, but for modeling shape variability.

Although shape analysis is a broad area that potentially involves comparisons of objects with different topologies as well as different geometries, we will restrict ourselves to geometric variability by choosing a fixed topology; specifically, we focus on shapes of genus-0 surfaces. This is still a very rich class of shapes with tremendous variability, and applications in many scientific disciplines. Hereafter, we will highlight the growing field of elastic shape analysis of genus-0 surfaces, and isolate an important problem that becomes the focus of this paper.

\subsection{Previous Methods}

Due to the importance of shape analysis, there is a large body of previous  work addressing it. However, most of this work shares a common drawback. Generally speaking, there are two subproblems of interest when comparing two shapes: {\bf registration}  and {\bf deformation}. Registration deals with complete or partial matching of points across objects, \ie deciding which point on one object matches which point on the other, while deformation is concerned with finding a continuous sequence of shapes starting from one shape and ending at the other. Most existing techniques treat registration and deformation as independent problems, despite their obvious interdependence; for example, the optimality of the solutions to each of the subproblems may be governed by different objective functions or metrics. We now discuss some examples of previous work, pointing out where this occurs.

A prominent statistical shape analysis framework, pioneered by Kendall's school~\cite{dryden:1998,allen:2003}, works with point sets that are already registered, and focuses only on deformations. Approaches such as medial surfaces~\cite{bouix:2001,gorczowski:2010} and  level sets~\cite{osher:2003} either presume registration, or solve for it using some independent pre-processing criterion such as MDL~\cite{davies:2010}. Kilian~\etalnospace~\cite{Kilian:2007} represent surfaces by discrete triangulated meshes and compute geodesic paths (deformations) between them, while assuming that the meshes are registered. Heeren~\etalnospace~\cite{heeren:2012} propose a method for computing geodesic-based deformations of thin shell shapes, with extensions for computing summary statistics in the shell space~\cite{zhang2015shell}, but with known registration.

Some other papers solve for registration~\cite{oliver:2011,zhang:2008} while ignoring deformation. Examples include Windheuser~\etalnospace~\cite{windheuser:2011}, who solves a dense registration problem, but uses linear interpolation between registered points in $\rthree$ to form deformations. Techniques such as SPHARM or SPHARM-PDM~\cite{brechbuhler:1995,styner:2006} seek uniform sampling on the domain to address the registration issue. This is a major restriction, as it limits registration of corresponding features across surfaces. In conclusion, a majority of the papers in the literature treat the registration and deformation problems in a disjoint fashion, each with their own optimality criteria. Due to this disconnection, the overall shape analysis pipeline becomes suboptimal.

\subsection{Elastic Shape Analysis and SRNFs}

A more comprehensive approach to shape analysis is to perform registration and deformation in a single {\bf unified} fashion. Elastic shape analysis (ESA) is a Riemannian approach that accomplishes exactly that, for objects  such as curves and surfaces. In ESA, one identifies an appropriate representation space for parameterized versions of the objects, and endows it with a Riemannian metric.\footnote{We refer the reader to
\url{http://stat.fsu.edu/~anuj/CVPR_Tutorial/} for a comprehensive tutorial on differential geometry methods in shape analysis.} The metric allows the definition of a \emph{geodesic}, \ie a shortest path, between two objects, which is by definition the optimal deformation under the metric. Registration is handled by finding the shortest path, not between the original two parameterized objects, but between all possible reparameterizations of those objects. This simultaneously identifies the optimal registration of the objects, and renders the resulting registration and deformation independent of the original parameterizations. The result is therefore a registration and deformation of the corresponding \emph{geometric} (\ie unparameterized) objects. The optimization over all possible reparameterizations can alternatively be viewed as finding a geodesic in a quotient of the original space by the group of reparameterizations, using a Riemannian metric derived from that on the original space. One can also choose to introduce further invariances to, for example, translations and rotations, thereby incorporating alignment into the deformation.

Central to this program is the definition of a Riemannian metric on the space of parameterized objects that is preserved by the action of the relevant transformations, typically reparameterizations, translations, rotations, and perhaps scale. Key to its success in practical terms is the definition of a Riemannian metric that `measures' the types of shape changes that are important, and that renders feasible computationally the corresponding geodesic calculations.

These goals have been achieved in the shape analysis of {\bf curves} by using a particular member of the family of {\em elastic metrics}, in conjunction with a representation called the {\em square-root velocity function (SRVF)}. The form of the elastic metric is extremely simple when expressed in terms of the SRVF: it becomes the $\ltwo$ metric~\cite{srivastava-PAMI:2011}. This greatly facilitates computations, enabling sophisticated statistical analyses that require many geodesic calculations; consequently, this method has been used in many practical problems~\cite{laga2014landmark,laga2012riemannian}. Critical to its utility is the fact that the mapping from the space of curves to the SRVF space is a bijection (up to a translation). Solutions found in SRVF space using the $\ltwo$ metric can be uniquely mapped back to the original curve space, using an analytical expression. This is significantly more efficient than performing analysis in the curve space itself.

To analyze genus-0 \textbf{surfaces}, Kurtek~\etalnospace~\cite{kurtek:2010,kurtek:2012} introduced the {\em square-root map} (SRM). Let $f:\stwo \to \rthree$ be a parameterized surface, and let $\Space{F}$ be the space of all such smooth mappings. Suppose $\stwo$ is parameterized by the pair $s \equiv (u,v)$ for $s\in \stwo$, then the SRM of $f$ is defined to be $q(s) = \Abs{n(s)}^{1/2} f(s)$, where $n(s) = f_u(s)\times f_v(s)$ is the unnormalized normal to the surface at $f(s)$. The metric is then taken to be the $\ltwo$ metric in SRM space. Unfortunately, this metric has several limitations, including that the metric distance between two shapes changes when they are both translated by the same amount.

Jermyn~\etalnospace~\cite{jermyn:2012} addressed these issues by defining a new representation termed the {\em square-root normal field} (SRNF). A parameterized surface $f \mapsto Q(f)$, where $Q(f)(s) \equiv n(s) /\Abs{n(s)}^{1/2}$. The authors showed that the Euclidean distance between two points in the space of SRNFs is equivalent to geodesic distance under an elastic metric on surfaces~\cite{jermyn:2012}, \ie the metric is interpretable as the weighted sum of the bending and stretching needed to deform the corresponding surfaces into one another. Thus, as in the case of the SRVF for curves, equipping the space of SRNFs, $\Space{Q}$, with the $\ltwo$ metric means that many computations, \eg those for the mean, covariance, and PCA, are greatly simplified.

We add some details to explain these ideas further. Let $\Gamma$ be the group of all orientation-preserving diffeomorphisms of $\stwo$, and $SO(3)$ be the group of all rotations in $\real^3$. The product group ${\cal G} \equiv SO(3) \times \Gamma$ helps align and register surfaces as follows. For any two parameterized surfaces $f_1, f_2 \in {\cal F}$, and $s \in \stwo$, the points $f_1(s)$ and $f_2(s)$ are considered paired or matched originally. However, for any  $(O, \gamma) \in {\cal G}$, the surface $O (f _2 \circ \gamma)$ denotes a rotated and re-parameterized version of $f_2$, and now $f_1(s)$ is paired with $Of_2(\gamma(s))$. Thus, the elements of ${\cal G}$ can be used to control the (rotational) alignment and registration of surfaces. The SRNF of the transformed surface is given by $O (Q(f), \gamma)$ where $(q, \gamma)(s) \equiv q(\gamma(s)) \sqrt{J_{\gamma}(s)}$ and $J_{\gamma}(s)$ denotes the determinant of the Jacobian of $\gamma$ at $s$. The most important property of SRNFs is that, for any $f_1, f_2 \in {\cal F}$ and $(O, \gamma) \in {\cal G}$, we have:
	\begin{equation}
		\| O(Q(f_1), \gamma) - O(Q(f_2), \gamma)\| = \| Q(f_1) - Q(f_2)\|
		\eqcomma
	\end{equation}

\noi where $\| \cdot \|$ denotes the $\ltwo$ norm. In other words, the group ${\cal G}$ acts by isometries on the $\ltwo$ metric in $\Space{Q}$, as required. Note that the SRNF representation is translation invariant by definition. As stated above, the $\ltwo$ metric on $\Space{Q}$ corresponds to a partial elastic Riemannian metric on $\Space{F}$~\cite{jermyn:2012}.

$\Space{Q}$ is often called {\it pre-shape} space, since many elements correspond to the same geometric object. For instance, $Q(f)$ and $O(Q(f), \gamma)$ are two different elements of $\Space{Q}$, but represent rotated versions of the same surface. To facilitate ESA, these representations are identified using an equivalence relation, each shape being represented by an orbit under ${\cal G}$:
	\begin{equation}
		[Q(f)] = \{ O(Q(f), \gamma) | (O, \gamma) \in {\cal G} \}
        \eqstop
	\end{equation}

\noi The set of all such orbits is the quotient space ${\cal S}  = Q({\cal F})/{\cal G}$, also called {\it shape space}. The computation of geodesics in ${\cal S}$ thereby corresponds to joint registration and deformation of surfaces.

\subsection{Open Issues and Problem Motivation}

The SRNF is very promising as a representation of surfaces for ESA. If geodesics, mean shapes, PCA, etc.\ could be computed in $\Space{Q}$ under the $\ltwo$ metric, and then mapped back to $\Space{F}$, just as is possible in the case of curves using the SRVF, there would be large gains in computational efficiency with respect to \eg~\cite{jermyn:2012,xie-iccv:2013,zhang2015shell}.  Unfortunately, the SRNF shares one difficulty with the SRM, and that is the problem of inversion. First, unlike the curve case, there is no analytical expression for $Q^{-1}$ for arbitrary points in ${\cal F}$. Moreover, the injectivity and surjectivity of $Q$ remain to be determined, meaning that for a given $q\in\Space{Q}$, there may be no $f\in\Space{F}$ such that $Q(f) = q$, and if such an $f$ does exist, it may not be unique.\footnote{Note that it is not simply a case of applying Bonnet's theorem, because in addition to $dn$, the second fundamental form involves the derivative $df$, which is the quantity we are trying to find.} If one cannot invert the representation, it is not clear how to transfer geodesics and statistical analyses conducted in $\Space{Q}$ back to $\Space{F}$, thereby removing one of the main motivations for introducing the SRNF.  One can always pull the $\ltwo$ metric back to $\Space{F}$ under $Q$ and perform computations there, as in ~\cite{xie-iccv:2013,tumpach2015gauge}, but this is computationally expensive, and rather defeats the purpose of having an $\ltwo$ metric in the first place.

In this paper, we address this problem by developing a method that, given $q\in\Space{Q}$, finds an $f\in\Space{F}$ such that $Q(f) = q$, if one exists, or an $f$ whose image $Q(f)$ is the closest, in the elastic metric, to $q$ if it does not. We achieve this by formulating SRNF inversion as an optimization problem: find an element $f \in \Space{F}$ whose image $Q(f)$ is as close as possible to the given $q\in \Space{Q}$ under the $\ltwo$ norm. We then propose an efficient numerical procedure to solve this problem. In particular, we show that by carefully engineering an orthonormal basis of $\Space{F}$, and combining it with a spherical-wavelet based multiresolution and multiscale representation of the elements of $\Space{F}$, SRNF maps can be inverted with high accuracy and efficiency, with robustness to local minima, and with low computational complexity.  We also show that an \emph{analytical} solution to the inversion problem exists  for star-shaped surfaces, \ie those  whose enclosed volumes are star domains, a large family of surfaces found in many real problems.

Using the proposed methods for inverting the SRNF map, we are able to compute geodesics, transfer deformations, perform statistical analyses, and synthesize random shapes sampled from the space of genus-0 surfaces, in a very efficient manner:  the {\bf computational cost is reduced by  an order of magnitude} compared to previous work. We demonstrate the utility of the proposed tools using complex genus-0 surfaces exhibiting complex bending and stretching deformations, taken from the TOSCA dataset, the SHREC07 watertight database~\cite{giorgi2007shape}, and the human shape database from~\cite{hasler2009statistical}.

The rest of the paper is organized as follows. Section~\ref{sec:method} describes the inversion problem, its formulation as an optimization problem, and its numerical solution, as well as the analytic solution in the case of star-shaped surfaces. Section~\ref{sec:analysis} lists some statistical tasks to demonstrate the ESA framework, and compares the algorithms for these tasks under previous and the proposed methods. Section~\ref{sec:results} presents experimental results on computing geodesics, transferring deformations, computing statistical summaries, and performing classification of generic 3D shapes and of subjects with Attention Deficit Hyperactivity Disorder using the shapes of brain subcortical structures.

\section{The Inversion Problem}
\label{sec:method}

In order to take full advantage of the simplicity of the metric in the SRNF representation, we need to be able to find, given $q \in \Space{Q}$, a surface $f$ such that $Q(f) = q$. This is the inversion problem. Let $g(s)$ denote the $2 \times 2$ matrix $\left[ \begin{array}{cc}
\inner{f_u}{f_u} & \inner{f_u}{f_v} \\
\inner{f_u}{f_v}  & \inner{f_v}{f_v} \end{array} \right]$ at each $s \in \stwo$. The inversion problem
can be formulated in several different ways:
\begin{enumerate}
\item

Given a map $\ntilde: \stwo\to\stwo$ and a function $\area: \stwo\to\real_{+}$, when are these the Gauss map $\ntilde = n/\Abs{n}$ and area form $\area = \Abs{n}^{1/2} = \det(g(s))^{1/2}$ of an immersion $f: \stwo\to\rthree$? If such an $f$ exists, is it unique(up to translation)

\item

Given $q :\stwo\to\rthree$, does the differential equation $n = \Abs{q}q$ have a (global) solution, and is it unique (up to translation)~\cite{Michor:2013}?

\end{enumerate}

In the first case, it is known that if we are given ($\tilde{n}, g)$, rather than just $(\tilde{n}, \area)$, then $f$, if it exists, can be reconstructed up to translation and rotation~\cite{abe:1975,eschenburg:2010}. This is also true when we are given the conformal class of $g$~\cite{hoffman:1985}, \ie we know $(\tilde{n}, \tilde{g})$, where $\tilde{g}(s)  = g(s) a(s)$, and $a(s) \in \real_+$. Unfortunately, little seems to be known about the nature of the solution in the case at hand. The problem was stated by Arnold in 1990~\cite{Arnold1990-jm}, but in a 2004 book listing Arnold's problems and their subsequent elucidation or solution~\cite{Arnold2004-bz}, the same problem still appears, without commentary. 

In the second case, a global solution clearly does not always exist, at least in degenerate cases (if $g$ is constant, for example~\cite{Michor:2013}). The discussion in~\cite{Michor:2013} does lead to the conclusion that the equation is locally solvable, but this is not sufficient for us: we require a global solution. Even if a global solution does exist, it may not be unique. In degenerate cases, any flat area of the surface could be subjected to an area-preserving diffeomorphism, while for open surfaces, one may find distinct, non-degenerate cases that share the same Gauss map and area form~\cite{Klassen:2011}. If the solution is to be unique, then, it can be so only for closed surfaces.

We will manage these difficult questions about the injectivity and surjectivity of $Q$ by formulating inversion as an optimization problem. We define an energy function $E_{0}:\mathcal{F}\rightarrow \real_{\geq 0}$ by
	\begin{equation}
		E_{0}(f;q) = \min_{(O, \gamma) \in {\cal G}} 					\normtwo{Q(f) - O(q, \gamma)}
		\eqcomma
		\label{eq:energy_to_minimize}
	\end{equation}

\noi where $(q, \gamma) = (q \circ \gamma) \sqrt{J_{\gamma}}$, and $J_{\gamma}$ is the determinant of the Jacobian of $\gamma$.  Finding an $f \in \Space{F}$ such that $Q(f)=q$ is then equivalent to seeking zeros of $E_{0}$. We denote by $f^*$ an element of $\Space{F}$ satisfying $E_{0}(f^{*}; q) = \min_{f\in\Space{F}} E_{0}(f;q)$.

The minimization will be performed using a gradient descent approach. From a computational point of view, it will be easier to deal with deformations of a surface, rather than the surface itself. We therefore set $f = f_0 + w$, where $f_0$ denotes the current estimate of $f^*$, and $w$ is a deformation of $f_0$. Then, we minimize
	\begin{equation}
		E(w;q) = \min_{O, \gamma} \normtwo{Q(f_0 + w) -O(q, \gamma)}
		\eqcomma
		\label{eq:energy_2}
	\end{equation}

\noi with respect to $w$. One can view $f_0$ as an initial guess of the solution or a known surface with shape similar to the one being estimated. If no initial guess is possible, one can initialize $f_0$ as a unit sphere or even set $f_0 = 0$.

Formulating the inversion problem as energy minimization has two purposes. First, by starting gradient descent for similar $q, q'\in \Space{Q}$ from similar initial points $f_{0}, f_{0}'\in\Space{F}$, we can help to ensure that the inverted surfaces we find are, if not uniquely determined, at least consistent with each other. This will be important when we wish to invert, for example, whole geodesics. Second, because there is no guarantee that $Q$ is surjective, and, equally, no guarantee that a geodesic between two points in the image of $Q$ lies wholly in the image of $Q$, we have to be able to `invert' points in $\Space{Q}$ that do not lie in the image of $Q$. Energy minimization achieves this by finding an $f^{*}$ whose image under $Q$ is as close as possible to the point whose inverse is desired. This process is equivalent to projecting points in $\Space{Q}$ to $Q(\Space{F})$.

\subsection{Numerical Optimization}
\label{sec:numerical_optimization}

We solve the optimization problem of Eqn.~\eqref{eq:energy_2} by iterating over the following  two steps:
\begin{enumerate}

\item  Given $f_0$ and $w$,  find a rotation $O$ and a diffeomorphism
    $\gamma$ such that the energy $E$ is minimized. Note this step is not
    strictly necessary, since $\Space{F}$ is closed under the action of
    these groups. Separating out these transformations, however, enables
    large `leaps' in $\Space{F}$, helping the algorithm to search more
    globally.

\item Given a fixed $O$ and $\gamma$, optimize $E$ with respect to $w$
    using gradient descent.
\end{enumerate}

\noindent The first step is solved using the approach described in~\cite{jermyn:2012}. That is, we find the best rotation $O$ using SVD, given a fixed $\gamma$. Then we find the best $\gamma$ by a search over the space of all diffeomorphisms using a gradient descent approach, see~\cite{jermyn:2012}. In practice, however, we found that restricting the search over $\Gamma$ to the space of rigid rotations, $\Gamma_0  \equiv SO(3)$, is sufficient to achieve good reconstruction accuracy while being computationally more efficient. In other words, we restrict to only the rigid re-parameterizations of a surface during the inversion process. (We clarify that we still use the full $\Gamma$ for the registration step when computing geodesics between shapes.)

To minimize $E$ with respect to $w$ using a gradient descent approach, we need the directional derivative of $E$. Since $\Space{F}$ is an infinite-dimensional vector space, we will approximate the derivative using a finite basis for $\Space{F}$. We therefore set $w =\sum_{b \in {\cal B}} \alpha_b b$, with $\alpha_b \in \real$, and where ${\cal B}$ forms an orthonormal basis of $\Space{F}$. The directional derivative of $E$ at $f_0+w$ in the direction of $b$, $\nabla_b E(w;q)$, is then given by:
    \begin{align}
        \nabla_b E(w;q,f_0)
        & =
        \Deps{\normtwo{ Q(f_0 + w+\epsilon b) - q }}
        \non\\
        & =
        2 \inner{Q(f_0+w)-q}{Q_{*,f_0+w}(b)}
        \eqstop
        \label{eqn:derivative}
    \end{align}

\noi (Note that here, for notational simplicity,  we have dropped $O$ and $\gamma$.) Here $Q_{*,f}$ denotes the differential of $Q$ at $f$. This can be evaluated using the following expression: for all $s\in \stwo$,
    \begin{equation}
        Q_{*,f}(b)(s)
        =
        {n_b(s) \over \sqrt{|n(s)|}} - { n(s)\cdot n_b(s) \over 		2|n(s)|^{5/2}} n(s)
        \eqcomma
        \label{eqn:differential}
    \end{equation}

\noi where $n_b(s) = (f_u(s)\times b_v(s)) + (b_u(s)\times f_v(s))$. To improve numerical accuracy, the second term can be replaced by the more stable ${\tilde{n}(s)\cdot n_b(s) \over 2 \sqrt{ |n(s)|}} \tilde{n}(s)$, resulting in
    \begin{equation}
        Q_{*,f}(b)(s)
        =
        {1 \over \sqrt{|n(s)|} } \left( n_b(s) - { \tilde{n}(s) \cdot n_b(s) 		\over 2} \tilde{n}(s) \right)
        \eqstop
        \label{eqn:simplify}
    \end{equation}

\noi Finally, the update is determined by the gradient $\nabla E(f_0;q) = \sum_{b\in\mathcal{B}} \left( \nabla_b E(b;q,f_0) \right) b$ obtained using Eqn.~\ref{eqn:derivative}, \ref{eqn:differential}, and~\ref{eqn:simplify}.

A naive implementation of this procedure results in the algorithm becoming trapped in  local minima. We show, however, that this problem can be overcome by carefully engineering the orthonormal basis ${\cal B}$ of $\Space{F}$ (Section~\ref{sec:basis_formation}) and combining it with a multiscale and multiresolution representation of the elements of $\Space{F}$ and $\Space{Q}$ (Sections~\ref{sec:multiscaleSRNF} and~\ref{sec:multiresolution_inversion}).

\subsubsection{Basis Formation}
\label{sec:basis_formation}

An important ingredient of our approach is the representation of the vector space $\Space{F}$  using an orthonormal basis set ${\cal B}$. Since elements of $\Space{F}$ are smooth mappings of the type $f : \stwo \to \real ^3$, we can use spherical harmonics (SHs), in each coordinate, to form the basis set.
Let $\{Y_i(s)\}$, for $i$ in some index set, be the real-valued SH functions. Then, a basis for representing surfaces $f : \stwo \to \real^3$  can be constructed as $\{Y_i e_1\} \cup \{Y_j e_2\} \cup \{Y_k e_3\}$, where ${e_1, e_2, e_3}$ is the standard basis for $\real^3$. We will refer to the resulting basis as $\mathcal{B} = \{b_j\}$, for $j$ in some index set. $\mathcal{B}$ is a generic basis that can be used to represent arbitrary spherically-parameterized surfaces. However, complex surfaces will require a large number of basis elements to achieve high accuracy and capture all the surface details. We show that by using a spherical wavelet-based multiresolution approach~\cite{laga2006spherical}, the optimization problem of Eqn.~\eqref{eq:energy_to_minimize} can be solved in a computationally efficient manner.

Alternatively, depending on the context and the availability of training data, it may be more efficient to use a principal component basis instead of a generic basis.  For instance, in the case of human shapes, there are several  publicly available data sets that can be used to build a PCA basis. Thus, given a set of training samples, we can compute its PCA in $\Space{F}$ and use the first $N$ components to construct the basis. In our experiments, the number of PCA basis elements required is of the order of $100$, significantly smaller than the number of SH elements (more than $3000$).

Finally, note that one can also combine the PCA and SH bases in order to account for shape deformations that have not been observed in the training dataset. It is also possible to use different bases for the surfaces (elements of $\Space{F}$) and the deformations (elements of tangent spaces $T_f ( \Space{F}))$ to improve efficiency, although we do not explore that here.

\subsubsection{Multiscale representation of SRNFs}
\label{sec:multiscaleSRNF}

The gradient descent  method at a fixed resolution finds it difficult to reconstruct complex surfaces that contain elongated parts and highly non-convex regions, such as the ones shown in Fig.~\ref{fig:reconstruction_real}.  We observe however that such parts often correspond to the high frequency components of the surface, which can be easily captured using a multiscale analysis.

Given an SRNF $q$ of an unknown surface $f$, and an SRNF $q_s$ of a unit sphere $f_s$, the geodesic between them under the $\ltwo$ metric is the linear path $\beta(\tau) = (1-\tau)q_s + \tau q, \tau \in [0, 1]$; let the corresponding path in  ${\cal F}$ be $\alpha(\tau)$ such that $f_s = \alpha(0)$ and $f = \alpha(1)$.  One can then treat $\beta$ as a multiscale representation of the SRNF $q$. In practice, we divide the linear path $\beta$ into $m+1$ equidistant points $q^i, i=0\dots, m$, such that $q^0 = q_s$,  $q^m = q$, and $\norm{q^i - q^{(i-1)}} = \Delta > 0$.  We refer to $\textbf{q} = (q^1, \dots, q^m)$ as the multiscale SRNF of an unknown multiscale surface $\textbf{f} = (f^1, \dots, f^m)$. (Here,  $f^0 = f_s$ and $q^0 = q_s$ are not included in the representation.)

With this representation, the SRNF inversion problem can be reformulated as follows. Given a multiscale SRNF $\textbf{q} = (q^1, \dots, q^m)$, the goal is to find a multiscale surface $\textbf{f} = (f^1, \dots, f^m)$ such that $Q(f^i)$ is as close as possible to $q^i$, \ie $E_0(f^i; q^i)$ is minimized. Gradient descent procedures are known to provide good results if the initialization is close to the solution. We thus use the following iterative procedure: in step $1$, we find a surface $\tilde{f}^1$ that minimizes $E_0(f; q^1)$, initializing gradient descent with a sphere $f^{0}$. Then, at step $i$, we find the surface $\tilde{f}^i$ that minimizes $E_0(f; q^i)$, initializing gradient descent with $\tilde{f}^{(i-1)}$. Fig. 3 of the supplementary material shows a few examples of SRNF inversion using this multiscale approach.

\subsubsection{Multiresolution Inversion Algorithm}
\label{sec:multiresolution_inversion}

While this allows us to invert a single SRNF, by allowing the gradient descent-based optimization procedure to converge to the desired solution, it can be computationally very expensive when dealing with high resolution surfaces.  In practice, the computation time can be significantly reduced by adopting a multiresolution representation of the elements of $\Space{F}$, and subsequently the elements of $\Space{Q}$. Specifically, given a surface $f: \stwo \to \real^3$, we use spherical wavelet analysis to build  a multiresolution representation $\textbf{f} = (f^1, \dots, f^m = f)$ of $f$. The  surface $f^i$ at level $i$ is a smoothed, sub-sampled, and thus coarse version of the surface $f^{i+1}$. With a slight abuse of notation, we also define the multiresolution SRNF map sending $\textbf{f} \mapsto Q(\textbf{f}) = \textbf{q} = (q^{i})$, where $q^i = Q(f^i)$.

To build the multiresolution representation, we apply an analysis filter $\tilde{h}_i$ to the spherical surface $f$ via spherical convolution, resulting in the output $w_i = f \star \tilde{h}_i$. We then convolve $w_i$ with another spherical filter $h_i$, a synthesis filter, producing a coarse surface $f^i$ that is an approximation  of the surface $f$ at scale $i$. By defining the analysis filters $\{\tilde{h}_i\}$ as dilated versions of a template filter $\psi$, \ie $~\tilde{h}_i = D_i\psi$,  the coefficients $\{w_n\}$ and thus the reconstructed surfaces $\{f^i\}$ capture the original surface properties at multiple scales.  For details of these filters, we refer the reader to~\cite{yeo2008construction}. In practice, we choose the template wavelet $\psi$ to be the Laplacian-of-Gaussian and we compute the wavelet decomposition of each coordinate function using convolutions with the analysis-synthesis filters in the SH domain.  Since surfaces at large scales require a smaller number of samples, we start with a spherical grid of resolution $128\times128$ and  at each  wavelet decomposition we subsample the obtained coarse surface by halving the spherical resolution.  We have found that four decomposition levels are  sufficient to achieve high accuracy inversion.

\begin{figure}[t]
	\centering 
	\includegraphics[width=.45\textwidth]{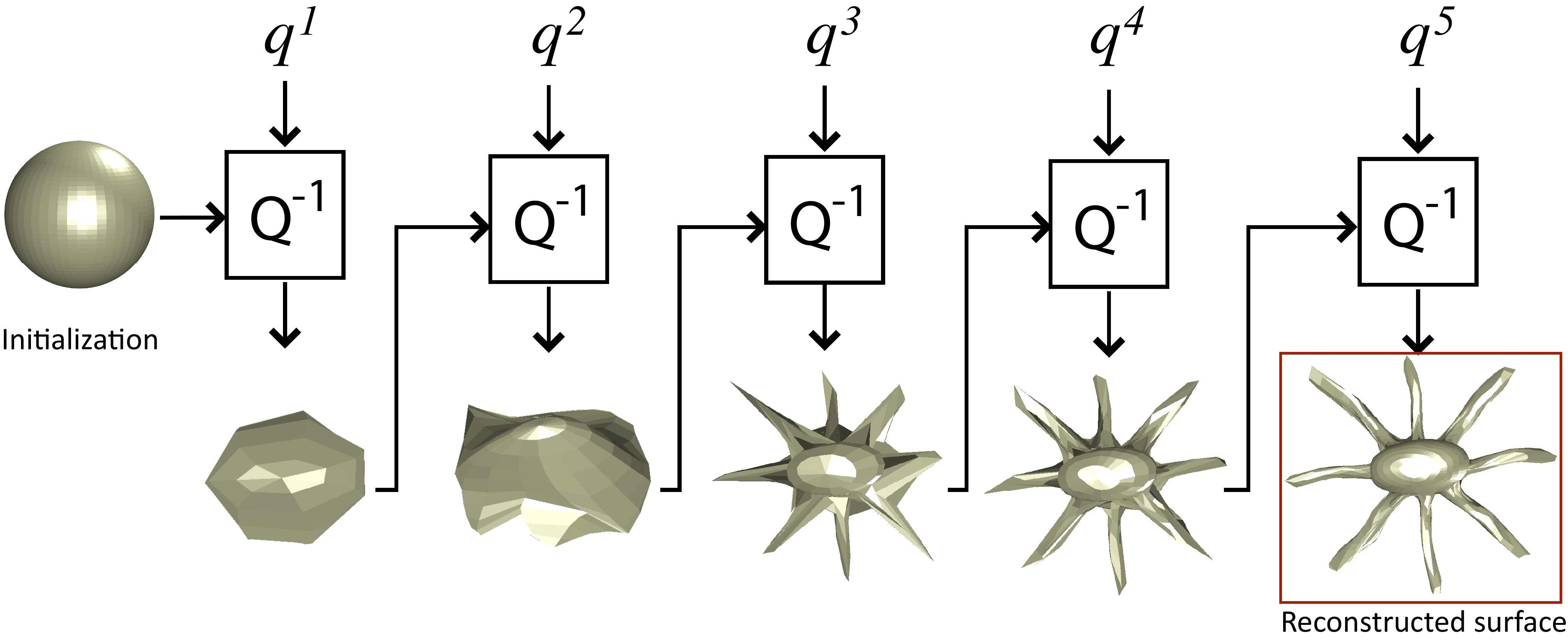}\\
	\small{(a) Multiresolution SRNF inversion.}

	\begin{tabular}{c@{}c}
	\includegraphics[width=.22\textwidth]{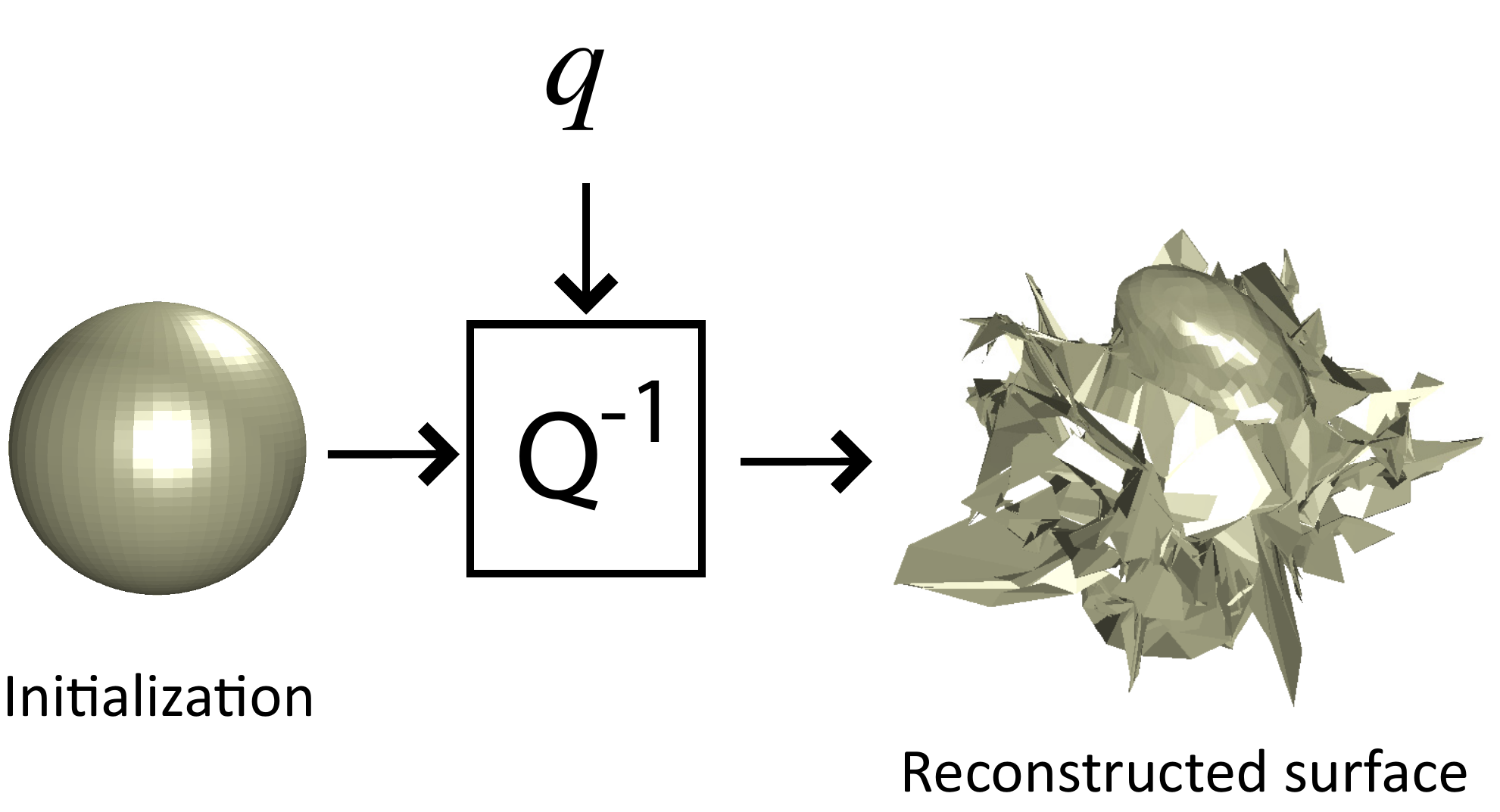}	 & \includegraphics[width=.15\textwidth]{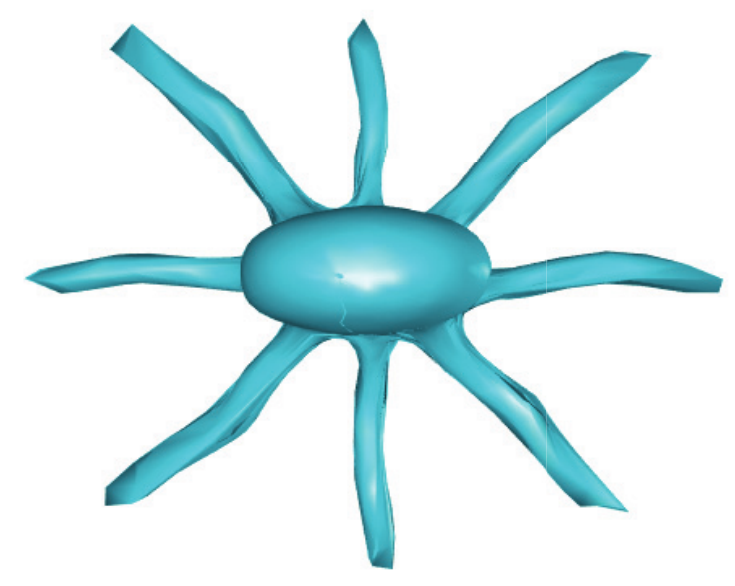}	  \\
	\small{(b) Inversion at single resolution.}  & \small{ (c) Ground truth. } 	\end{tabular} 	\caption{Comparison between the proposed multiresolution approach (a) 	with the single resolution version (b). The latter fails to  recover 	 complex surfaces.} 	\label{fig:spherical_wavelets}\label{fig:multires_inversion}
\end{figure}

With this representation, a multiresolution SRNF can be inverted using the same procedure as the one described in Section~\ref{sec:multiscaleSRNF}. This time, the surface obtained at each iteration is first up-sampled and then used to initialize optimization at the next resolution. Fig.~\ref{fig:multires_inversion}(a) shows a reconstruction of the octopus surface from its SRNF using the proposed multiresolution inversion procedure.    Fig.~\ref{fig:multires_inversion}(b) displays the inversion result when Eqn.~\ref{eq:energy_2} is solved using a single-resolution representation. The latter results in a degenerate surface since the initialization is too far from the correct solution. The proposed multiresolution procedure converges to the correct solution since at each optimization step, the initialization is close to the correct solution. Note also that the multiresolution procedure significantly reduces computational cost: although the optimizations at coarser scales require a large number of iterations, they operate on spherical grids of low resolution (\eg $16\times 16$) and are thus very fast. At the finest scale, although the resolution is high ($128\times128$), gradient descent only requires a few iterations since the initialization is already very close to the solution.

We point out that our approach assumes that $\textbf{q}$, the multiresolution representation of the SRNF $q$ of an unknown surface $f$, is available. This is not a restriction, since in practice, we are interested in computing geodesics between known surfaces $f_1$ and $f_2$, deforming a given surface $f_1$ along a given geodesic direction, or computing summary statistics of a given set of surfaces. In the first application, for example, one can: (1) build a multiresolution representation of the surfaces $f_1$ and $f_2$, hereinafter denoted by $\textbf{f}_1$ and $\textbf{f}_2$, respectively, using the spherical wavelet decomposition described in this section; (2) compute their multiresolution SRNFs $\textbf{q}_1$ and $\textbf{q}_2$; and finally (3) invert the multiresolution SRNF $\textbf{q} = (q^1, \dots, q^m)$ where $q^i = (1 - \tau)q_1^i + \tau q_2^i, \tau \in [0, 1]$.  Section~\ref{sec:analysis}  elaborates this point further in different contexts.

\subsection{Star-Shaped Surfaces}
\label{sec:starshapedsurfaces}

Here we consider a special subset of genus-0 surfaces, {\it star-shaped} surfaces, which are of great relevance to many applications, \eg the diagnosis of attention deficit hyperactivity disorder (ADHD) using MRI scans. Remarkably, in this case an analytic solution to the inversion problem exists. By a star-shaped surface, we mean a parameterized surface $f\in\Space{F}$ that, up to a translation, can be written in the form $f(u,v) = \rho(u,v) e(u, v)$, where $\rho(u,v) \in \real_{\geq 0}$, and $e(u, v)\in\stwo$ is the unit vector in $\rthree$ given in Euclidean coordinates by $e(u, v) = (\cos(u)\sin(v), \sin(u)\sin(v), \cos(v))$. It can be seen by inspection that in spherical coordinates, $f$ is given by $(r, \theta, \phi) = (\rho(u, v), u, v)$.

Note that the volume enclosed by a star-shaped surface is a star domain (\ie there exists a point in the enclosed volume such that the straight line segments from that point to every point on the surface all lie entirely in the enclosed volume), but that in addition to this purely geometric property, we demand that the surface have a particular parameterization and that the `center' be located at $r = 0$.

For this type of surfaces, the map $Q$ can be \emph{analytically} inverted, as follows. The radial component of the normal vector $n$ of an star-shaped surface is, by definition, given by
    \begin{equation}
      n^{r}(u, v) = \left\langle n(u, v), e(\theta(u, v), \phi(u, v))
      \right\rangle
      \eqcomma
    \end{equation}
\noi since $e(\theta, \phi)$ is the radial unit vector in the direction in $\rthree$ defined by $(\theta, \phi)$. If the star-shaped surface were in general parametrization, we could not compute $n^{r}$ because we would not know $\theta$ and $\phi$, the angular coordinates of the surface we are trying to recover. In the special parameterization, the expression becomes: $n^{r}(u, v) = \left\langle n(u, v), e(u, v) \right\rangle$, and this can be calculated. The result is very simple: $ n^{r}(u, v) = \rho^{2}(u, v)$. As a result, given an SRNF $q$ and a parameterization $e$, the star-shaped surface $\tilde{f}$ corresponding to this $q$, \ie such that $Q(\tilde{f}) = q$, takes the form:
    \begin{equation}
        \tilde{f}(u, v) = \left(\sqrt{\Abs{q(u, v)} q^{r}(u, v)}\right) e(u,
        v)
        \eqcomma
        \label{eqn:analytic}
    \end{equation}

\noi where $q^{r} = \left\langle q, e\right\rangle$ is the radial component of $q$.

Note that $\tilde{f}$ depends on both $q$ and a fixed parameterization $e(u, v)$. If both are known, then $Q$ can be analytically inverted, as above. If a surface encloses a star domain, but is in a general parameterization, one can still choose to apply Eqn.~\ref{eqn:analytic}. In this case, the resulting $\tilde{f}$ will not in general be the original surface $f$, but it may provide a good initialization for solving the reconstruction-by-optimization problem.

\subsection{Reconstruction examples}
\label{sec:reconstructionexamples}
\begin{figure}[t]
    \centering
    \begin{tabular}{@{}c@{}c@{}c@{}c@{}c@{}}
   		\includegraphics[trim=2cm 1cm 1cm 1cm, clip=true, width=.07\textwidth]{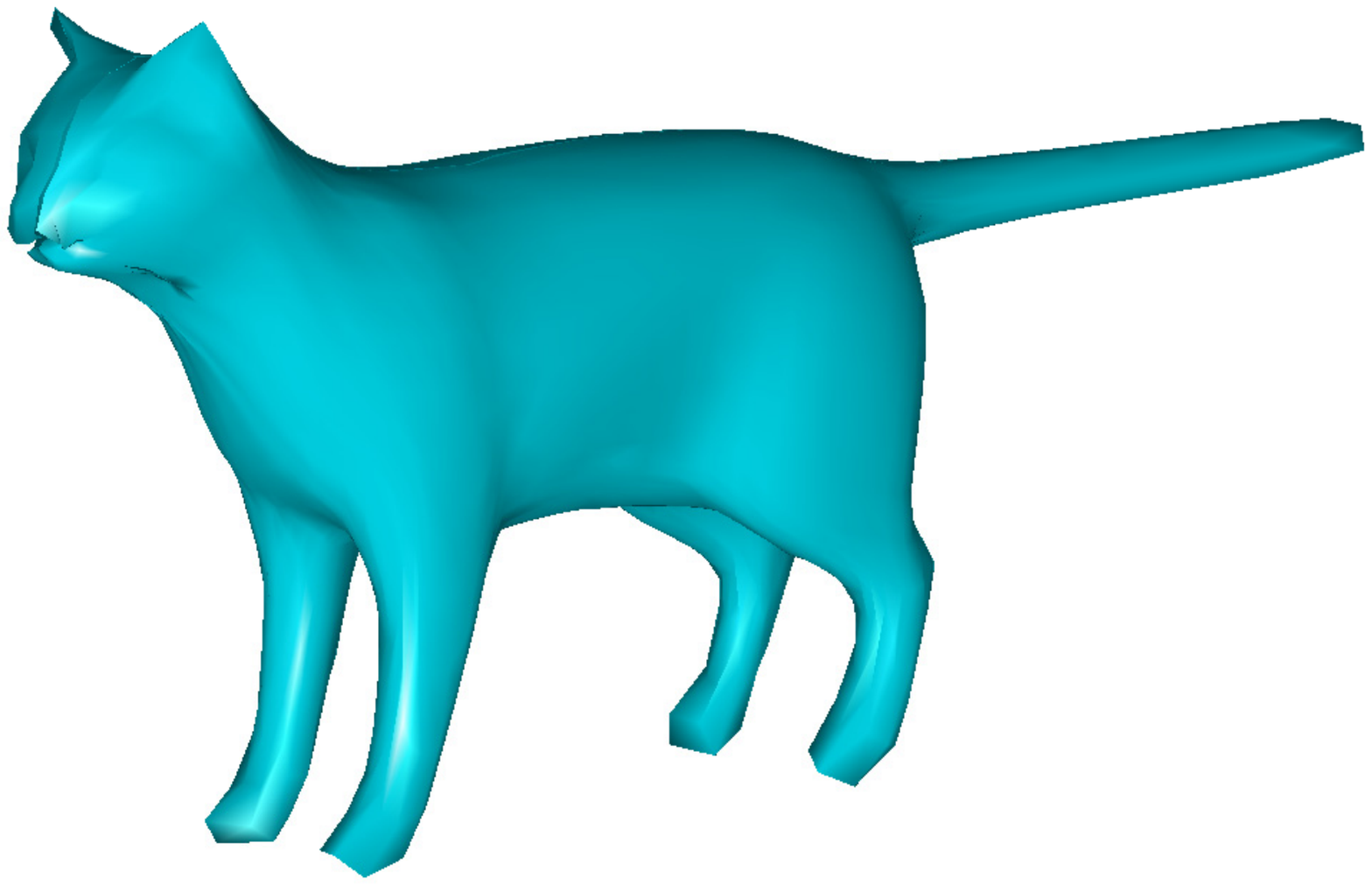} &
		 \includegraphics[trim=4cm 1cm 1cm 1cm, clip=true,width=.07\textwidth]{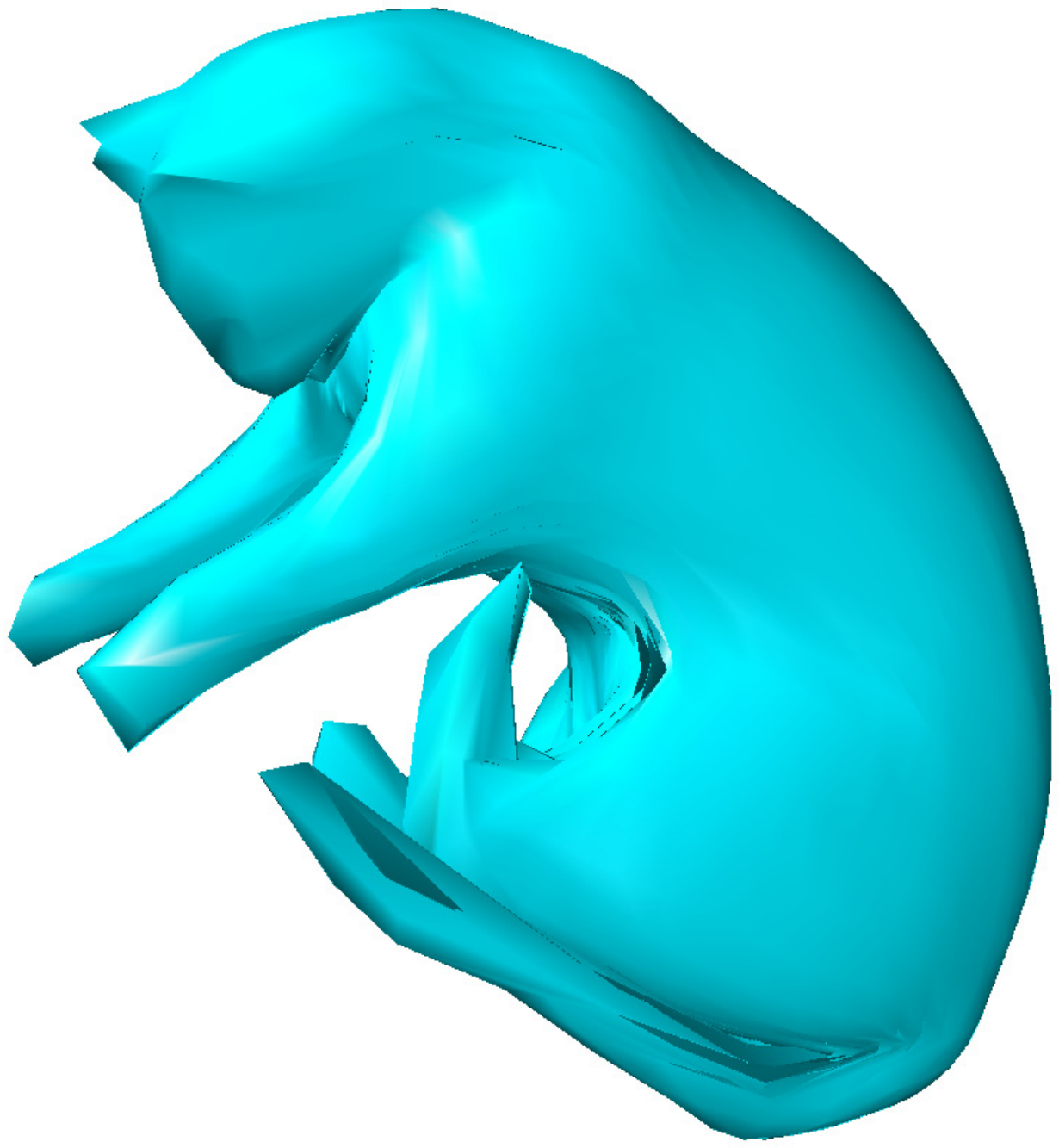} &
		 \includegraphics[trim=6cm 1cm 6cm 0cm, clip=true,width=.05\textwidth]{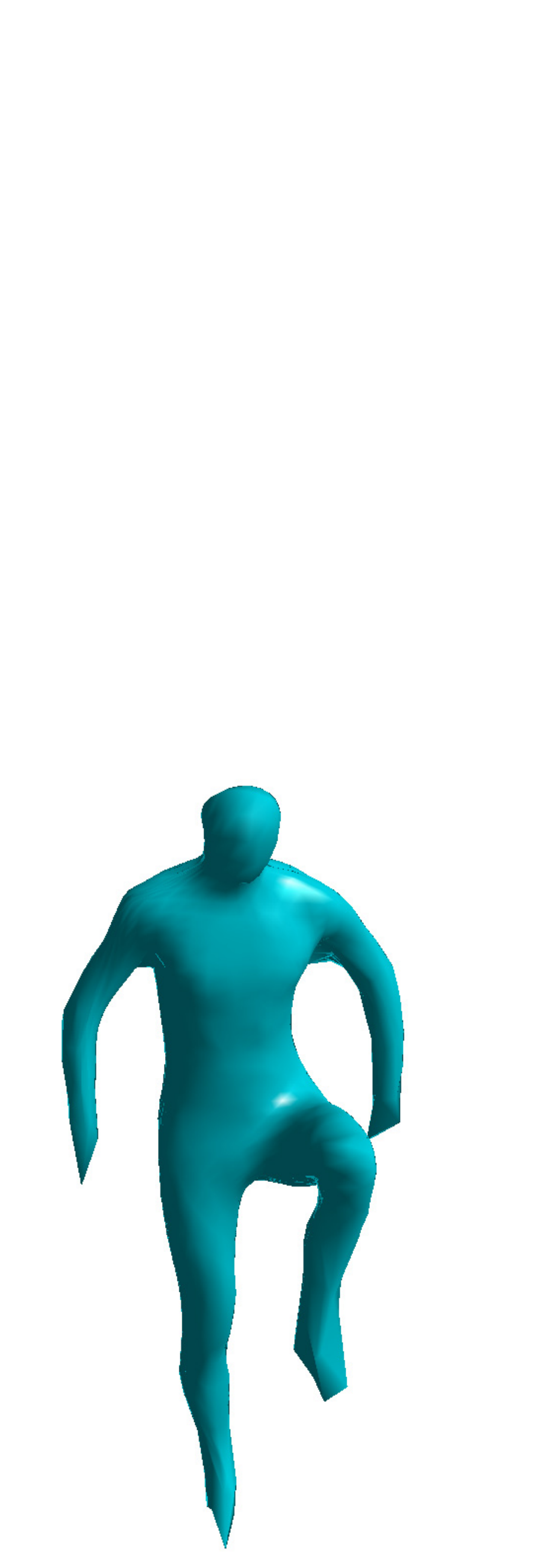}& 		 
		 \includegraphics[trim=6cm 1cm 6cm 0cm, clip=true,width=.05\textwidth]{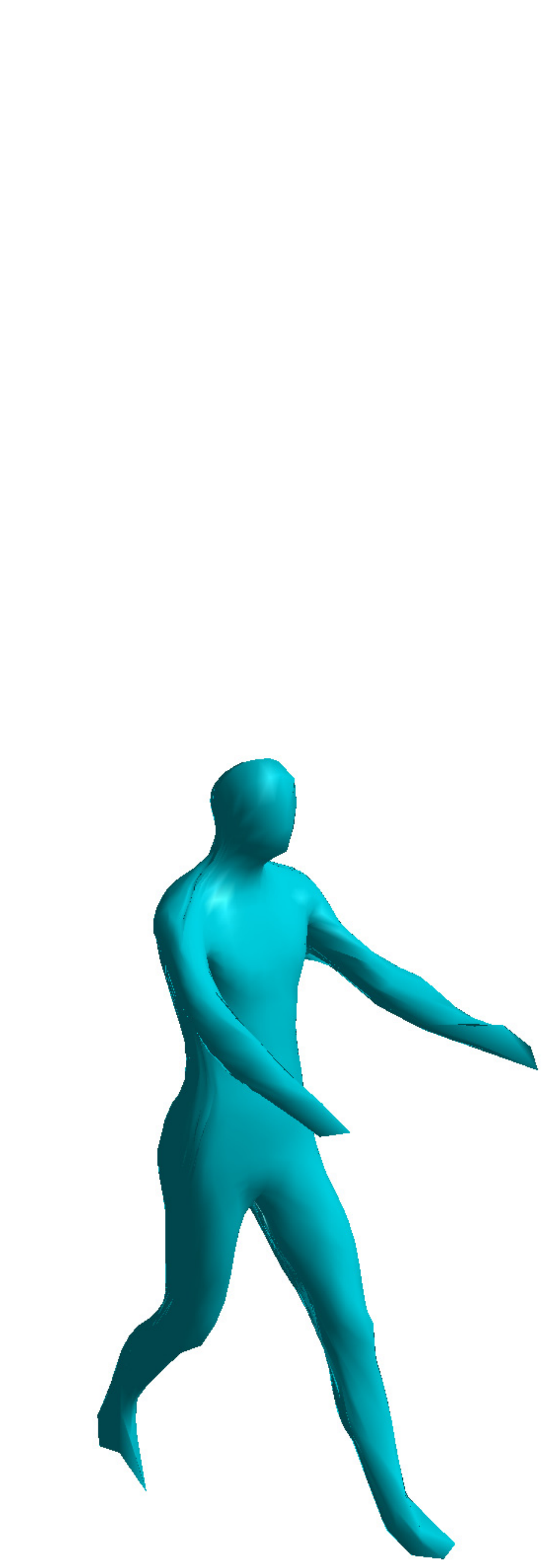}  &
		  \includegraphics[trim=1cm 1cm 1cm 1cm, clip=true,width=.07\textwidth]{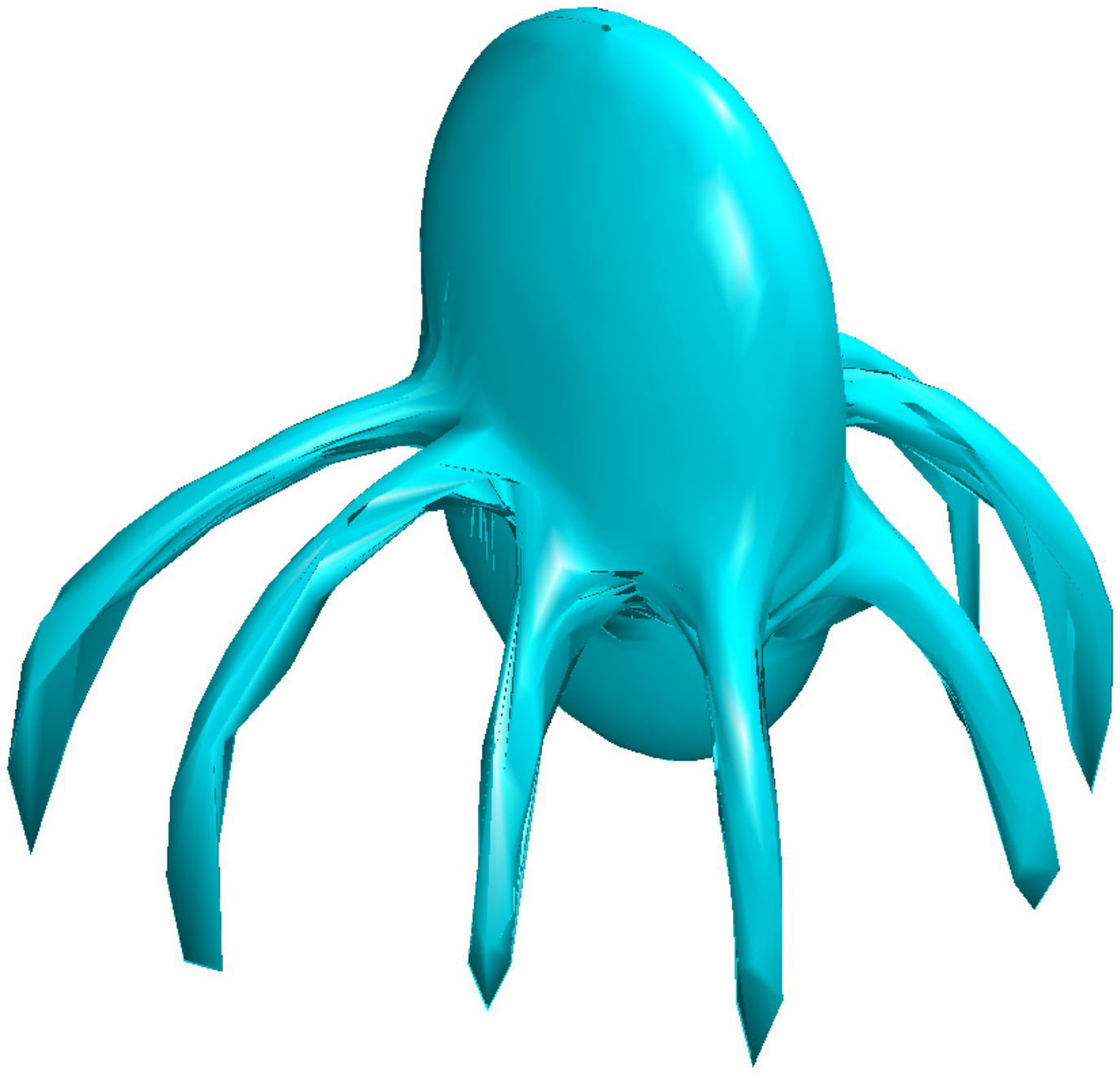}  \\

		\multicolumn{5}{c}{(a) The target surfaces $f_o$. }\\

		\includegraphics[trim=2cm 1cm 1cm 1cm, clip=true,width=.07\textwidth]{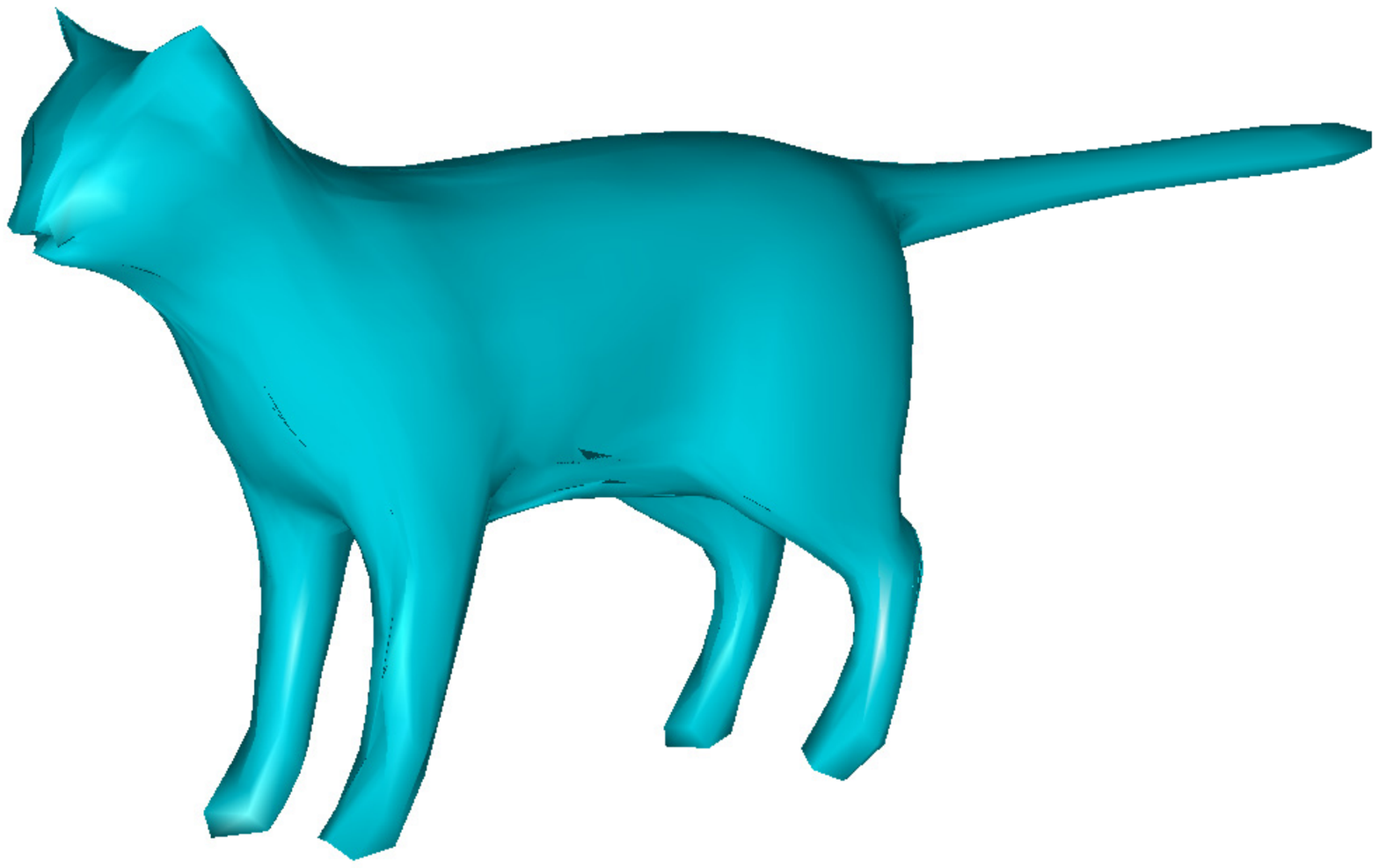}  &
		\includegraphics[trim=4cm 1cm 1cm 1cm, clip=true,width=.07\textwidth]{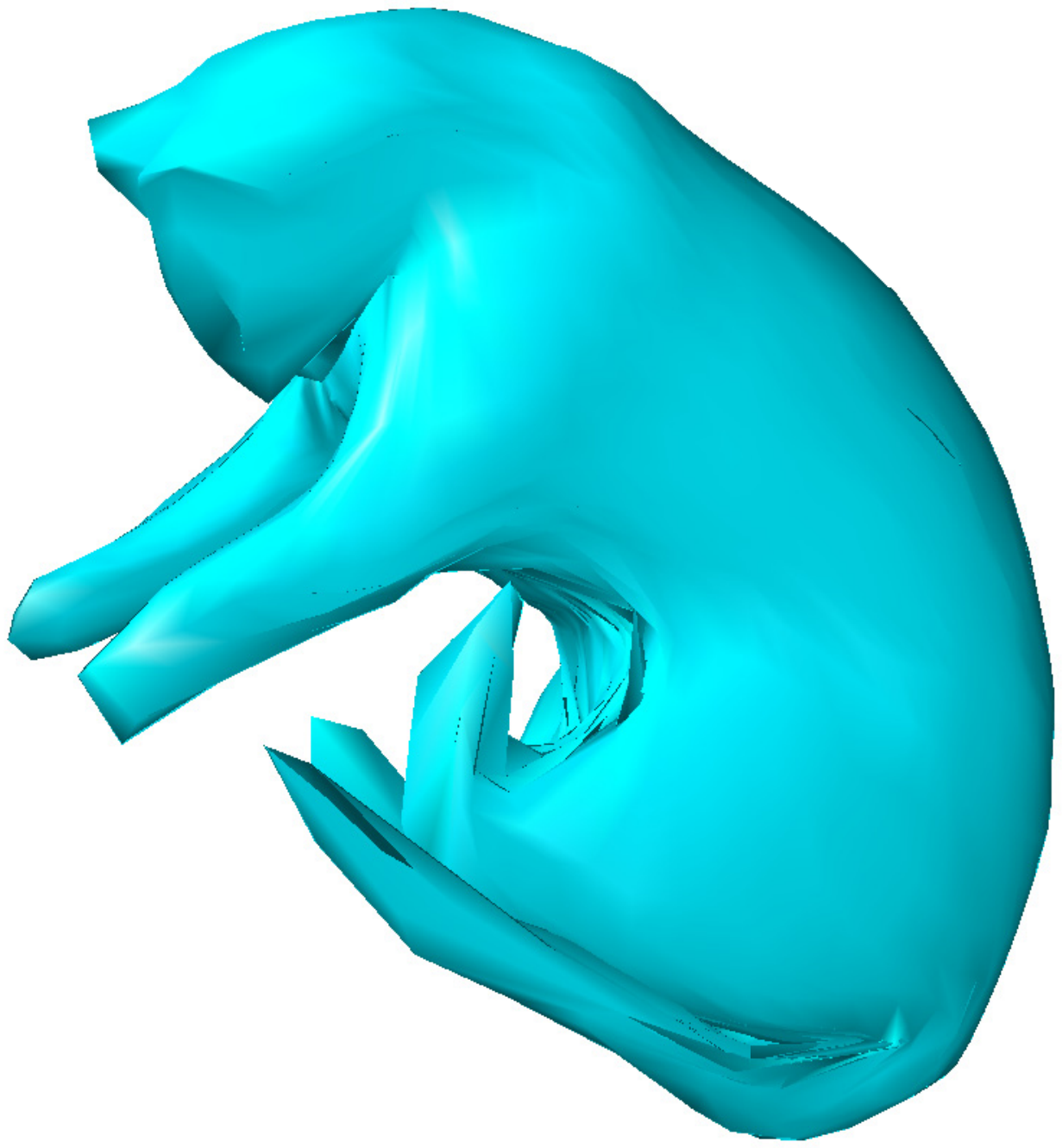} &
		 \includegraphics[trim=6cm 1cm 6cm 0cm, clip=true,width=.05\textwidth]{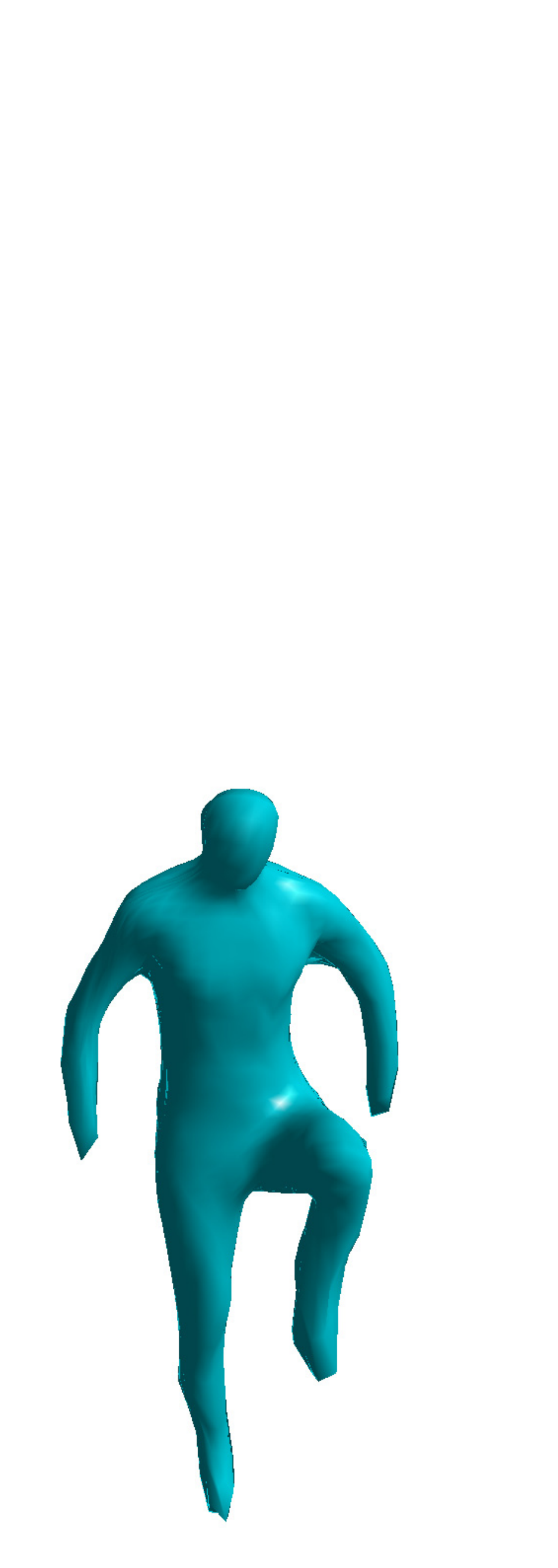}& 		
		 \includegraphics[trim=6cm 1cm 6cm 0cm, clip=true,width=.05\textwidth]{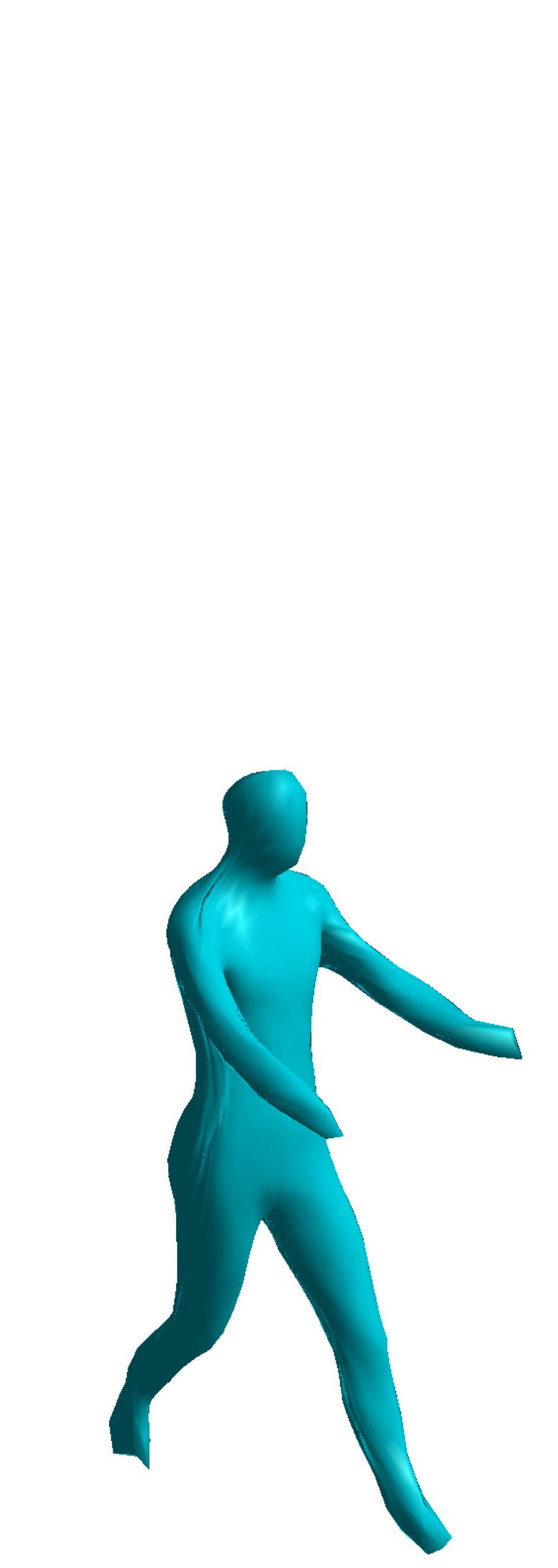} &
		 \includegraphics[trim=1cm 1cm 1cm 1cm, clip=true,width=.07\textwidth]{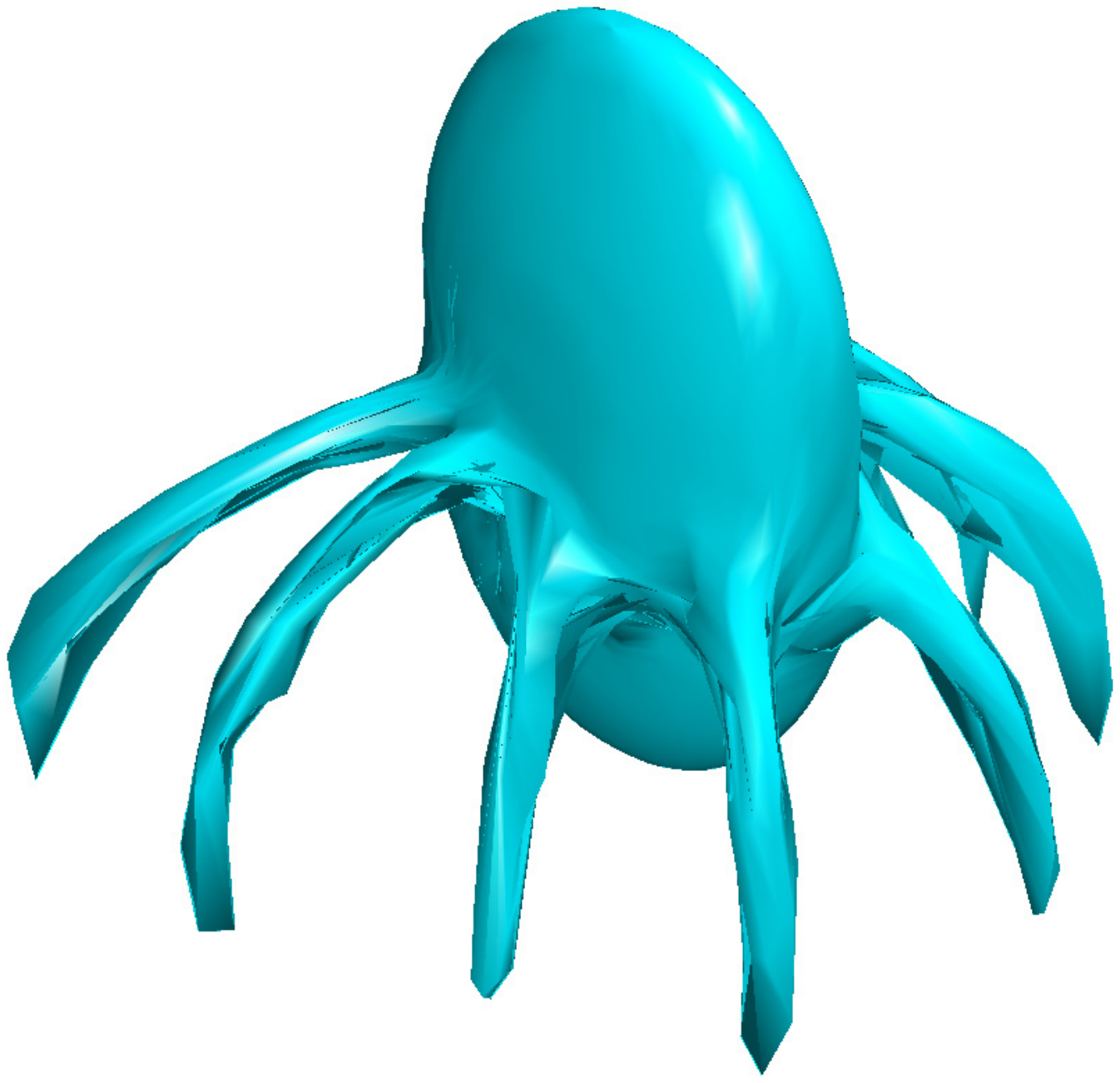} \\
		\multicolumn{5}{c}{(b) The reconstructed  surfaces $f^*$. } \\

		\includegraphics[width=.09\textwidth]{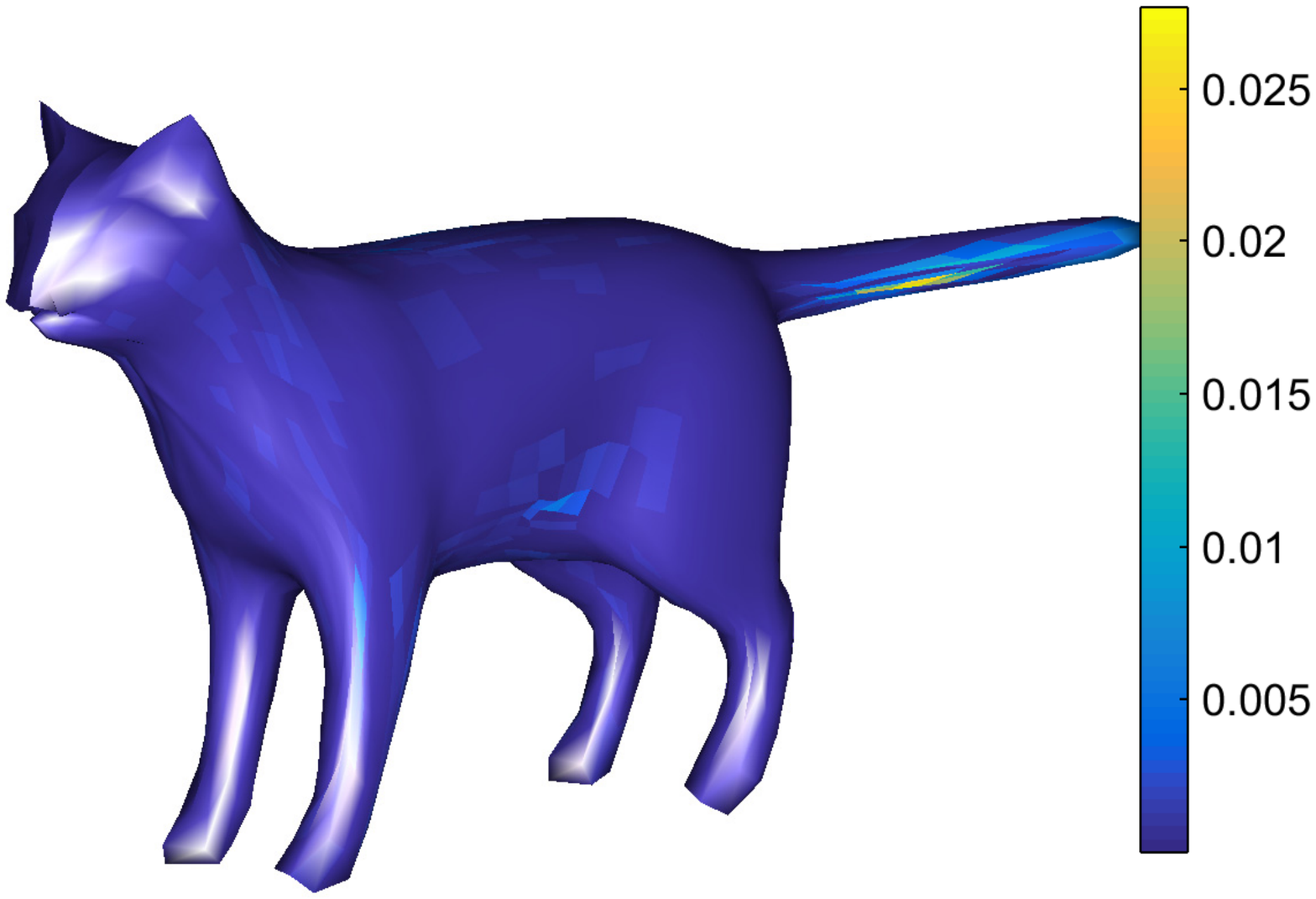}&
		\includegraphics[trim=3cm 0cm 0cm 0cm, clip=true, width=.09\textwidth]{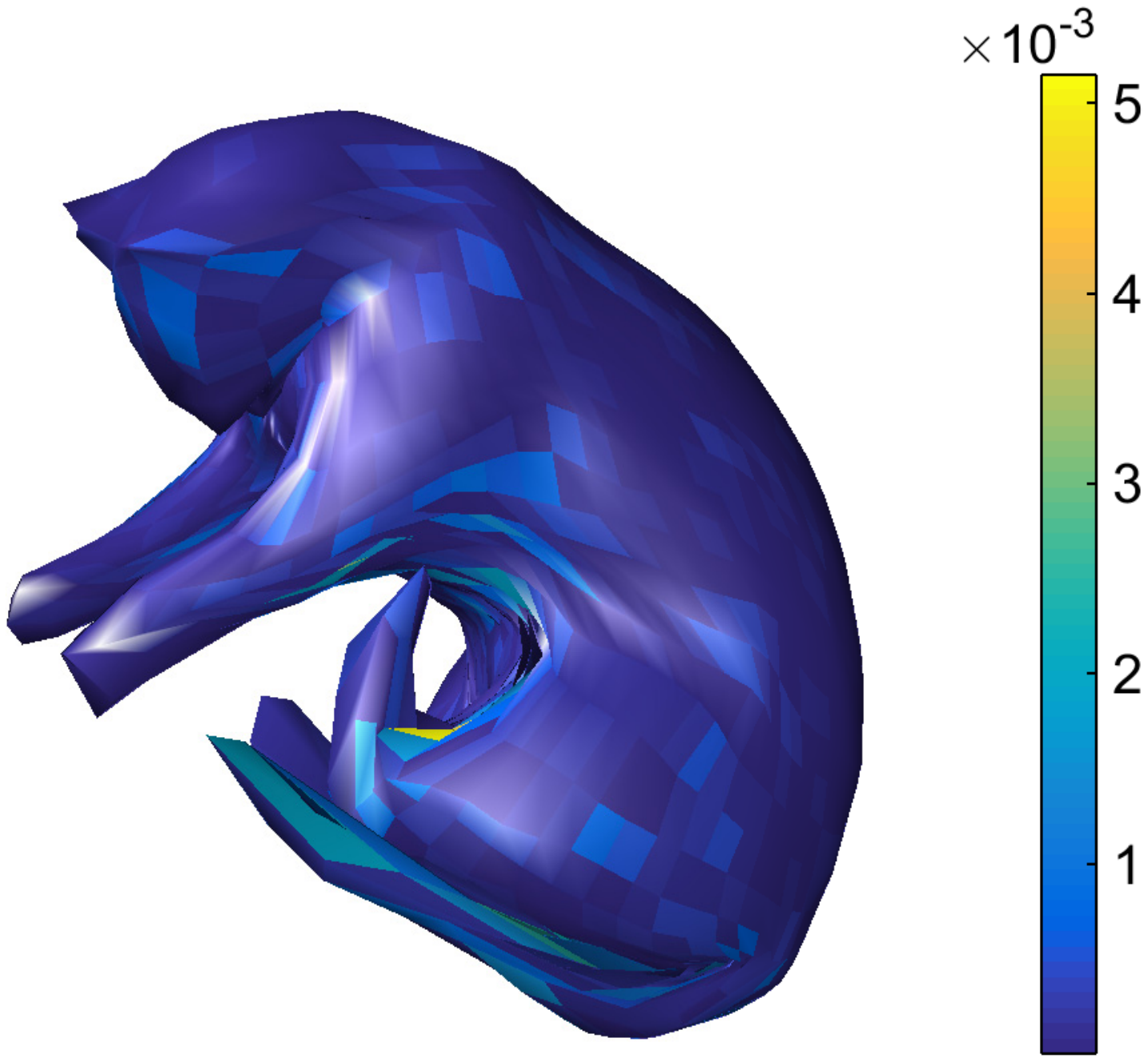}&
		 \includegraphics[trim=3cm 1cm 2cm 0cm, clip=true,width=.1\textwidth]{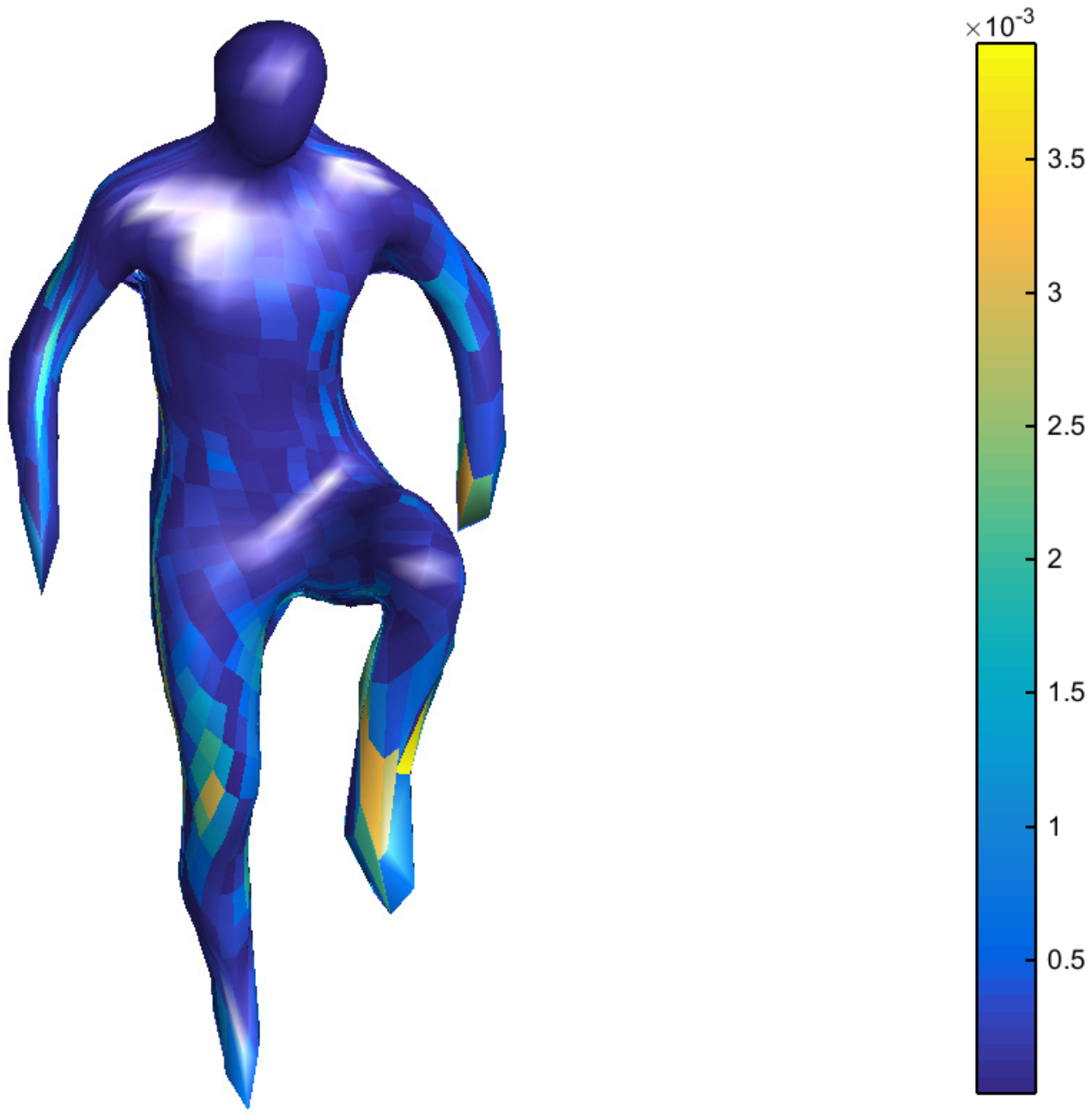}& 		
		 \includegraphics[trim=3cm 1cm 2cm 0cm, clip=true,width=.1\textwidth]{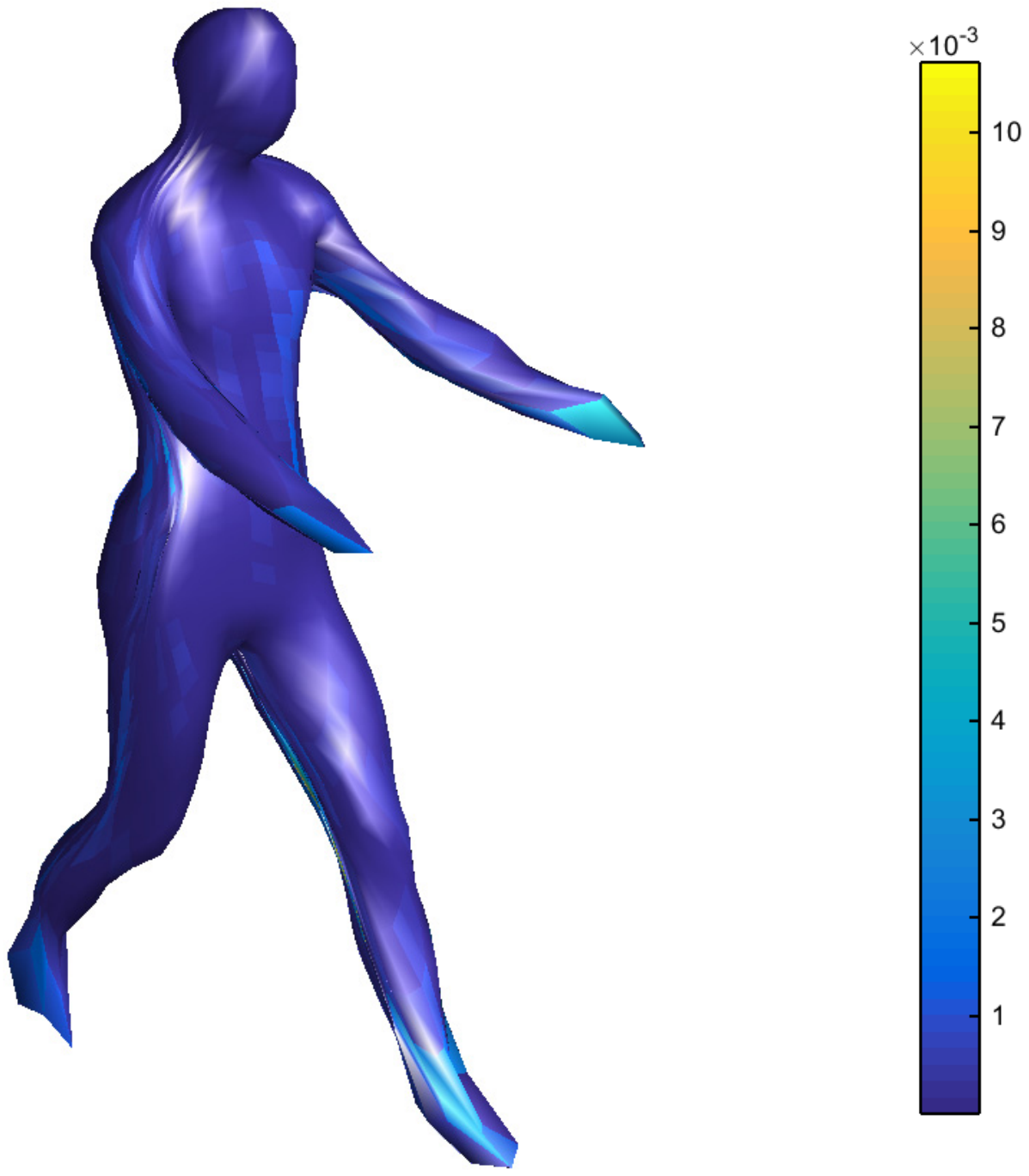}&
		\includegraphics[width=.09\textwidth]{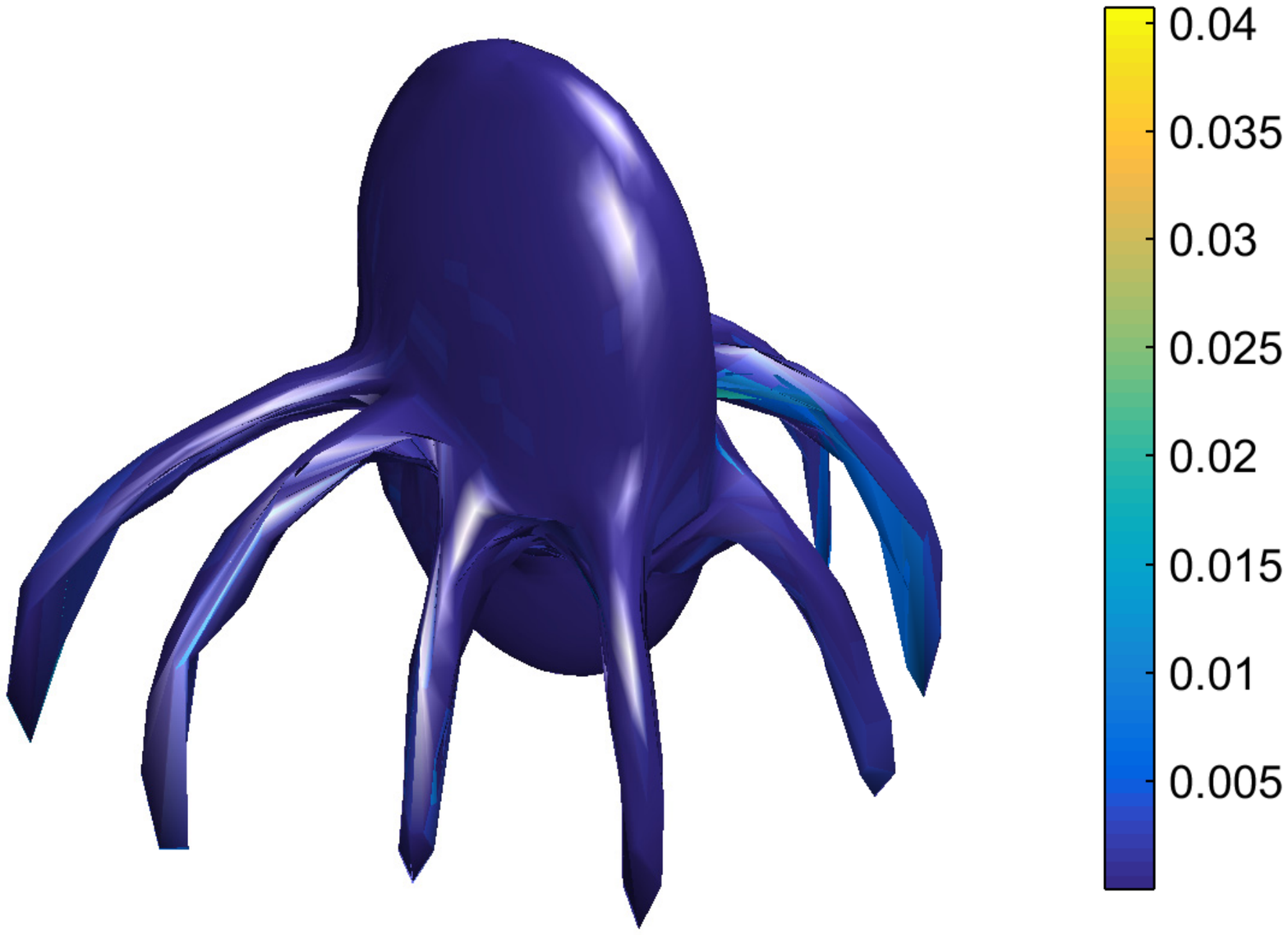}\\
		\multicolumn{5}{c}{(c)  Reconstruction errors $|f^*(s) - f_o(s)|$. }\\

	\end{tabular}

    \caption{ \label{fig:reconstruction_real} Reconstruction of
    surfaces from their SRNF representations,  using $3642$ SH basis elements.  Additional examples are shown in the supplementary file. }

\end{figure}

Fig.~\ref{fig:reconstruction_real} shows some examples of the reconstruction of  real objects. In each case, from the original surface $f_o$, we compute its multiresolution surface $\textbf{f}_o$, compute the multiresolution SRNF $\textbf{q}_o = Q(\textbf{f}_o)$, and then estimate  $f^*$ by SRNF inversion of $\textbf{q}_o$. We display the original surface $f_o$, the reconstructed surface $f^*$, and the reconstruction errors $|f^*(s) - f_o(s)|$. All these results were obtained using $3642$ SH basis elements.

An inspection of these results shows that the multiresolution inversion procedure is able to reconstruct the original surface with a very high accuracy. For the complex articulated surfaces of Fig.~\ref{fig:reconstruction_real}, the iterative optimization procedure reduces the energy by {\bf three orders of magnitude}. The plots of the reconstruction errors in Fig.~\ref{fig:reconstruction_real}(c)  show that most of the reconstruction errors occur at high curvature regions.

To quantify the impact of initialization on the convergence of the proposed algorithm, we tested it in a worst-case scenario. We took $13$ shapes classes from the SHREC07 watertight models (see Section~\ref{sec:classification}), with each class containing 20 models. For each model, we computed its SRNF, then re-estimated the original surface using the proposed inversion algorithm, initialized with a unit sphere (the worst-case scenario). Finally, we measured the reconstruction error. Median reconstruction error was $0.0008$ with a $25$th percentile of $0.0003$ and a $75$th percentile of $0.0022$. (All surfaces were rescaled to unit surface area.) The supplementary file includes reconstruction error analyses for each of the $13$ classes of shapes.

Table~\ref{tab:timing} summarizes average computation cost for inverting one SRNF on a 3.40Ghz i7-3770 CPU with 16GB RAM. In conjunction with Table~\ref{tab:algorithms}, it underlines the dramatic improvements achieved for the statistical shape analysis tasks listed in Section~\ref{sec:analysis}. In conclusion, this algorithm allows us to apply statistical tools for ESA of complex surfaces for the first time.

\begin{table}[t]
    \caption{Computation time, in seconds, on Intel i7-3770 CPU @ 3.40Ghz with 16Gb of RAM. }
    \label{tab:timing}
    \centering {
    \begin{tabular}{@{}l ll@{ }}
        \toprule
        				 & $3642$ Harmonic basis   & $100$  PCA basis \\ 
				 
        \midrule
        Grids of $64\times64$      & $16.5$ min  & $20.14$ sec \\
        \midrule
        Grids of $128\times128$ &  $-$ 	        & $2.5$ min\\

        \bottomrule
     \end{tabular}

}
\end{table}

\section{Development  of Statistical Tools}
\label{sec:analysis}

The ability to invert $Q$ enormously simplifies statistical shape modeling. In contrast with the approach in~\cite{xie-iccv:2013}, where analysis is performed in ${\cal F}$ under a complicated metric, the new framework performs analysis in the $\ltwo$ space of SRNFs, and transforms only the end results to ${\cal F}$. To illustrate the advantages of the new methods, we have selected as points of comparison, several algorithmic and computational tasks that are fundamental to statistical shape analysis. The basic algorithms for computing the Karcher mean shape, for parallel transport, and for transferring deformations from one shape to another, using both previous and the proposed methods, are described in Table~\ref{tab:algorithms}. Computationally intensive steps are underlined, and the computational complexity is indicated in boxes.  In the following subsections, we elaborate on the list of target analyses studied, and the mechanisms used to perform them using SRNF inversion.

\begin{table*}[t!]\footnotesize
\ra{0.9} \caption{Comparison of algorithms in terms of complexity and computation time.} \label{tab:algorithms} \centering
\begin{tabular}[!ht]{m{.5in}|c|c}
\hline
	& Previous~\cite{kurtek:2012,xie-iccv:2013} (spherical grids of $32\times 32$) & Proposed (spherical grids of $64\times 64$) \\
\hline Karcher Mean 	& \begin{minipage}[t]{.45\textwidth}\centering
\begin{myalgo}
	Let $\mu_f^0$ be an initial estimate. 	Set $j=0$.
\begin{enumerate}

\item Register $f_1,\dots,f_n$ to $\mu_f^j$.

\item For each $i=1,\dots,n$, construct a \underline{geodesic} to connect
    $f_i$ to $\mu_f^j$ and evaluate $v_i=\exp^{-1}_{\mu_f^j}(q_i)$.

\item Compute the average direction $\bar{v} = \frac{1}{n}\sum_{i=1}^n
    v_i$.

\item If $\left\|\bar{v}\right\|$ is small, stop. Else, update $\mu_f^{j+1}
    =
    \exp_{\mu_f^j}(\epsilon \bar{v})$ by shooting a geodesic,
    $\epsilon>0$,  small.

\item Set $j=j+1$ and return to Step 1.
\end{enumerate}
\label{algo:mean-old}
\end{myalgo}
\framebox{$n$ geodesics per iteration, $2$ hours per geodesic.}
\end{minipage}
& 	\begin{minipage}[t]{.42\textwidth}\centering
\begin{myalgo}
Let $\bar{q} = Q(\mu_f^0)$ with $\mu_f^0$ as an initial estimate. 	Set $j=0$.
\begin{enumerate}

\item Register $q_i=Q(f_i), i=1\dots, n$,  to $\bar{q}$.

\item Update the average $\bar{q} = \frac{1}{n}\sum_{i=1}^n q_i$.

\item If change in $\left\|\bar{q}\right\|$ is small, stop. Else, set
    $j=j+1$ and return to Step 1.
\end{enumerate}
Find $\mu_f$ by \underline{ inversion} s.t. $Q(\mu_f) = \bar{q}$. \label{algo:mean-new}
\end{myalgo}
	\framebox{ \parbox{180\unitlength}{1 inversion,  $16.5$ min (with harmonic basis)  or $20.14$ seconds (with PCA basis).}}
\end{minipage}
\\
\hline Parallel Transport & \begin{minipage}[t]{.45\textwidth}\centering
\begin{myalgo}
Find a \underline{geodesic} $\alpha(t)$ connecting $f_1$ to $f_2$. For $\tau = 1,\dots, m$, do the following.
\begin{enumerate}

\item \underline{Parallel transport} $V(\frac{\tau-1}{m})$ from
    $\alpha(\frac{\tau-1}{m})$ to $\alpha(\frac{\tau}{m})$ and
	name it  $V(\frac{\tau}{m})$.
\end{enumerate}
Set $v^{||} = V(1)$.
\end{myalgo}
\framebox{1 geodesic + $m$ parallel transports, $2$ hours per geodesic.}
\end{minipage}
&
\begin{minipage}[t]{.42\textwidth}\centering
\begin{myalgo}
Parallel transport on $\ltwo$ remains constant.
\begin{enumerate}

\item Compute $w = Q_{*,f_1}(v)$ (differential of  $Q$).

\item Find $f$ by \underline{inversion} s.t. $Q(f) = Q(f_2) + \epsilon w$,
    $\epsilon$ is small.

\item Evaluate ${f - f_2 \over \epsilon}$ and set it to be $v^{||}$.
\end{enumerate}
\end{myalgo}
	\framebox{ \parbox{180\unitlength}{1 inversion,  $16.5$ min (with harmonic basis)  or $20.14$ seconds (with PCA basis).}}
\end{minipage}
\\
\hline Transfer Deformation & \begin{minipage}[t]{.45\textwidth}\centering
\begin{myalgo}{\ \\}
\begin{enumerate}

\item Find a \underline{geodesic} $\beta(t)$ connecting $f_1$ to $h_1$ and
    evaluate $v = \exp^{-1}_{f_1}(h_1)$.

\item Find a \underline{geodesic} $\alpha(t)$ connecting $f_1$ to $f_2$.
    Set $V(0) = v$. For $\tau = 1,\dots, m$, do the following.
\begin{enumerate}
\item \underline{Parallel transport} $V(\frac{\tau-1}{m})$ from
    $\alpha(\frac{\tau-1}{m})$ to $\alpha(\frac{\tau}{m})$ and
	name it  $V(\frac{\tau}{m})$.
\end{enumerate}

\item Shoot a \underline{geodesic} $\beta'(t)$ from $f_2$ with velocity
    $v^{||}=V(1)$ and set $h_2 = \beta'(1)$.
\end{enumerate}
\end{myalgo}
	\framebox{3 geodesics + $m$ parallel transports, $2$ hours per geodesic.}
\end{minipage}
&
\begin{minipage}[t]{.42\textwidth}\centering
\begin{myalgo}
Parallel transport on $\ltwo$ remains constant.
\begin{enumerate}
\item Compute $v = Q(h_1) - Q(f_1)$. \item Find $h_2$ by \underline{inversion} s.t. $Q(h_2) = Q(f_2) + v$.
\end{enumerate}
\end{myalgo}
\framebox{ \parbox{180\unitlength}{1 inversion,  $16.5$ min (with harmonic basis)  or $20.14$ seconds (with PCA basis).}}
\end{minipage}
\\
\hline
\end{tabular}
\end{table*}

In Sections~\ref{sec:geodesic_path},~\ref{sec:geodesic_shooting}, and~\ref{sec:deformation_transfer}, we are given two surfaces $f_1$ and $f_2$, and their SRNF representations $q_1= Q(f_1)$ and $q_2=Q(f_2)$. We first perform an elastic registration by solving for the optimal rotation $O$ and optimal re-parameterization $\gamma$ such that $| q_1 - O(q_2, \gamma)|$ is minimized (see~\cite{jermyn:2012} for details). If  $O^*$ and $\gamma^*$ is the solution to this optimization problem, then  $\bar{f}_2= O^*(f_2\circ \gamma^*)$ and its corresponding SRNF is given by $\bar{q}_2 = O^*(q_2, \gamma^*)$. This is equivalent to performing the analysis in the quotient space ${\cal S}  = \Space{Q}/{\cal G}$.

\subsection{Geodesic Path Reconstruction}
\label{sec:geodesic_path}

Given two surfaces $f_1$ and $f_2$, one wants to construct a geodesic path $\alpha(t)$, such that $\alpha(0)=f_1$ and $\alpha(1) = \bar{f}_2$.   Let $\beta: [0,1] \to \ltwo(\stwo, \rthree)$ denote the straight line connecting $q_1$ and $\bar{q}_2$. Then, for any arbitrary point $\beta(\tau) \in \ltwo(\stwo, \rthree)$, we want to find a surface $\alpha(\tau)$ such that $\|Q(\alpha(\tau)) - \beta(\tau)\|$ is minimized. We will accomplish this using the following multiresolution analysis. Let $\textbf{f}_1$ and   $\bar{\textbf{f}}_2$ be the multiresolution representations of $f_1$ and $\bar{f}_2$, respectively. We then compute the multiresolution SRNFs of $\textbf{f}_1$ and $\bar{\textbf{f}}_2$, which we denote by $\textbf{q}_1$ and $\bar{\textbf{q}}_2$, respectively. Recall that $\textbf{q}_{i} = (q_i^1, \dots, q_i^m)$ where $q_i^j =  Q(f_i^j)$.  To compute the point $\alpha(\tau)$ on the geodesic path $\alpha$, we  compute $\alpha^j(\tau)$, for $j=1$ to $m$, by SRNF inversion of $\beta^j(\tau) = (1-\tau)q_{1}^j + \tau \bar{q}_{2}^j$, using $\alpha^{j-1}(\tau)$ as an initialization for the optimization procedure. The final solution $\alpha(\tau)$ is set to be $\alpha^{m}(\tau)$.

For the first iteration, one can set $\alpha^{0}(\tau)$ to be either a unit sphere or $f^{1}_{1}$. Note that Xie et al.~\cite{xie2014numerical} required building the geodesic path $\alpha$ sequentially. That is, for a small $\epsilon > 0$, Xie et al.\ start by solving for $\alpha(\epsilon)$, using $\alpha(0)=f_1$ as an initialization, and then use $\alpha(\epsilon)$ to initialize the next step for finding $\alpha(2 \epsilon)$,  and so on. This is no longer  a requirement with our approach; one can   compute $\alpha(\tau)$ for any arbitrary $\tau \in [0, 1]$ without computing the entire geodesic.

Finally, if $f_1$ and $f_2$ are star-shaped surfaces, one can use the analytic solution of Section~\ref{sec:starshapedsurfaces} together with numerical inversion to construct the geodesic path $\alpha$ as follows. First, find the geodesic in the quotient space  ${\cal S}  = Q({\cal F})/{\cal G}$ between the corresponding SRNFs, which is trivially a straight line. It is not guaranteed, however, that all the intermediate SRNFs correspond to star-shaped surfaces; thus the analytic form $\tilde{f}$ may not be the correct inversion. One can use $\tilde{f}$, however, as an initial guess for the inverse, thereby better initializing the reconstruction-by-optimization problem.

\subsection{Shooting Geodesics}
\label{sec:geodesic_shooting}

Given a surface $f$ and a tangent vector $v_0$ at $f$, one wants to construct a geodesic $\alpha(\tau)$ such that $\alpha(0)=f$ and $\dot{\alpha}(0) = v_{0}$. (Here $\dot{\alpha} = d\alpha/d\tau$.) Note that shooting a geodesic is essentially evaluating numerically the exponential map $\exp_{f}(\tau v_0) = \alpha(\tau), \tau \in [0,1]$. Let $\beta$ denote a straight line, \ie $\beta(\tau) = Q(f) + \tau Q_{*,f}(v_0)$, where $Q_{*,f}$ is the differential of $Q$ at $f$ given by Eq.~\eqref{eqn:simplify}. Then the desired geodesic $\alpha(\tau)$ is of the form $Q(\alpha(\tau)) = \beta(\tau)$. This path $\alpha(\tau)$ is computed using the same approach as the one described in Section~\ref{sec:geodesic_path}. Some statistical analyses computed using shooting geodesics are presented in Section~\ref{sec:analysisresults}.

\subsection{Transferring Deformation between Shapes}
\label{sec:deformation_transfer}

Given surfaces $f_1$, $h_1$ and $f_2$,  we are interested in estimating the deformation from $f_1$ to $h_1$ and then applying this deformation to $f_2$. We first optimally register $h_1$ and $f_2$ onto $f_1$ using the approach described above. To simplify the notation, we also use $f_1$, $h_1$ and $f_2$ to denote the optimally aligned and re-parameterized surfaces.

\begin{figure}[t]
    \centering 
    \subfloat[$f_1\rightarrow
    h_1$]{\includegraphics[width=.3\linewidth]{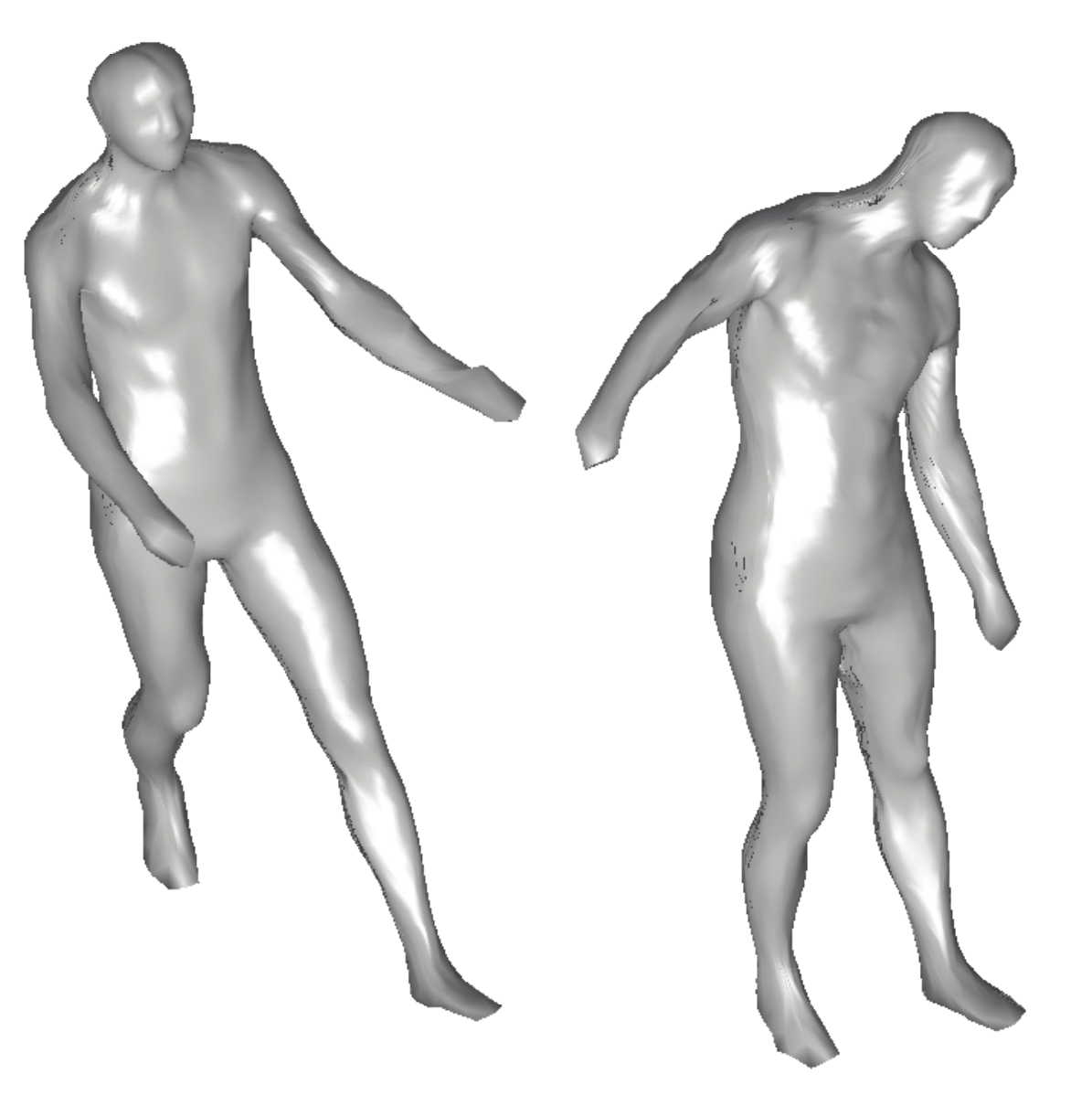}
    } \hspace{1cm}
    \subfloat[$f_2\rightarrow
    h_2$]{\includegraphics[width=.3\linewidth]{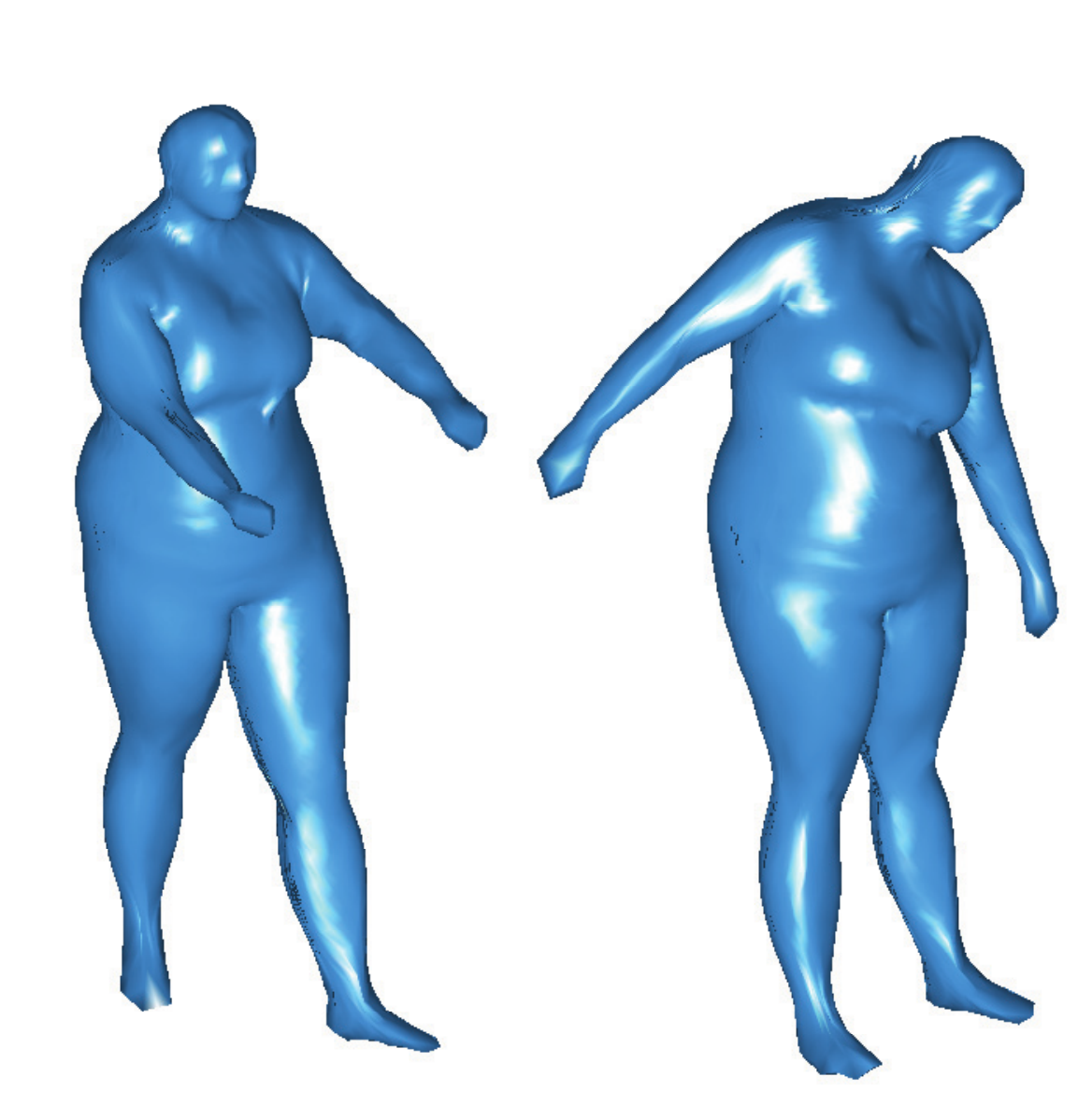}}
    \caption{Deformation transfer: surfaces $f_1$, $h_1$
    and     $f_2$ are given. Deformation from $f_1$ to $h_1$ is learnt and used to
    deform
    $f_2$ to get the new surface $h_2$.}
    \label{fig:transfer}
\end{figure}

The deformation transfer task can be decomposed into three steps: compute the deformation $v$ from $f_1$ to $h_1$; transfer $v$ at $f_1$ to $f_2$ resulting in $v^{||}$; and deform $f_2$ into $h_2$ using $v^{||}$. The first and third steps imply constructing geodesics, while the second step uses a parallel transport. With the proposed framework, $h_2$ can be computed by inversion of $Q(f_2) + \left(Q(h_1) - Q(f_1) \right)$, which is computationally more efficient than previous work, since all the computations are done in $\Space{Q}$. The algorithm is detailed in Table~\ref{tab:algorithms}. Fig.~\ref{fig:transfer} shows an example of transferring a deformation from one surface to another using the proposed approach.

\subsection{Statistical Summaries of Shapes}
\label{sec:statistical_summaries}

Given a sample of observed surfaces $f_1,\dots,f_n$, one wants to estimate the mean shape and principal directions of variation. The mean shape $\mu_f$ is computed using Algorithm~2 of Table~\ref{tab:algorithms}. Let $q_i, i=1,\dots, n$ be the SRNFs of the aligned and registered surfaces in the sample; let $\mu_q$ be their average and let $u_q^k$ be the $k^{\text{th}}$ principal component. The $k^{\text{th}}$ principal mode of variation for the SRNFs is given by $\mu_q \pm \lambda u_q^k$, $\lambda \in \real^+$.  While these summary statistics are computed using Principal Component Analysis (PCA) in $\Space{Q}$, to visualize them, one must map them back to $\Space{F}$. Thus the mean shape $\mu_f \in \Space{F}$ is computed by finding $\mu_f$ such that $Q(\mu_f) = \mu_q$. To visualize the principal directions in $\Space{F}$, we need to find $f^k$ such that $Q(f^k_{\lambda}) =\mu_q \pm \lambda u_q^k$. This is  a geodesic shooting problem, which requires one SRNF inversion for each $\lambda$.

In terms of computational complexity, computing the mean shape only requires the inversion of one multiresolution SRNF, while  previous techniques, \eg~\cite{kurtek:2010,kurtek:2012,xie-iccv:2013} are iterative, and required the computation of $n$ geodesics per iteration, with $n$ being the number of shapes to average, and each geodesic being computed using the expensive pullback metric.

\subsection{Random Sampling from Shape Models}
\label{sec:random_sampling}

Given a sample of observed surfaces $f_1,\dots,f_n \in \Space{F}$, one wants to fit a probability model to the data. Estimating probability models on nonlinear spaces like $\Space{F}$ is difficult: classical statistical approaches, developed for vector spaces, do not apply directly. By first computing the SRNF representations of the surfaces, one can use all the usual statistical tools in the vector space $\Space{Q}$, and then map the results back to $\Space{F}$ using the proposed SRNF inversion algorithms.

Let $q_1,\dots,q_n$ be the SRNFs of the registered surfaces $f_1,\dots,f_n$ and $G(q)$ be the model probability distribution fitted to $\{q_1,\dots,q_n\}$. (Recall that registration and alignment is performed using the SRNF framework described above and detailed in~\cite{jermyn:2012}.) A random sample $q_s$ can be generated from $G$. We then find $f_s$ such that $Q(f_s) = q_s$. Note that $G$ can be an arbitrary distribution, parametric or non-parametric. In this article, we use a wrapped truncated normal distribution, which can be learned using Principal Component Analysis on $\Space{Q}$. We caution the reader that the distribution is defined and analyzed in SRNF space; a distribution on $\Space{Q}$ induces a distribution on $\Space{F}$, but we have not derived it explicitly. We compute $f_s$ for the purposes of display and validation only.

\section{Experimental Results}
\label{sec:results}

Sections~\ref{sec:geodesic_results} to~\ref{sec:analysisresults} present experimental results for the shape analysis tasks described in Section~\ref{sec:analysis}. Section~\ref{sec:applications} describes two applications: the classification of generic 3D shapes, and the diagnosis of attention deficit hyperactivity disorder (ADHD) using MRI scans.

\subsection{Geodesic paths}
\label{sec:geodesic_results}
\begin{figure}[t]
    \centering
    \begin{tabular}{@{}ccc@{}}
		 \includegraphics[width=.09\textwidth]{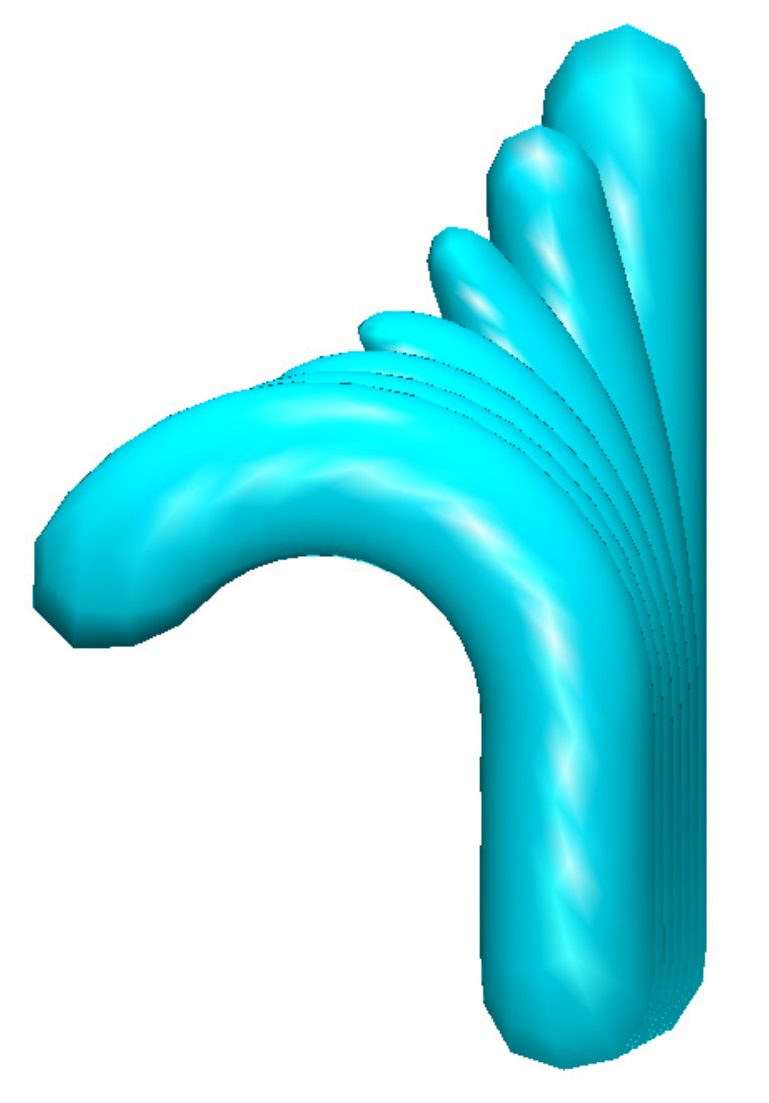}& 		 
		 \includegraphics[width=.09\textwidth]{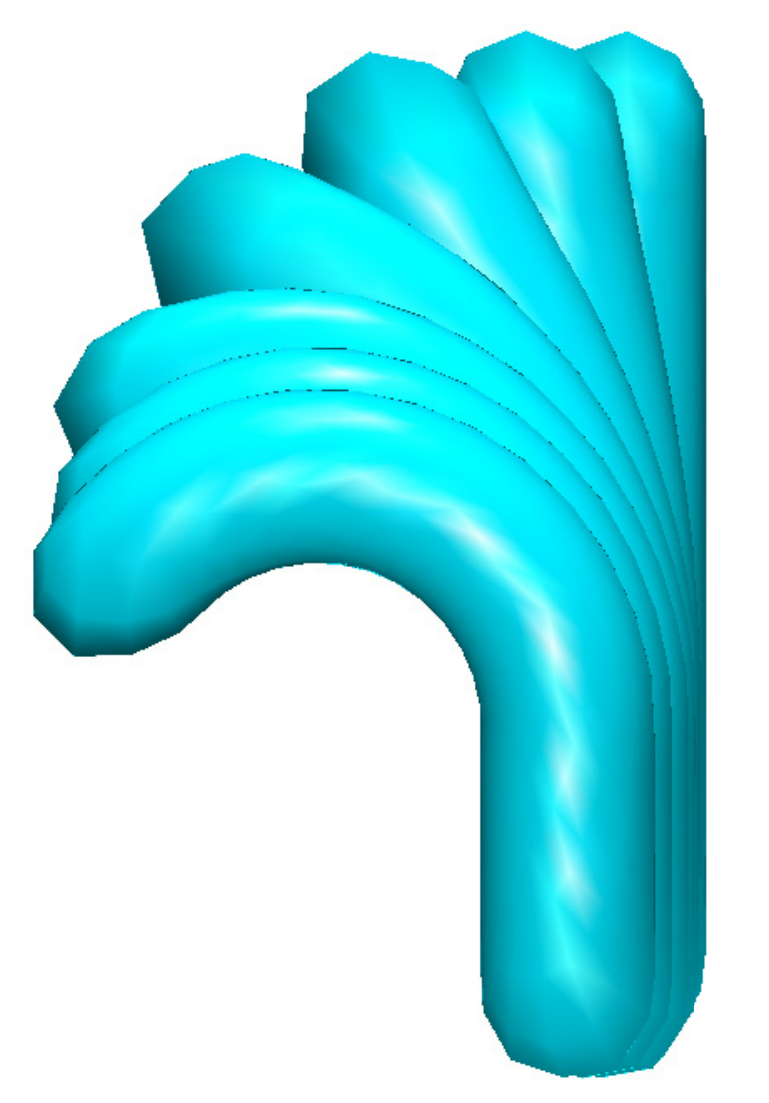}&
		 \includegraphics[width=.15\textwidth]{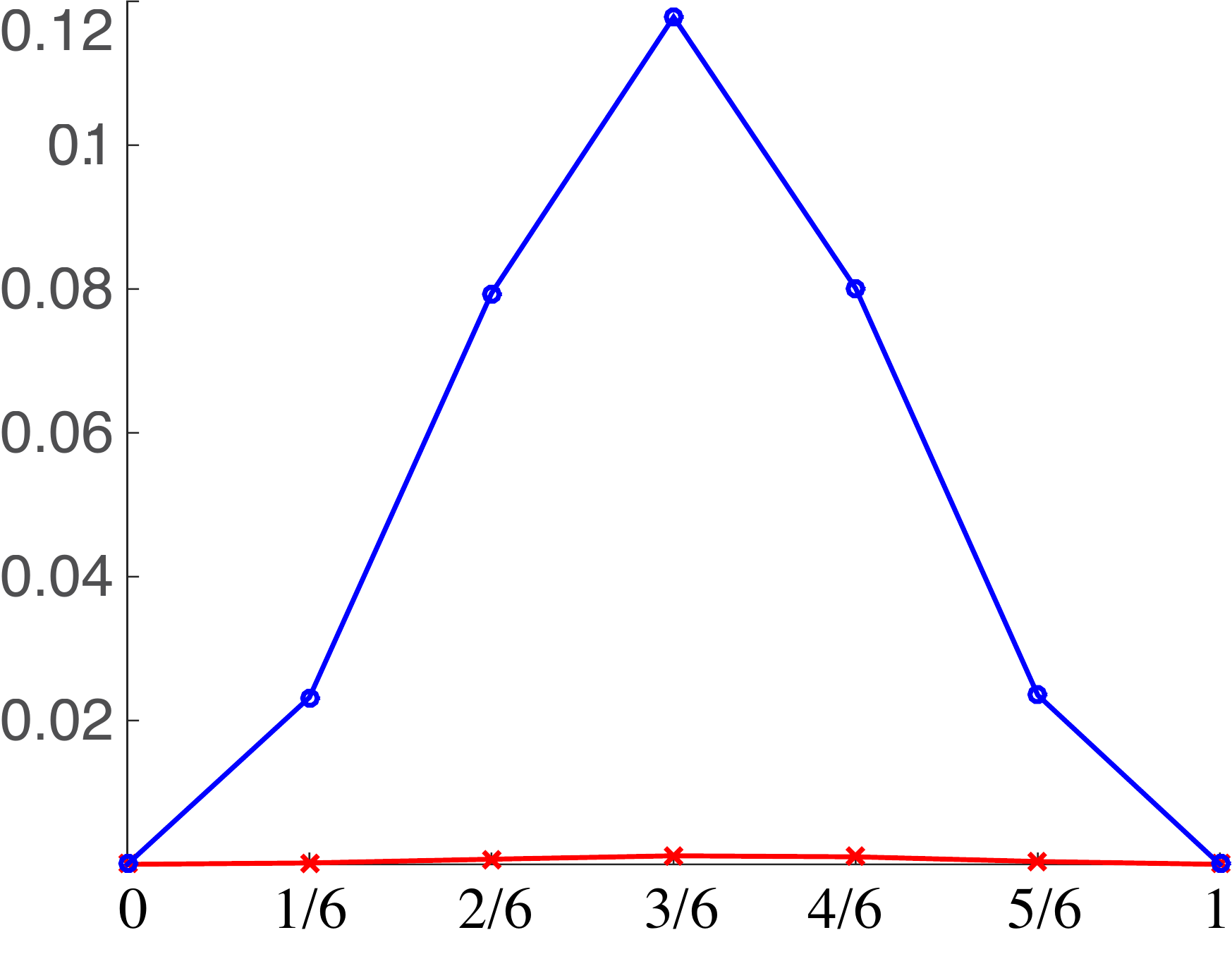}\\

		\small{(a) Linear path}    & \small{(b) Geodesic path} & \small{(c) Energy paths.}  \\
		\small{$(1-t)f_1 + tf_2$}  & \small{by SRNF inversion }\\
    \end{tabular}
    \caption{\label{fig:geod-syn} Comparison between linear interpolation in
    $\mathcal{F}$ and geodesic path by SRNF inversion. The red and blue curves in (c) correspond  to the energy path along the geodesic and along the linear path, respectively. The energy decreases from the order
    of $10^{-2}$ to $10^{-4}$.  }
\end{figure}

Fig.~\ref{fig:geod-syn} shows an example of a geodesic path between $f_1$ and $f_2$, where $f_1$ is a straight cylinder and $f_2$ is a bent cylinder. Fig.~\ref{fig:geod-syn}(a) shows  the linear path between $f_1$ and $f_2$ in ${\cal F}$ (\ie each pair of corresponding points is connected by a straight line). Observe how the intermediate shapes along this path shrink unnaturally. Fig.~\ref{fig:geod-syn}(b) shows the geodesic path computed by SRNF inversion where the cylinder bends quite naturally without shrinking.
Here the inversion procedure uses SH basis and is initialized with a sphere.
We also plot in  Fig.~\ref{fig:geod-syn}(c)  the energy of Eq.~\eqref{eq:energy_to_minimize} for each estimated surface along the geodesic path obtained by SRNF inversion (the red curve) and obtained by linear interpolation in $\mathcal{F}$. Observe that the energy of the geodesic path obtained by numerically inverting SRNFs is of  order $10^{-4}$, which is significantly lower  than the energies of the linear path (of order $10^{-2}$).

\begin{figure}[t]
    \centering

		  \includegraphics[width=.5\textwidth]{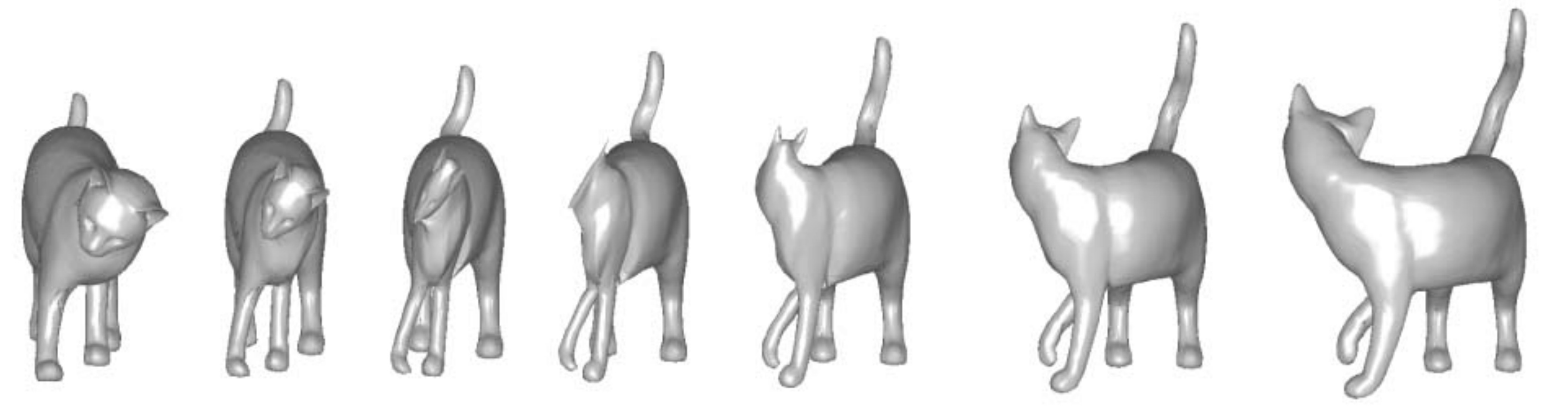}\\
		  \small{(a) Linear path $(1-t)f_1 + tf_2$}\\
		  \small{(registration computed with SRNF)}\\
		   \includegraphics[width=.5\textwidth]{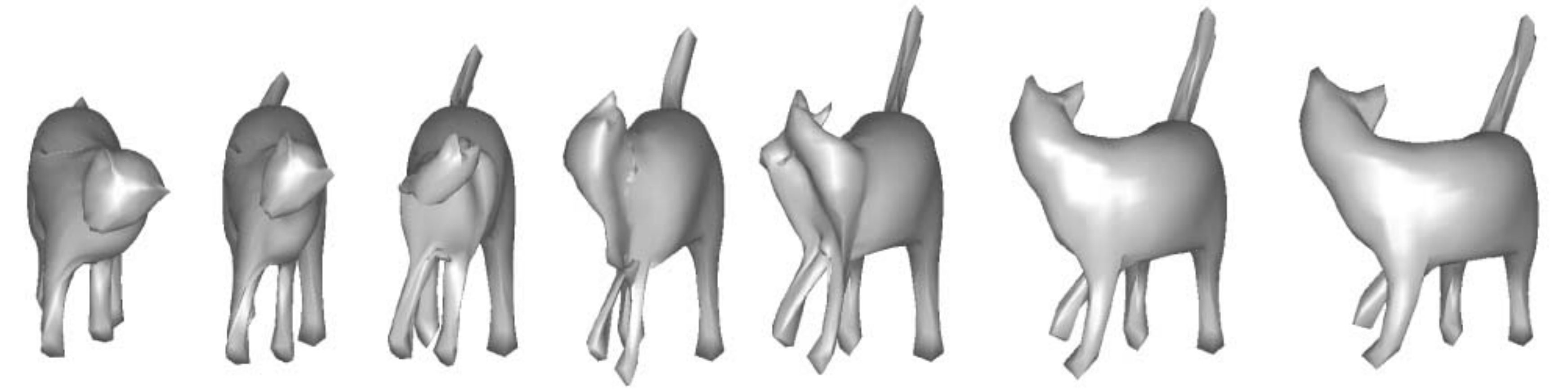}\\			  
		    \small{(b) Geodesic path using pullback elastic metric~\cite{kurtek:2012}.    }\\
		   \includegraphics[width=.5\textwidth]{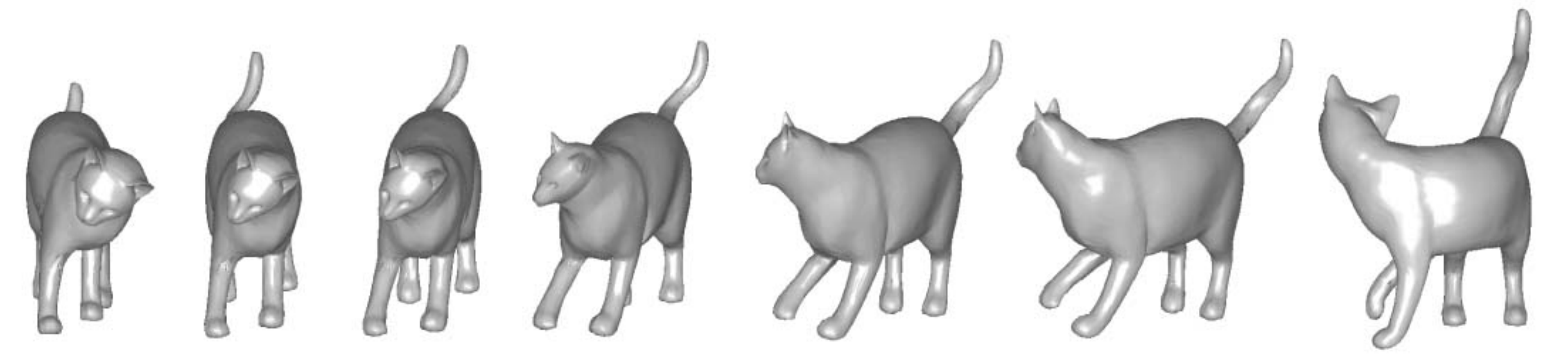}\\
		    \small{(c) Geodesic path using SRNF inversion proposed here.}

    \caption{\label{fig:geodesics_tosca_cat} Comparison between (a) linear
    interpolation in $\mathcal{F}$, (b) geodesic path estimated using the pullback metric of Kurtek et al.~\cite{kurtek:2012}, and (c) geodesic path computed using the proposed SRNF inversion.   }
\end{figure}

\begin{figure}[t]
    \centering

		  \includegraphics[width=.5\textwidth]{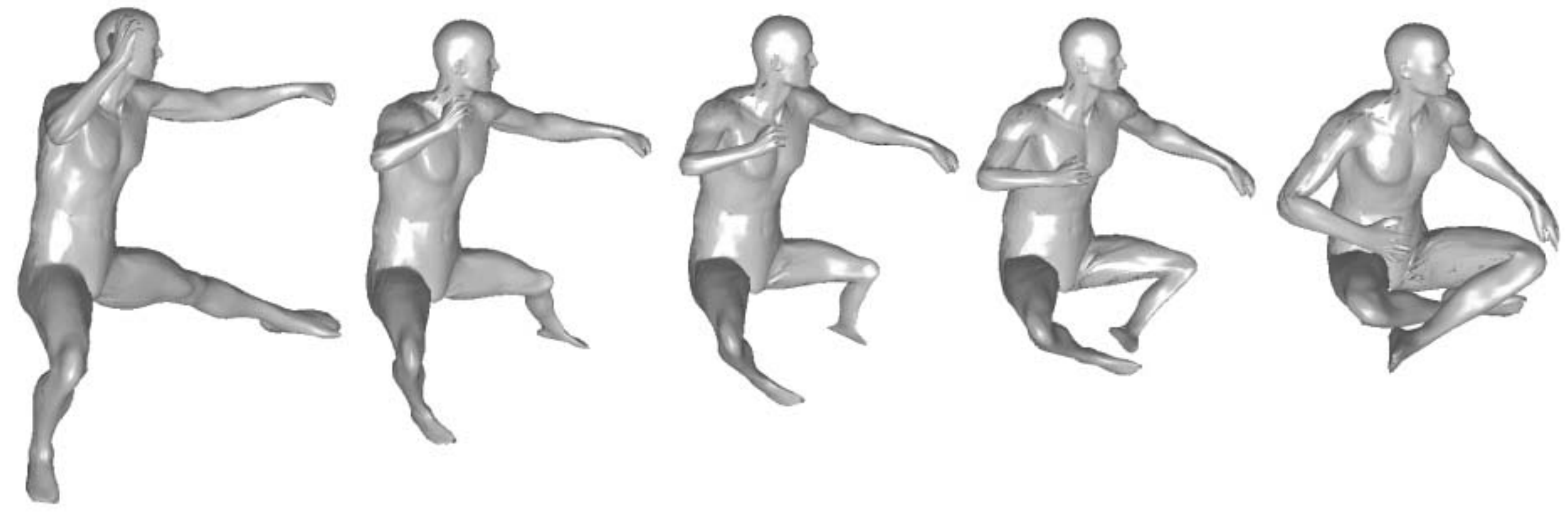}\\
		  \small{(a) Linear path $(1-t)f_1 + tf_2$}\\
		 \small{(registration computed with SRNF)}\\
		 
		  \includegraphics[width=.5\textwidth]{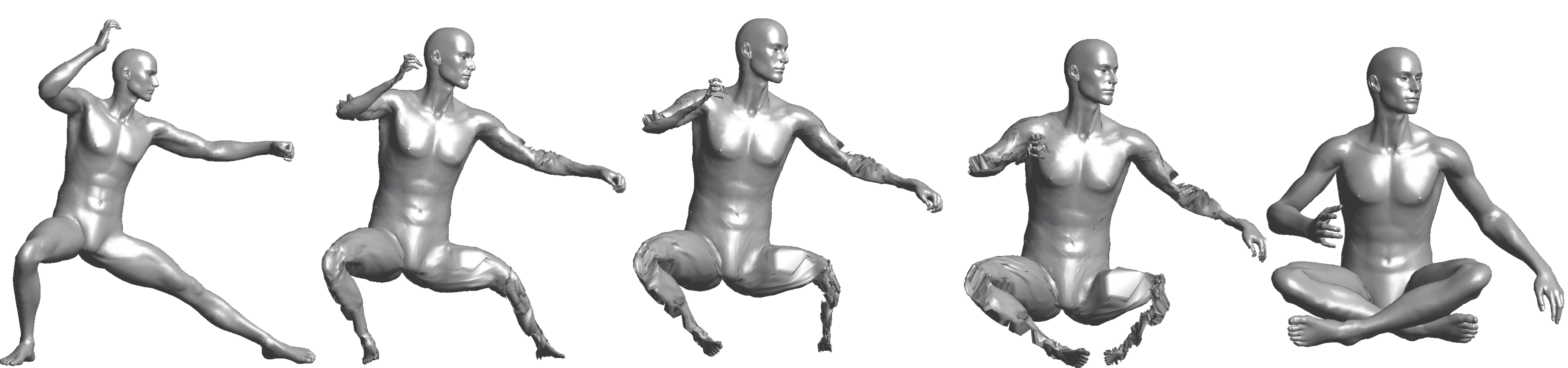}\\
		  \small{(b) Linear path  with functional maps registration~\cite{ovsjanikov2012functional}.} \\
		  
		   \includegraphics[width=.5\textwidth]{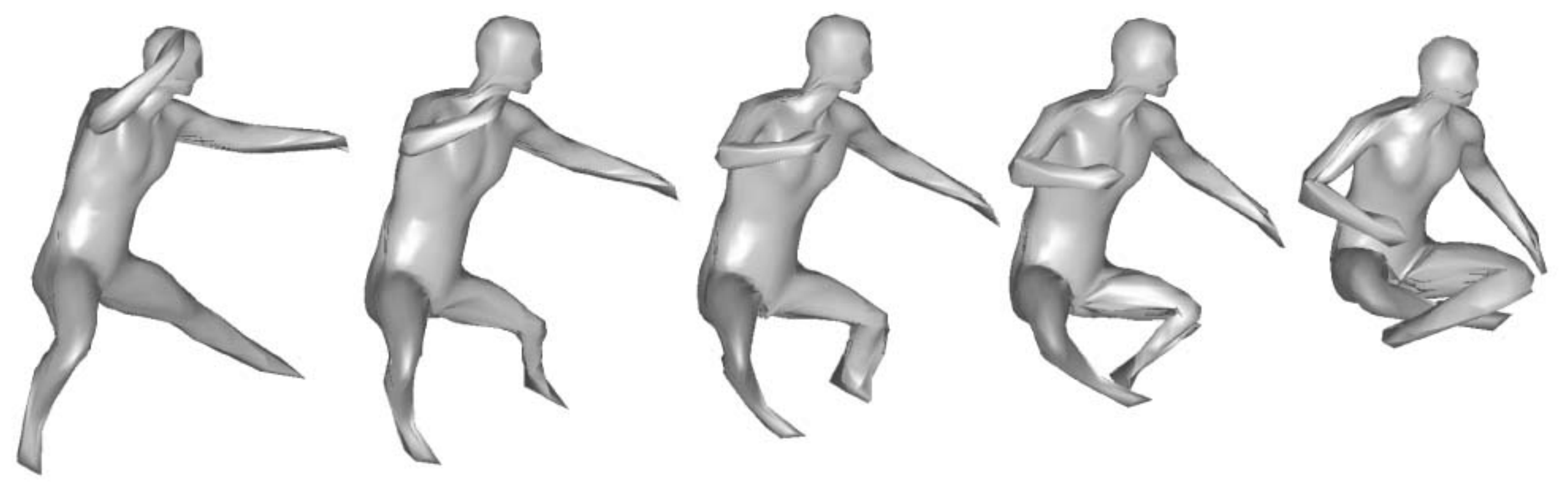}\\
		  \small{(c) Geodesic path using pullback elastic metric~\cite{kurtek:2012}.  }\\

		  \includegraphics[width=.5\textwidth]{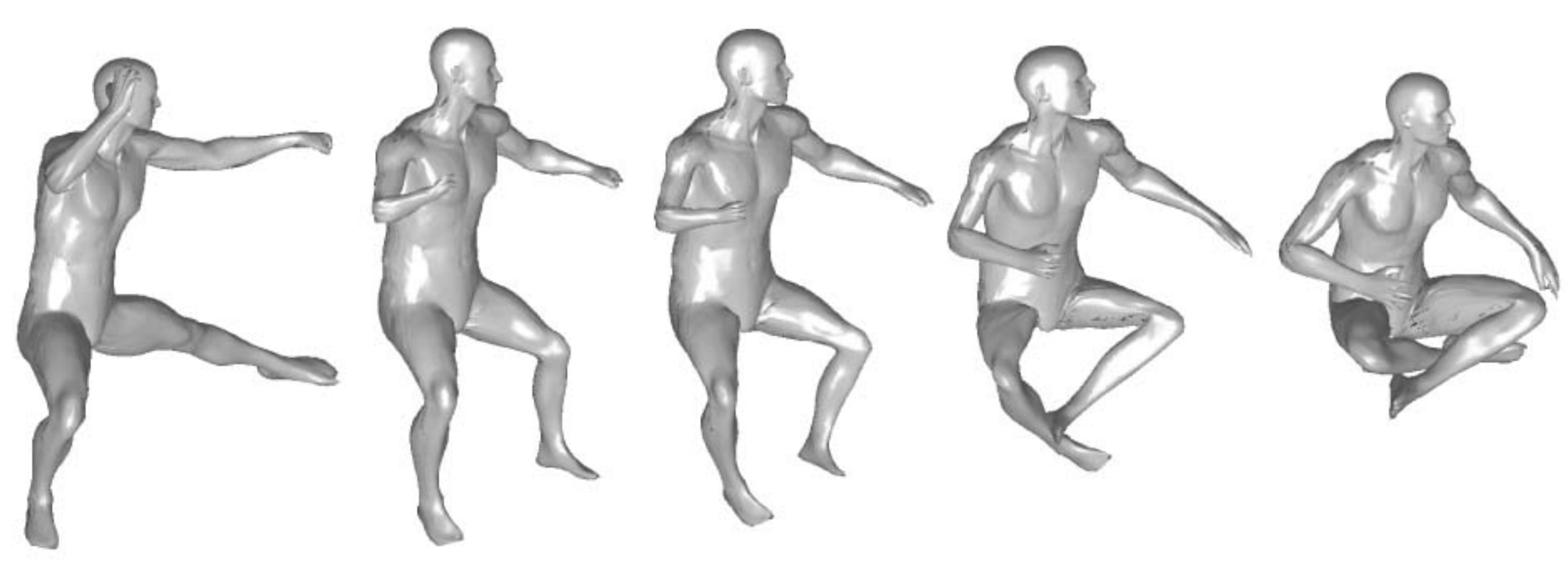}\\
		  \small{(d) Geodesic path using SRNF inversion proposed here.}

    \caption{\label{fig:geodesics_tosca_functional_maps} Comparison of our approach with different frameworks
     Observe that the in-between shapes in (a) contain artifacts due to the inaccurate correspondences computed using functional maps.  }
\end{figure}

\begin{figure}[t]
    \centering

		  \includegraphics[width=.5\textwidth]{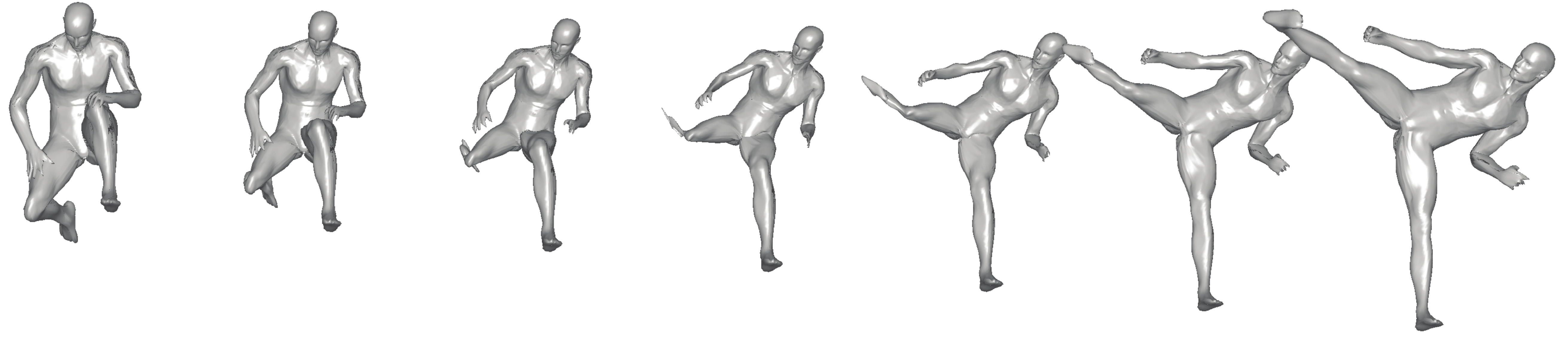}\\				  
		   \small{(a) Linear path  $(1-t)f_1 + tf_2$ }\\
		    \small{(registration computed with SRNF)}\\
		   \includegraphics[width=.5\textwidth]{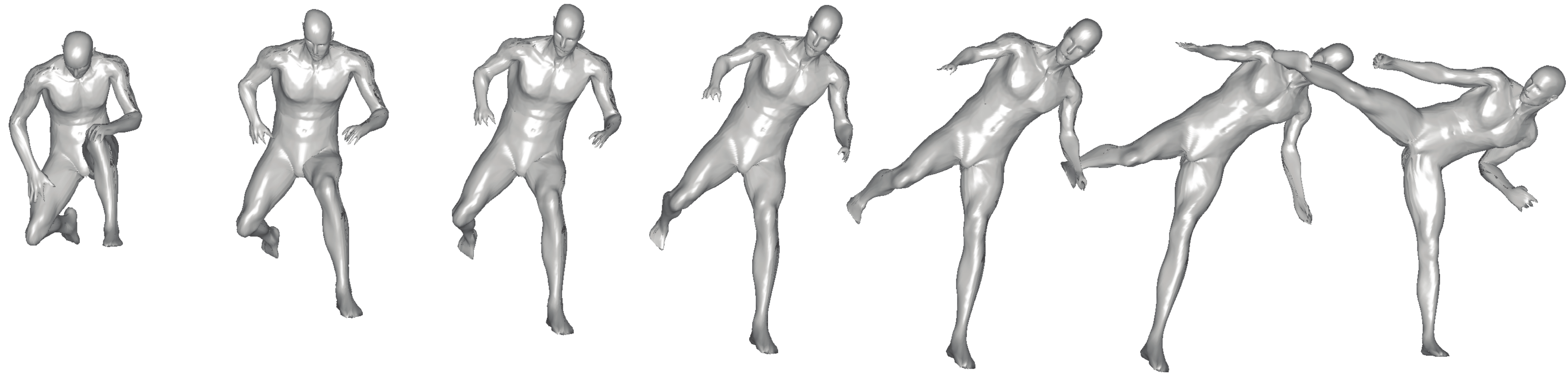}\\
		   \small{(b) Geodesic path using SRNF inversion proposed here.}\\
    \caption{\label{fig:geodesics_tosca_human} Comparison between linear interpolations in $\mathcal{F}$ and geodesic paths by SRNF inversion.   }
\end{figure}

\begin{figure}[tb]
    \centering
    		\begin{tabular}{@{}c@{}}
   		 \includegraphics[width=.45\textwidth]{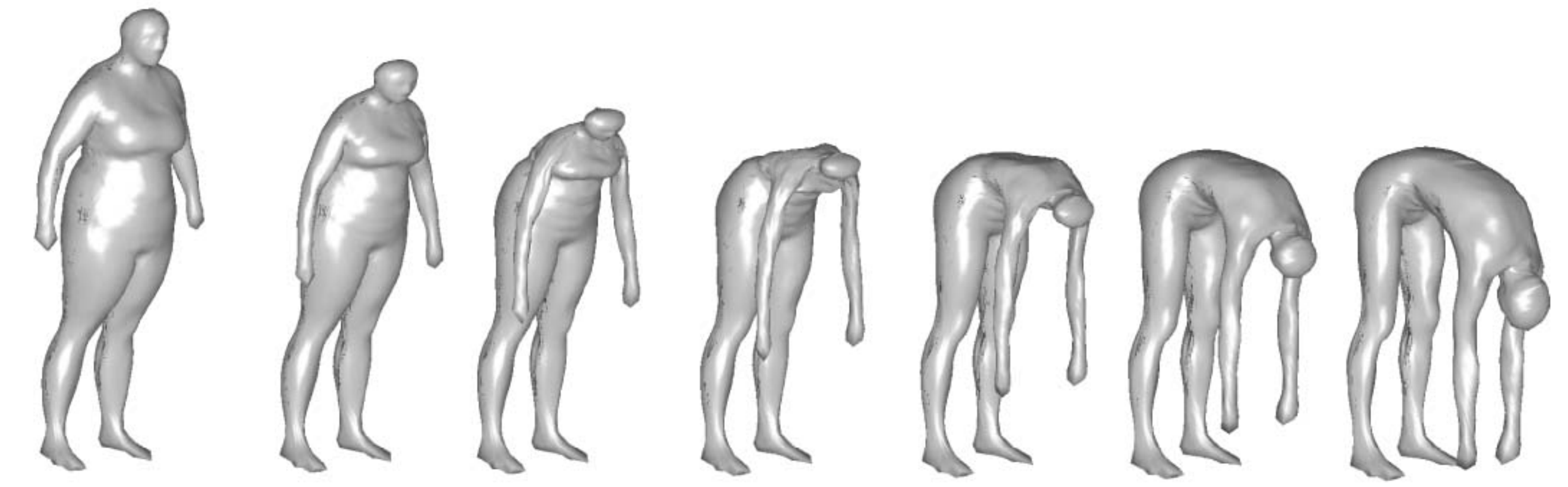} \\
		  (a) Linear path with SRNF registration. \\
		  \includegraphics[width=.45\textwidth]{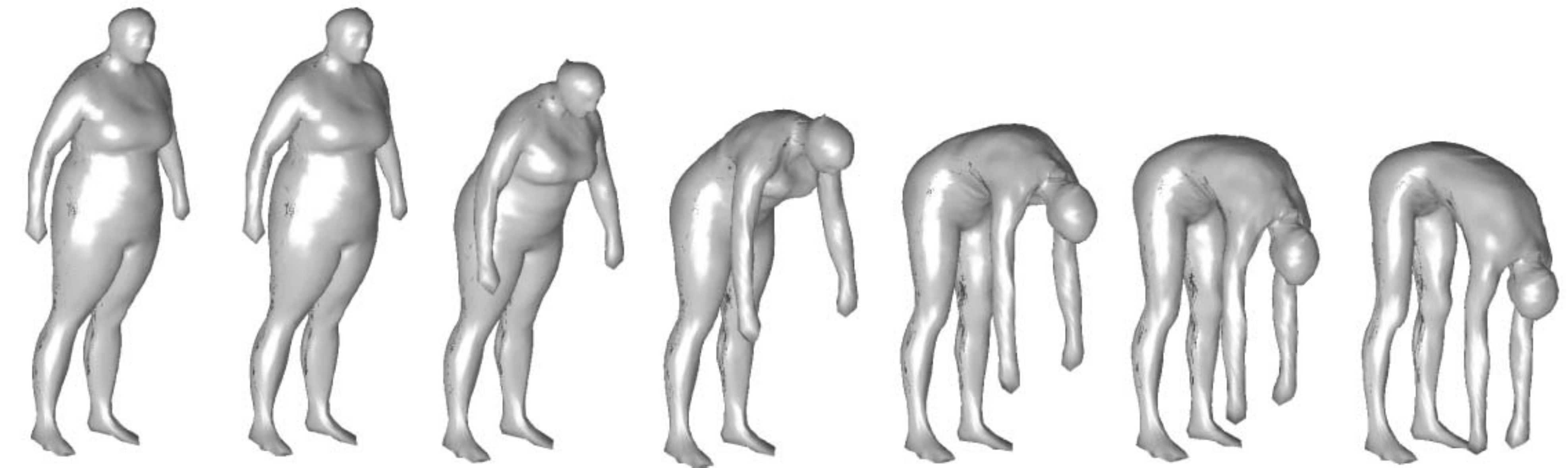}\\
		(b) Geodesic path using SRNF inversion proposed here.\\

   		 \includegraphics[width=.45\textwidth]{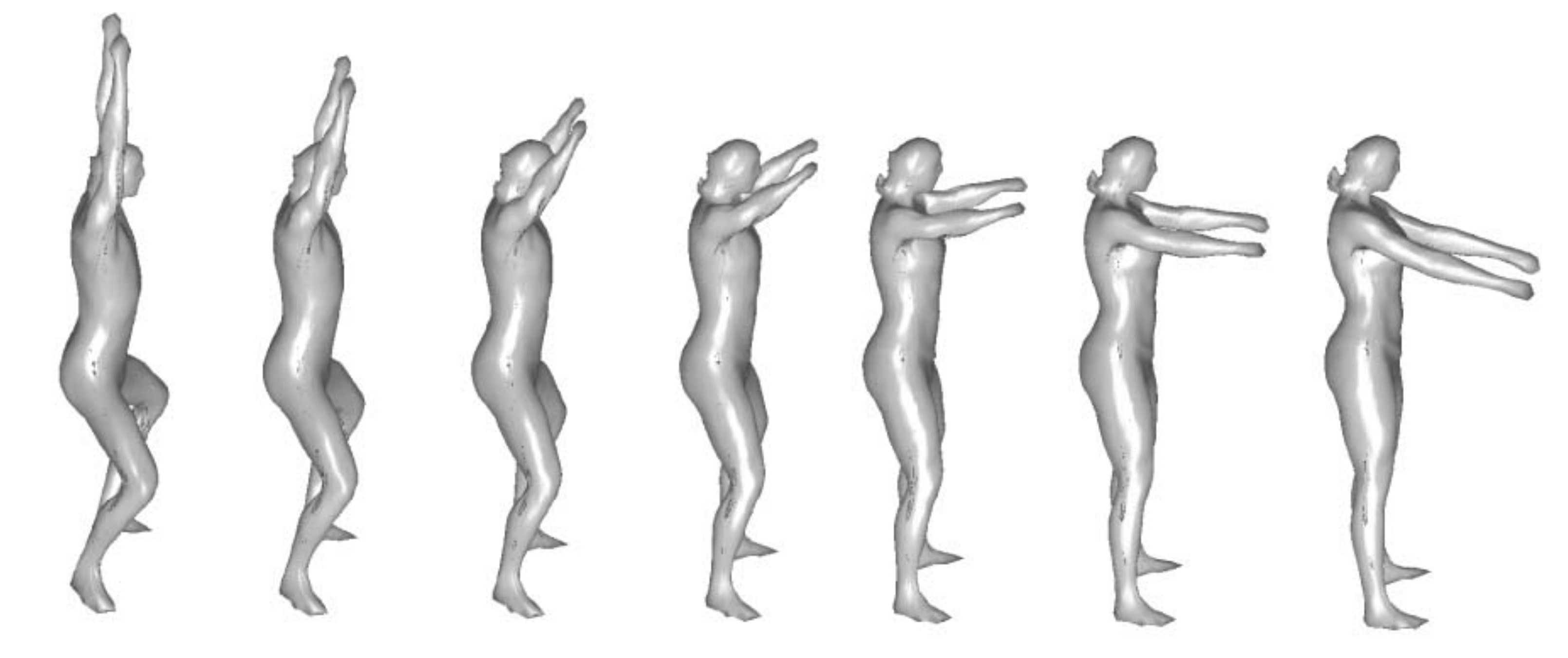} \\
		  (a) Linear path with SRNF registration. \\
		  \includegraphics[width=.45\textwidth]{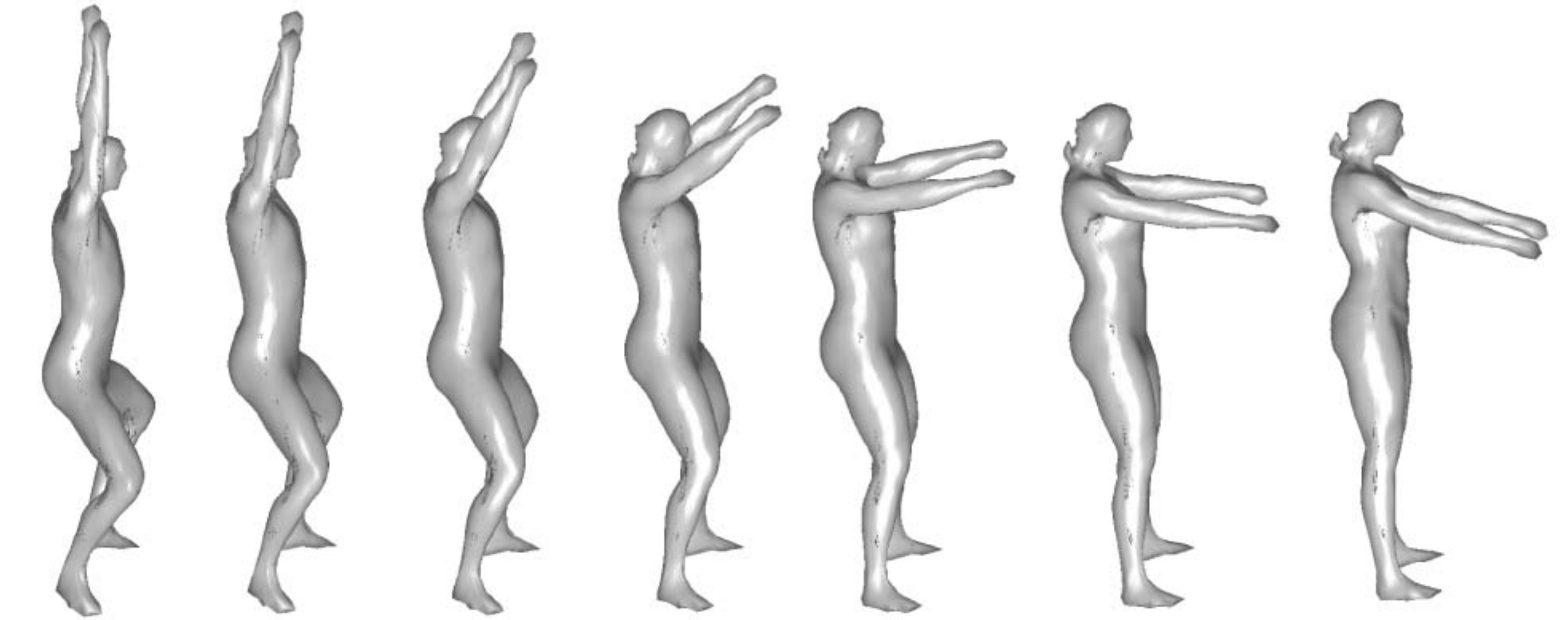}\\
		(b) Geodesic path using SRNF inversion proposed here.\\			\end{tabular}

    \caption{\label{fig:geodesics_humanshapes} Comparison between linear
    interpolations in $\mathcal{F}$ and geodesic paths by SRNF inversion (obtained using $100$ PCA
    basis elements).}
\end{figure}

Next, we present geodesics involving highly articulated surfaces undergoing complex deformations: cat shapes undergoing isometric (\ie bending)  motion (Fig.~\ref{fig:geodesics_tosca_cat}),  and human shapes in different poses undergoing both isometric and elastic deformations (Figs.~\ref{fig:geodesics_tosca_functional_maps},~\ref{fig:geodesics_tosca_human}, and~\ref{fig:geodesics_humanshapes}).
 In these examples (Figs.~\ref{fig:geodesics_tosca_cat},~\ref{fig:geodesics_tosca_functional_maps}, and~\ref{fig:geodesics_tosca_human}),
 the SRNF inversion procedure uses $3642$ SH basis
elements, while for Fig.~\ref{fig:geodesics_humanshapes}, inversion uses only $100$ PCA basis elements computed
from a collection of $398$ human shapes. We also compare our results with:\\
\noindent 	1. {\bf Linear interpolation in $\mathcal{F}$ with SRNF Registration}: Despite good quality registration between shapes, 	this approach leads to unnatural deformations, 	as shown in Figures~\ref{fig:geodesics_tosca_cat}(a), ~\ref{fig:geodesics_tosca_functional_maps}(a),~\ref{fig:geodesics_tosca_human}(a), and~\ref{fig:geodesics_humanshapes}(a).

\noindent 2. {\bf Linear interpolation in $\mathcal{F}$ with Functional Map Registration~\cite{ovsjanikov2012functional}}: As shown in Fig.~\ref{fig:geodesics_tosca_functional_maps}(b), functional maps do not produce correct one-to-one correspondences. As a result, the deformation between surfaces contains many artifacts and degeneracies. In addition, functional maps are: (1) limited to isometric surfaces, i.e. surfaces that only bend; (2) require pre-segmented shapes; and (3) require part-wise correspondences as input~\cite{ovsjanikov2012functional}. Our approach is fully automatic, and  finds both registrations and deformations between surfaces that undergo large bending and stretching.

\noindent 3.  {\bf Pullback Metric of Kurtek et al.~\cite{kurtek:2012}}: Computationally very expensive, in particular when computing summary statistics, see the discussion of Table~\ref{tab:algorithms}. The underlying metric is not translation invariant and can easily be trapped in local minima when dealing with complex surfaces: see Figs.~\ref{fig:geodesics_tosca_cat}(b) and~\ref{fig:geodesics_tosca_functional_maps}(c).

In all examples, our approach produces better results. In particular,  when the deformations between the source and target surfaces are significant, the intermediate shapes produced by linear interpolation in $\mathcal{F}$ contain self-intersections,  and unnatural shrinkage of the parts that bend.

%

\subsection{Geodesic shooting and deformation transfer}

%
%

\begin{figure*}
    \centering
    \ra{0.9}
    		\begin{tabular}{@{}c@{}c | @{ }c@{ } | c@{}}
		 \includegraphics[width=.07\textwidth]{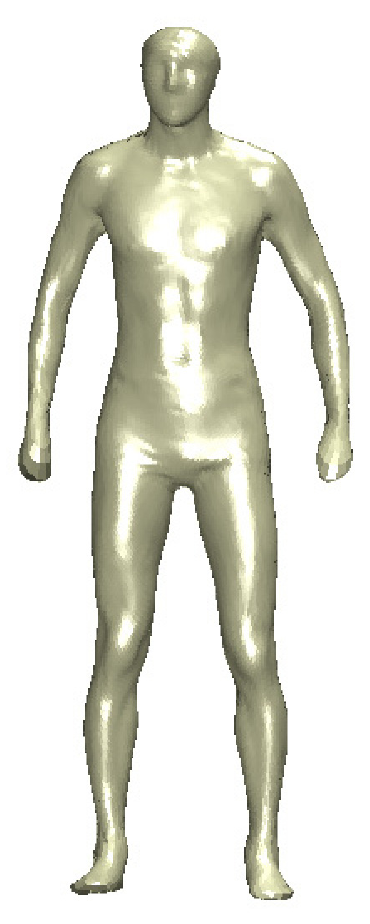}& 		
		  \includegraphics[width=.07\textwidth]{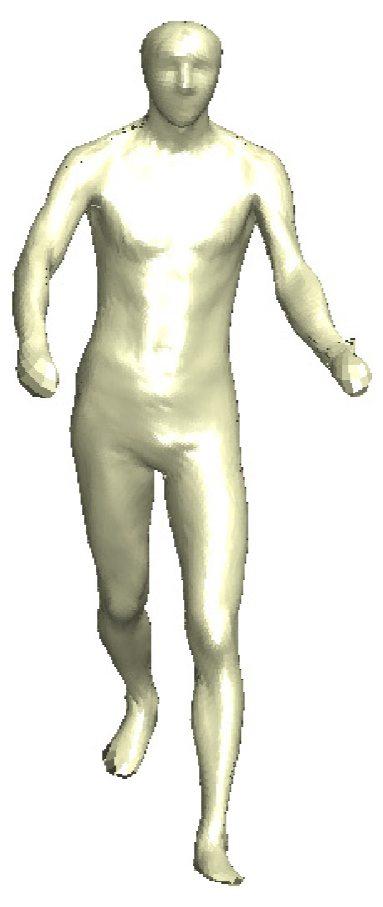}&
   		 \includegraphics[width=.4\textwidth]{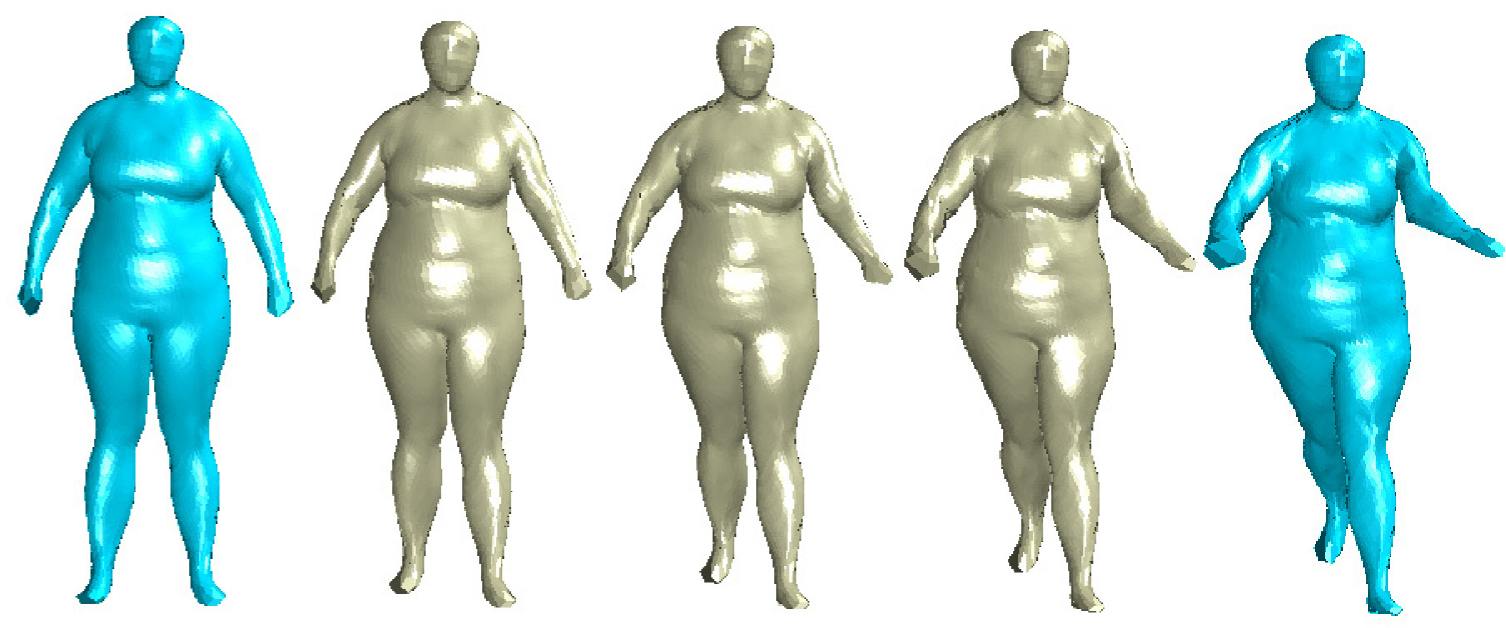} &
		 \includegraphics[width=.4\textwidth]{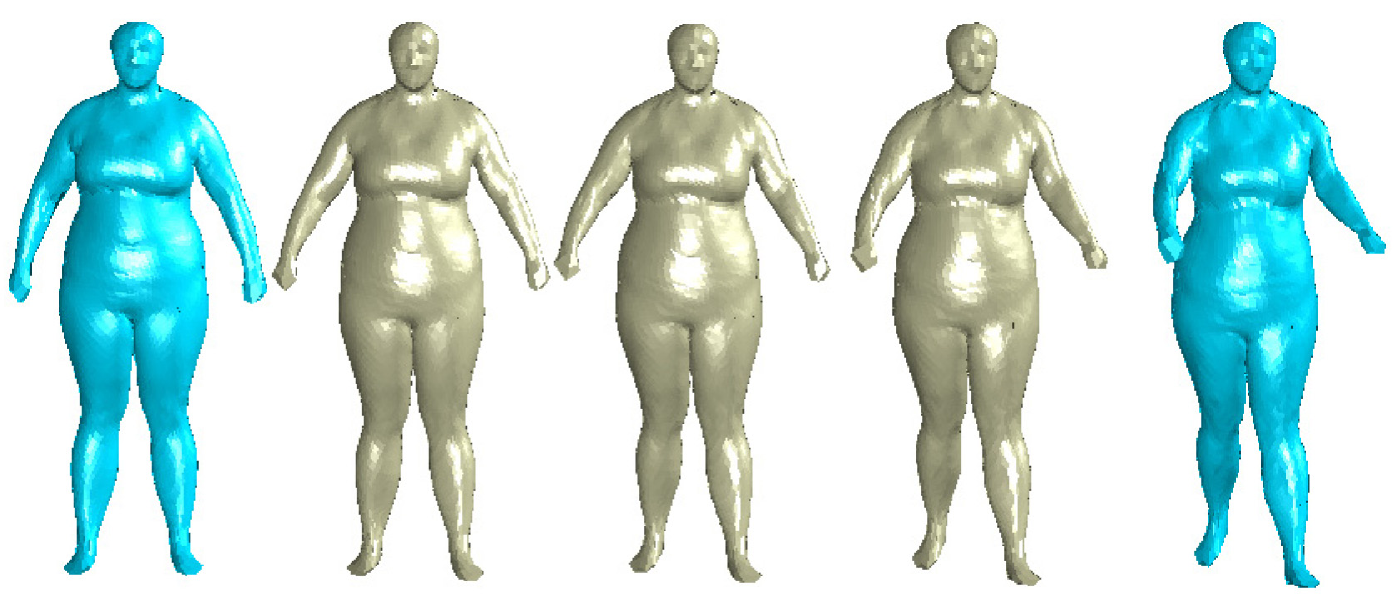} \\

		   \includegraphics[width=.07\textwidth]{s1p33}& 		 
		   \includegraphics[width=.08\textwidth]{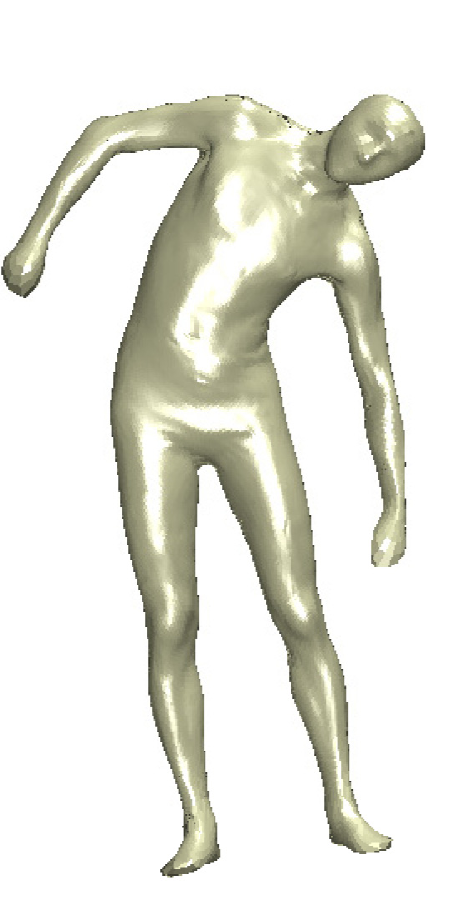}&
   		 \includegraphics[width=.4\textwidth]{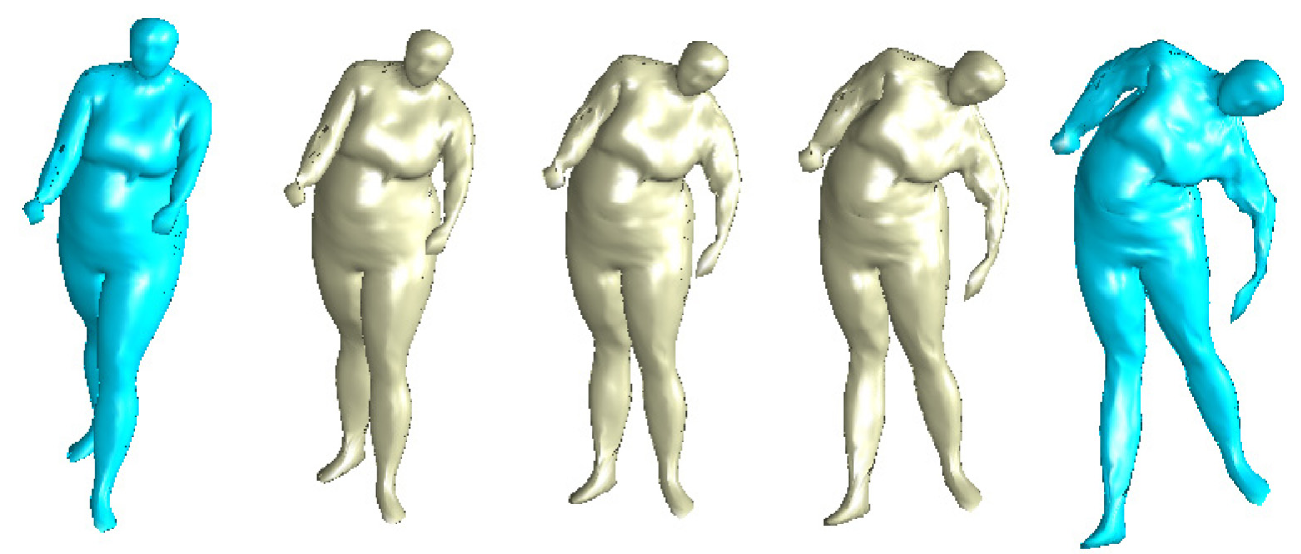} &
		  \includegraphics[width=.4\textwidth]{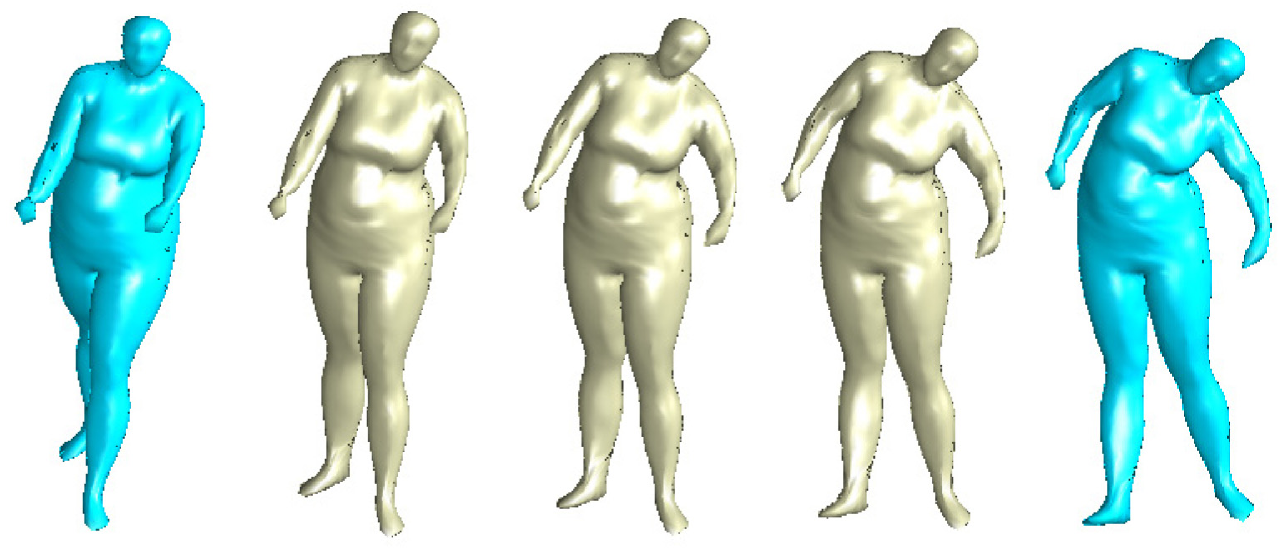} \\

		   \includegraphics[width=.07\textwidth]{s1p33}& 		
		    \includegraphics[width=.085\textwidth]{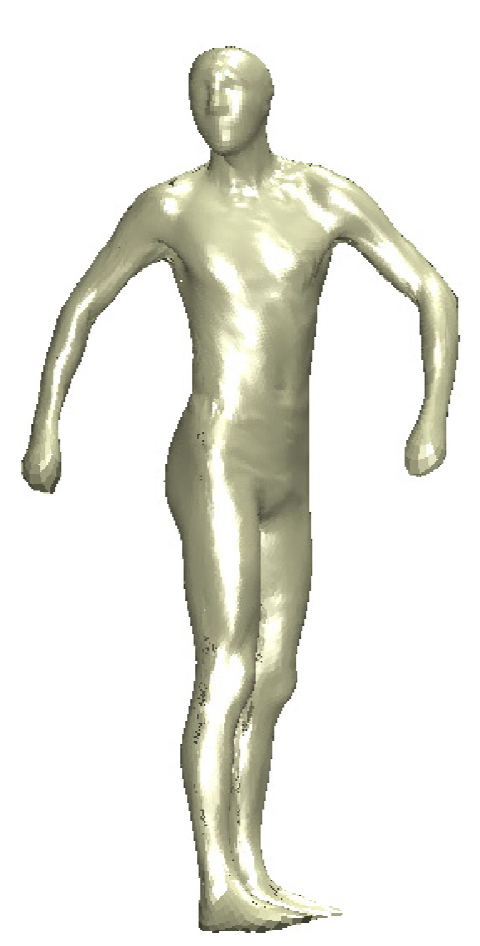}&
   		 \includegraphics[width=.4\textwidth]{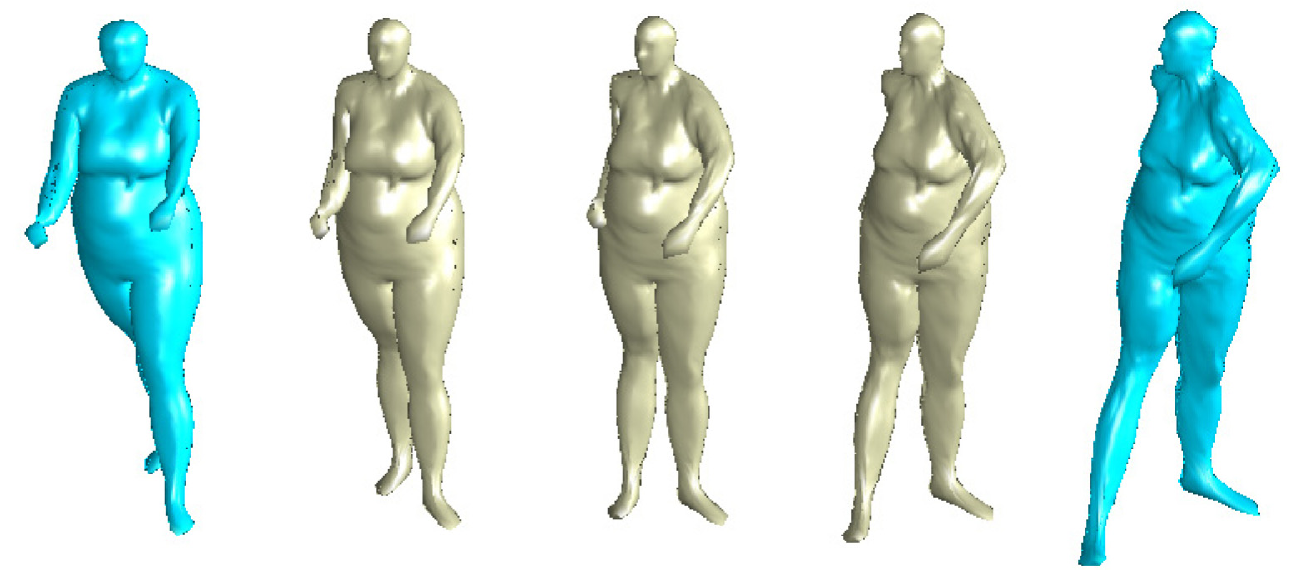} &
		  \includegraphics[width=.4\textwidth]{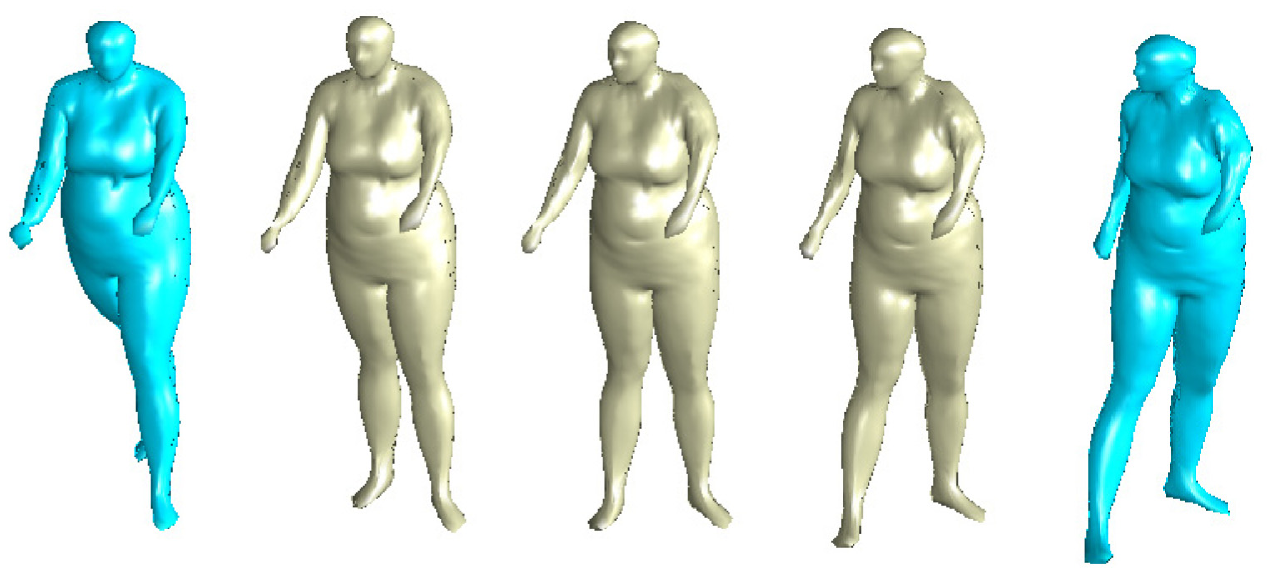} \\

		   \includegraphics[width=.08\textwidth]{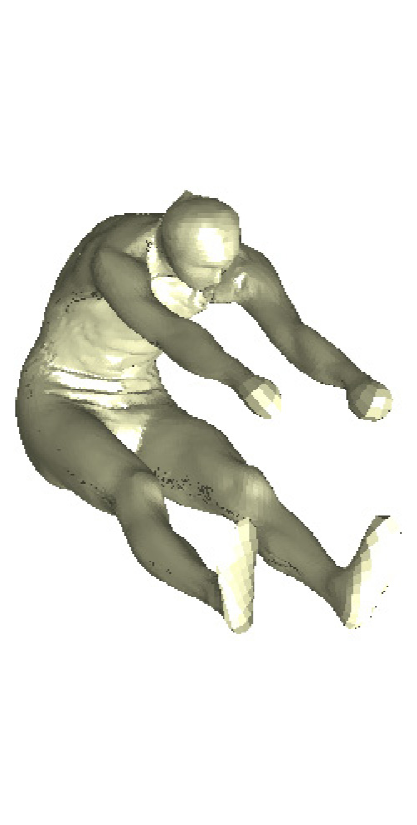}& 		
		    \includegraphics[width=.07\textwidth]{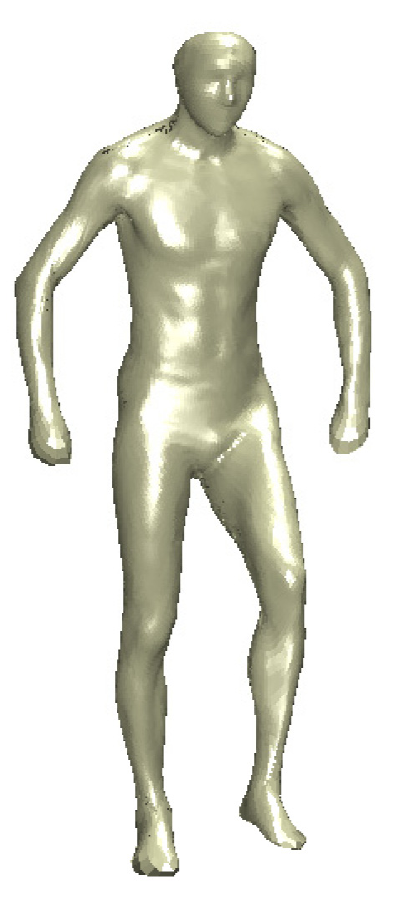}&
   		   \includegraphics[width=.4\textwidth]{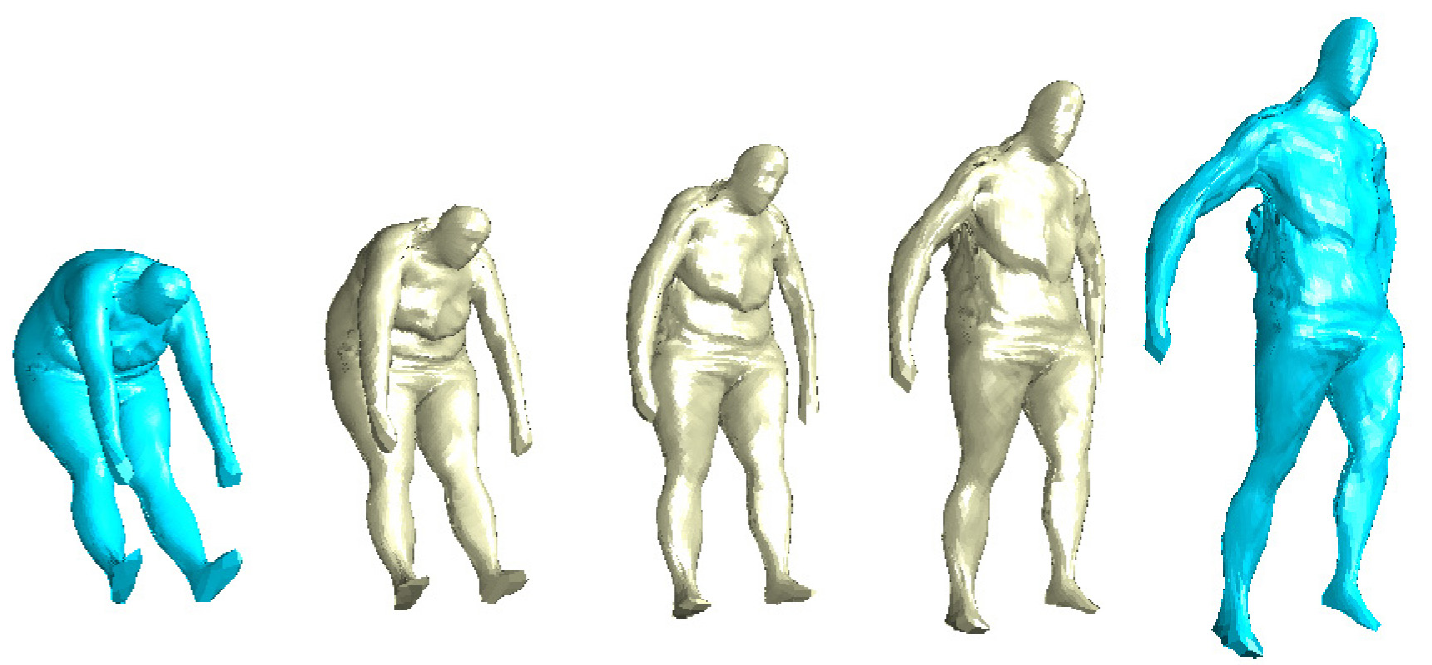} &
		  \includegraphics[ width=.4\textwidth]{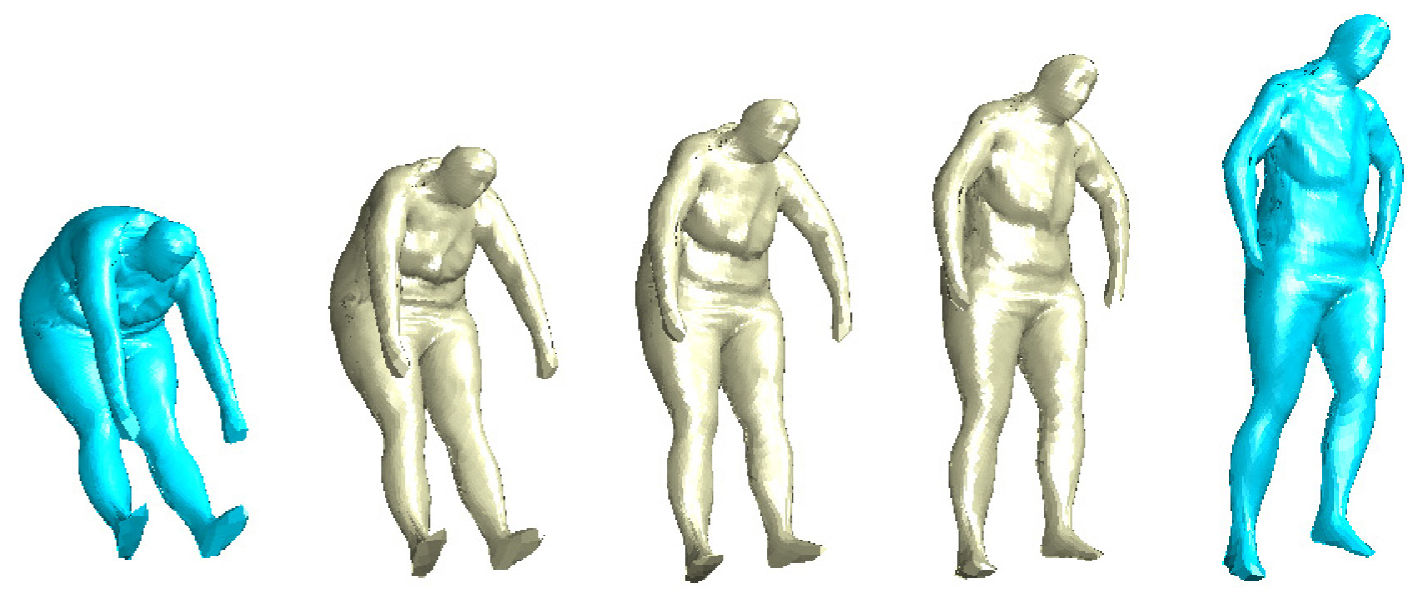} \\

		$f_1$ & $h_1$ & $f_2$ \hspace{3.5cm} $h_2$ & $f_2$ \hspace{3.5cm}
$h_2$  \\

		 \multicolumn{2}{c|}{\small{(a) Source } }&  \small{(b)
Deformation transfer by linear } & \small{(c) Deformation transfer by SRNF } \\
		 \multicolumn{2}{c|}{\small{ deformation}}  & \small{extrapolation $ h_2 = f_2 + \alpha(h_1 - f_1)$} & \small{ inversion $Q(h_2) = Q(f_2) +\alpha(Q(h_1) - Q(f_1))$}.

		\end{tabular}
    \caption{\label{fig:deformationtransfer_humanshapes} Examples of
    deformation transfer using geodesic shooting in the space of SRNFs.  The
    middle shapes, in gray, are intermediate shapes along the deformation
    paths.  }
\end{figure*}

Fig.~\ref{fig:deformationtransfer_humanshapes} shows  several examples of deformation transfer across human body shapes using geodesic shooting and parallel transport (Sections~\ref{sec:geodesic_shooting} and~\ref{sec:deformation_transfer}). The source deformations, \ie the surfaces $f_1$ and $h_1$, are shown in Fig.~\ref{fig:deformationtransfer_humanshapes}(a) while Fig.~\ref{fig:deformationtransfer_humanshapes}(c) shows these deformations transferred onto another subject using geodesics obtained by SRNF inversion. For comparison, we also show in Fig.~\ref{fig:deformationtransfer_humanshapes}(b) the deformations transferred by linear extrapolation in $\mathcal{F}$. Observe that the deformations transferred by SRNF inversion look very natural and are free from artifacts, self-intersections, and unnatural elongations, contrary to the deformations obtained by linear extrapolation.


\subsection{Statistical analysis of surfaces}
\label{sec:analysisresults}

We  consider a collection of $398$ human shapes~\cite{hasler2009statistical} composed of multiple subjects in different poses. The first subject has $35$ different poses including a neutral one. All other subjects have a neutral pose and a few other poses.
We first compute the SRNF representations of these surfaces, perform statistical analysis in the quotient space, ${\cal S}  = \Space{Q}/{\cal G}$, using standard linear statistics such Principal Component Analysis, and finally map the results to the surface space, $\Space{F}$, using the proposed SRNF inversion algorithm.

\begin{figure}[t]
    \centering
    \begin{tabular}{@{}cc@{}}
    	\includegraphics[width=.23\textwidth]{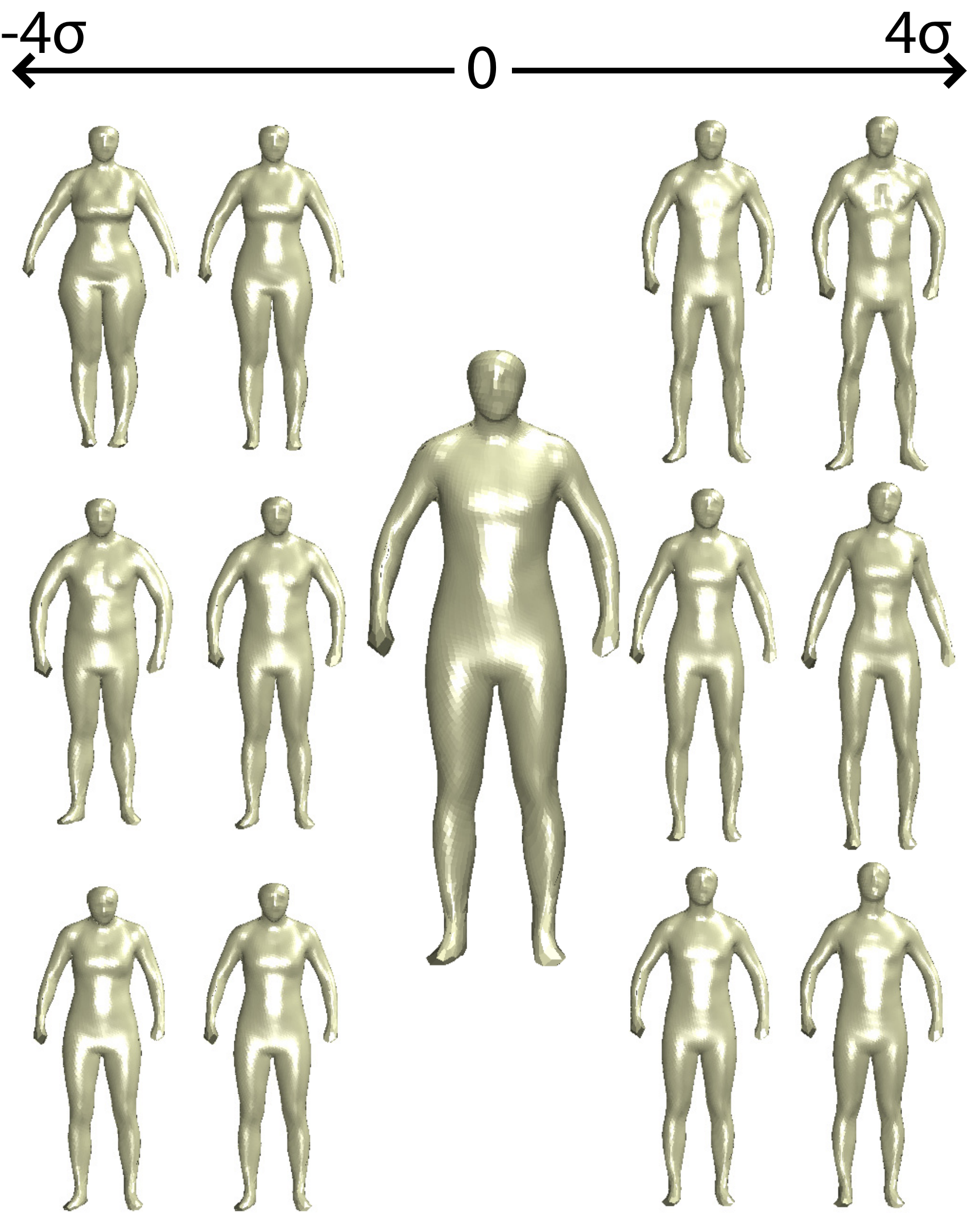} & 
	\includegraphics[width=.23\textwidth]{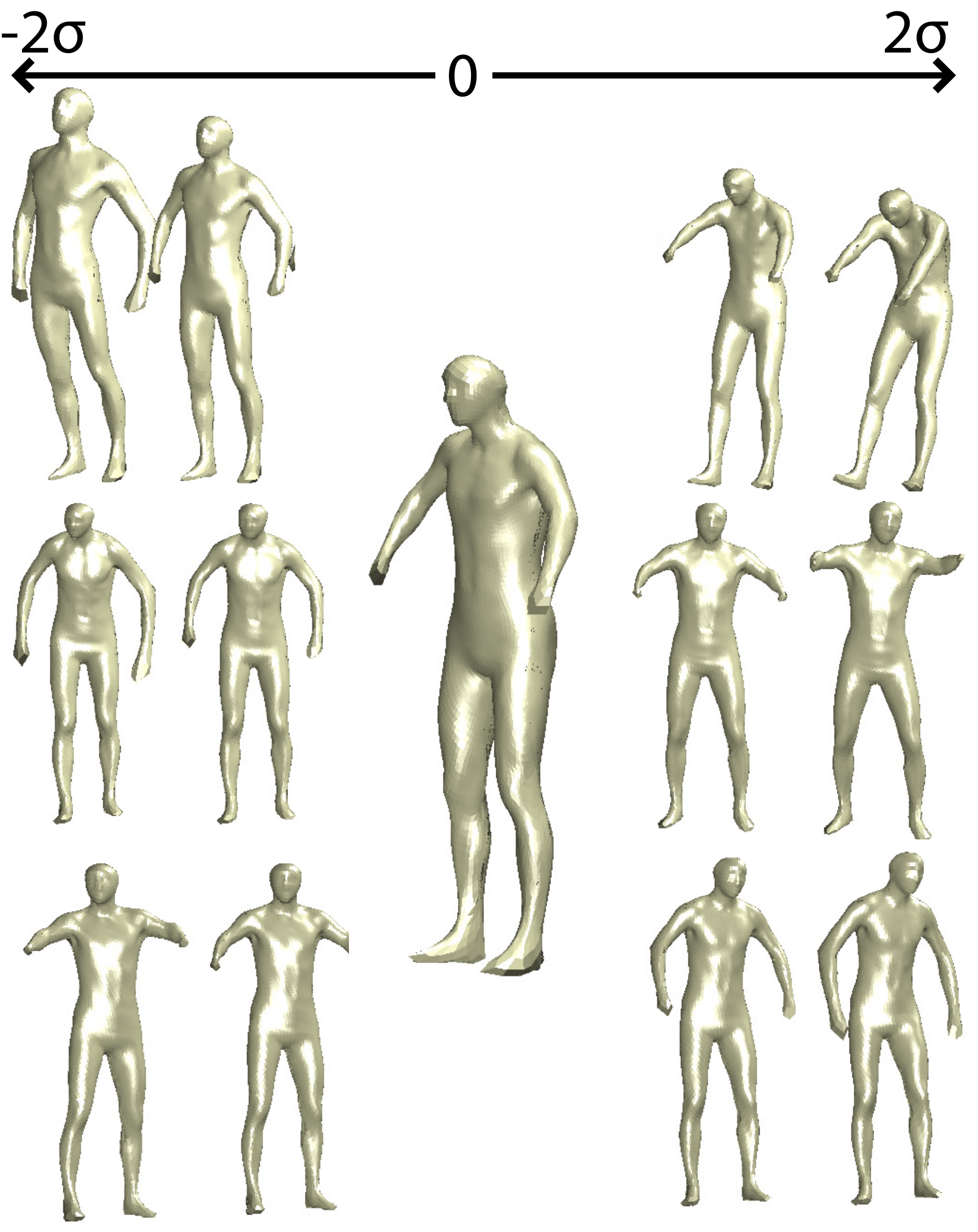} \\
	\small{(a) Mean shape and its top } & \small{(b) Mean pose  and  its top }\\
	\small{three modes of variation.} & \small{three modes of variation.}\\
   \end{tabular}

   \begin{tabular}{c}
	 \includegraphics[width=.4\textwidth]{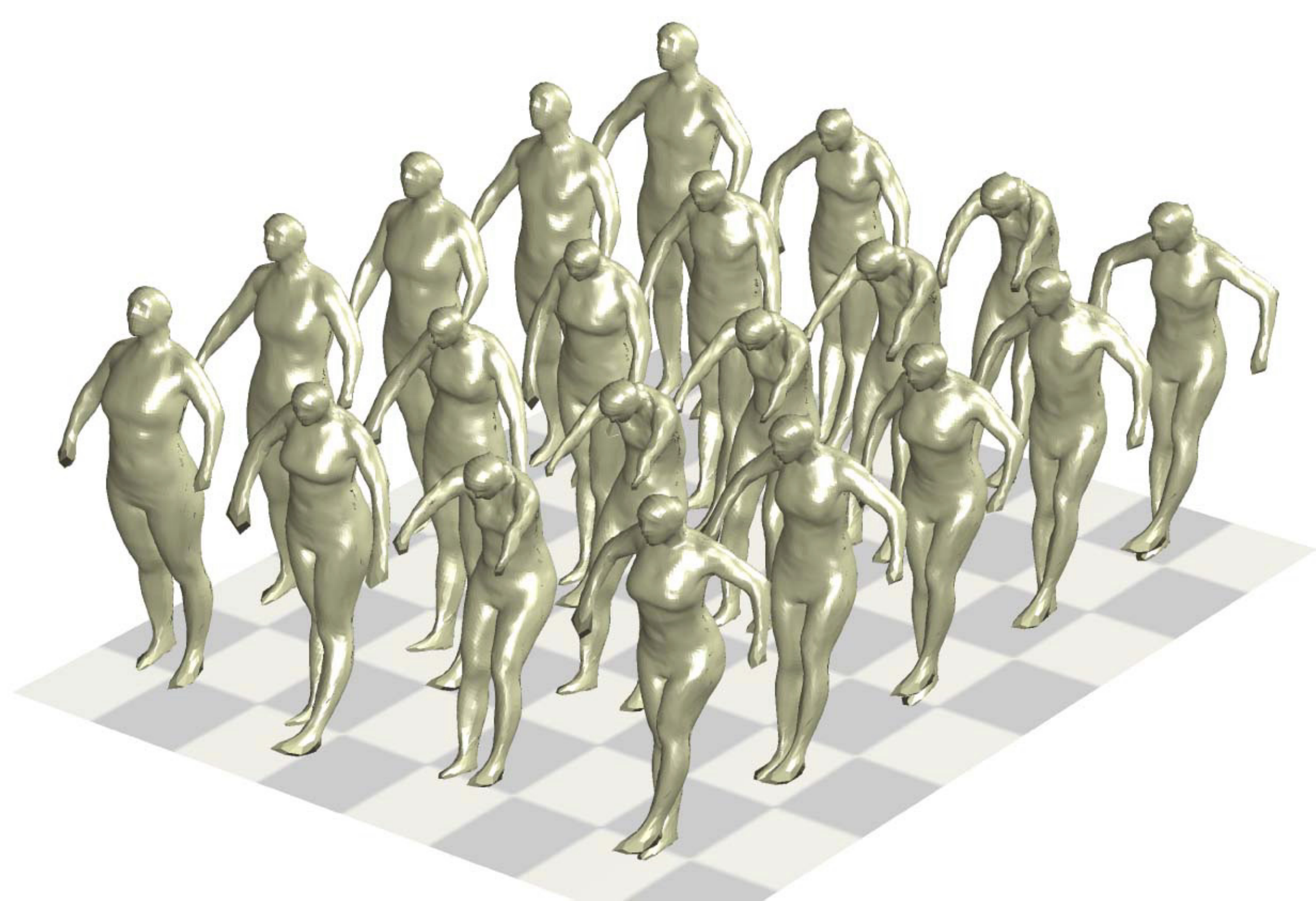} \\
	\small{(c) Random samples.}
    \end{tabular}
    \caption{\label{fig:mean_modes_shape} Statistical analysis of 3D human body shapes using the SRNF inversion proposed in this article.}
\end{figure}

Fig.~\ref{fig:mean_modes_shape}(a) shows the mean shape and the first three modes of variation computed using all the subjects in a neutral position. We refer to this statistical model as  {\it pose-independent}. Similarly, we consider the $35$ different poses of the first subject and compute their mean shape and modes of variation (Fig.~\ref{fig:mean_modes_shape}(b)). We call the resulting model the {\it pose shape model}.

We use these two statistical models to synthesize random human shapes in arbitrary poses. Let $f_1$ be the neutral pose in the $35$ models we used to learn the pose shape model. First, we randomly generate a surface $f_2$ that is in a neutral pose using the pose-independent statistical model. We then generate an arbitrary pose $h_1$ using the pose shape model. Finally, we use the deformation transfer procedure (Section~\ref{sec:deformation_transfer}) to deform $f_2$ in the same way that $f_1$ deforms into $h_1$. This results in a new random human shape $h_2$ in  an arbitrary pose. Fig.~\ref{fig:mean_modes_shape}(c)  shows $20$ arbitrary human shapes generated with this method. This approach uses two separate models to capture inter- and within-subject variabilities; used with the deformation transfer mechanism, this provides greater control over the generation of different subjects and poses. One could also can consider all the human shapes together irrespective of their pose and learn a single statistical model from which random instances can be sampled.

Table~\ref{tab:algorithms} summarizes the computational benefits of our framework compared to previous techniques. For example, for~\cite{kurtek:2012}, computing the mean shape of $N$ objects would require $N\times m$ geodesics, where $m$ is the number of iterations in their iterative algorithm. This is very expensive. This is also true for methods such as \cite{berkels2013discrete,zhang2015shell}, which have a similar approach for geodesic construction. In addition, \cite{berkels2013discrete,zhang2015shell} assume that the input surfaces are fully registered, using another, unrelated, method. Thus, these prior approaches, based on finding individual  geodesics using the pullback metric in $\mathcal{F}$, do not scale to large datasets. Our approach simultaneously finds  correspondences and geodesics.  Its computational cost  is limited to the cost of finding a Euclidean mean and  inverting  one SRNF. Thus, it is computationally very efficient and also scalable.

\subsection{Classification Applications}
\label{sec:applications}\label{sec:classification}

\begin{table}[t]
\centering{ \caption{\label{tab:shrec2007_results} Classification performance, in $\%$,  on the genus-0 watertight 3D models of SHREC07~\cite{giorgi2007shape}.} 	\begin{tabular}{@{}lcccccc@{}} 	\toprule
					&  Gauss & Gauss & 1NN & 3NN & 1NN & 3NN\\
					&  SRNF & $\mathcal{F}$ & SRNF & SRNF &  $\mathcal{F}$  & $\mathcal{F}$ \\
	\midrule
	w/o regist. 	& $52.42$ & $44.49$ & $52.86$ & $50.66$ & $56.38$ & $52.42$ \\
	\midrule
	ICP~\cite{besl:1992} 				&$-$ & $-$ &$-$ &$-$ &  $82.81$ & $82.37$ \\

	\midrule
	Elastic regist.   & $\textbf{96.47}$ & $91.18$ & $\textbf{100}$ & $\textbf{99.56}$ & $96.91$ & $96.91$\\
	\bottomrule		\end{tabular}

	\vspace{4pt} 	\begin{tabular}{@{}lcccc@{}} 	\toprule
	Other 	 & BoC 									   & Spatial BoC 							     & Hybrid BoW & Tabia et al. \\
	methods	 & ~\cite{tabia2014covariance, tabia2015covariance} &~\cite{tabia2014covariance, tabia2015covariance}  & \cite{lavoue2012combination} & \cite{tabia2011new} \\
	\midrule
	1NN  & 93.25										  & 92.5 &  91.8 & 85.3\\
	\bottomrule		\end{tabular}

}
\end{table}

\subsubsection*{Generic 3D shape classification. }

We apply the proposed approach to the important problem of classifying generic 3D shapes. We use  the SHREC07 watertight 3D model benchmark~\cite{giorgi2007shape}, which is composed of $400$ watertight 3D models evenly divided into $20$ shape classes. We only consider the $13$ classes that are composed of genus-0 surfaces. First, we compute spherical parameterizations~\cite{kurtek2013landmark}; then we map them to $\Space{Q}$. To classify a shape, we use:
\begin{itemize}
	\item Leave-One-Out $k$-nearest  neighbor classifier  using the elastic metric defined on the shape space ${\cal S}  = Q({\cal F})/{\cal G}$  (kNN-SRNF).
	\item The Gaussian classifier (Gauss-SRNF) on principal components, in 	the space of SRNFs as described in Section~\ref{sec:analysis}. We
	use the Leave-One-Out (LOO) strategy. 	We  classify each query to the 	shape category whose mean is the closest to the query in terms of 	Mahalanobis distance.
\end{itemize}

\noi We also compare our results to those obtained using:
\begin{itemize}
\item

The Gaussian classifier and the kNN classifier, but in the surface space $\mathcal{F}$, with and without elastic registration using the proposed metric;

\item

The iterative closest point (ICP) algorithm~\cite{besl:1992};

\item

The covariance descriptors~\cite{tabia2014covariance,tabia2015covariance}, the hybrid Bag of Words~\cite{lavoue2012combination} and the approach of~\cite{tabia2011new}.

\end{itemize}

\noi Table~\ref{tab:shrec2007_results} reports the classification rates, and shows that the best performance is attained by the proposed kNN-SRNF and Gauss-SRNF classifiers. The results suggest that the parametrization-invariant metric and probability models in $\Space{Q}$ improve surface matching, resulting in better 3D shape classification.

\subsubsection*{ADHD classification}

\begin{figure}[!ht]
    \centering
        {\includegraphics[width=.45\textwidth]{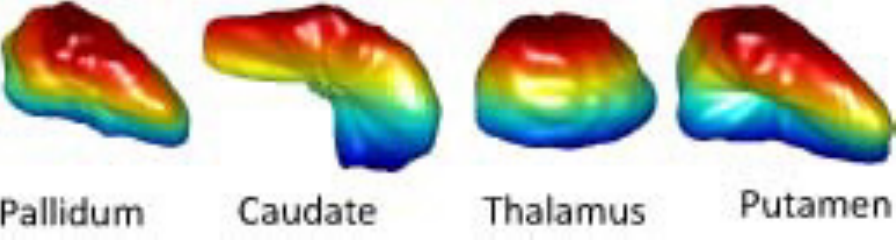} }
      \caption{Left anatomical structures in the brain.}
    \label{fig:braindata}
\end{figure}

Finally, we apply our methods to shape-based  diagnosis of attention deficit hyperactivity disorder (ADHD) using MRI scans. The surfaces of brain structures used here were extracted from T1 weighted brain MR images of young adults aged between 18 and 21. These subjects were recruited from the Detroit Fetal Alcohol and
Drug Exposure Cohort \cite{jacobson:2004,burden:2010}. Among the  
34 subjects, 19 were diagnosed with ADHD and the remaining 15 were controls (non-ADHD). Some examples of left structures are shown in Fig.~\ref{fig:braindata}. First we register the extracted surfaces as described in~\cite{jermyn:2012}, then map them into $\Space{Q}$ using $Q$. In order to distinguish ADHD and control samples, we use the Gaussian classifier on principal components  $\Space{Q}$   as described in Section~\ref{sec:analysis}.
\begin{table}[t]
	\ra{.8} 	\caption{Classification performance, in  $\%$,  for six different 	techniques.} 	\label{tab:classification} 	\centering
	{ 		\begin{tabular}{@{}l@{}cccccc@{}} 			\toprule
			& SRNF    & SRM                    & SRM  & Har-& ICP& SPHARM  \\
			& Gauss  &  Gauss		   & NN       & monic                 &
			\cite{besl:1992}  & PDF~\cite{styner:2006}  \\
			& {\bf Proposed} & \cite{kurtek-ipmi:2011} & ~\cite{kurtek:2010}
			&  & &  \\
			\midrule
			L. Caudate & \textbf{67.7}	& -  		& 41.2	& 64.7	& 32.4	& 61.8 \\
			L. Pallidus & 85.3	& \textbf{88.2} & 76.5	&79.4	& 67.7	& 44.1 \\
			L. Putamen & \textbf{94.1}	&82.4  	& 82.4	& 70.6	& 61.8	& 50.0 \\
			L. Thalamus & \textbf{67.7}	& -	& 58.8	& \textbf{67.7}	& 35.5	& 52.9 \\
			\midrule
			R. Caudate & 55.9	& -	& 50.0	& 44.1	& 50.0	& \textbf{70.6} \\
			R. Pallidus & \textbf{76.5} 	& 67.6 	& 61.8	& 67.7	& 55.9	& 52.9 \\
			R. Putamen & 67.7	&\textbf{82.4} 	 	& 67.7	& 55.9	& 47.2	& 55.9 \\
			R. Thalamus & \textbf{67.7}	&- 	& 58.8	& 52.9	& 64.7	& 64.7 \\
			\bottomrule 		\end{tabular} 	}
\end{table}

Table~\ref{tab:classification} shows the  LOO nearest neighbor classification rate of our approach compared to five other state-of-the-art techniques~\cite{kurtek-ipmi:2011,kurtek:2010,besl:1992,styner:2006}.  The best performance is attained using the proposed SRNF Gaussian classifier between left putamen surfaces. These results suggest that the parametrization-invariant metric and the probability models in our approach provide improved matching and modeling of the surfaces, resulting in a superior ADHD classification. In summary, our method is not only more efficient---the computational cost is an order of magnitude less than SRM and related ideas---it also provides significantly-improved classification performance.

\section{Conclusions}

The SRNF representation is potentially an important tool in the statistical shape analysis of surfaces. Since SRNF space is a vector space with the $\ltwo$-metric, statistical analysis in this space is simple and low-cost compared to previous methods that performed analysis directly in surface space. Use of the SRNF has been limited, however, by the absence of a method for reconstructing the corresponding surface. We have introduced methods for approximating this inverse mapping, thereby removing the obstacle. As a result, by adopting the proposed framework, the computational cost of algorithms for the statistical analysis of surface shape can be reduced by an order of magnitude. Experimental results show the use of the proposed framework in various 3D shape analysis tasks, such as computing geodesics, transferring deformations, computing statistical shape summaries, and fitting and sampling from probabilistic models. We have also demonstrated that the method achieves state-of-the-art performance in the classification of generic 3D shapes, and in the analysis of brain structures arising from ADHD data.

\ifCLASSOPTIONcompsoc
  \section*{Acknowledgments}
\else
  \section*{Acknowledgment}
\fi

The authors would like to thank: Nils Hasler for providing us with the 3D human shape models;  Sebastian Kurtek for sharing programs for registration of SRNFs under the $\ltwo$ metric; and  Raif Rustamov and Maks Ovsjkanikov for the discussion about functional maps and for sharing their data and results.

\ifCLASSOPTIONcaptionsoff
  \newpage
\fi

\bibliographystyle{IEEEtran}
\bibliography{references}

\end{document}